\documentclass[11 pt,a4paper]{article}
\usepackage{amsmath}
\usepackage[pdftex]{graphicx}
\usepackage{epsfig}
\usepackage{epstopdf}
\graphicspath{{pictures/}}
\usepackage{geometry}
\usepackage{cite}
\usepackage{amssymb}
\usepackage{amsfonts}
\usepackage{amsthm}
\usepackage{hyperref}
\date{}
\begin{document}
	\title{{\bf  Restrain  the losses of the entanglement and the non-local advantage of quantum coherence for accelerated quantum systems}}
	\author{A. R. Mohammed $ ^{1} $\footnote {E-mail a.radwan@azhar.edu.eg}~T. M. El-Shahat $ ^{1} $\footnote {E-mail el\_shahat@yahoo.com}~and N. Metwally $ ^{2,3} $\footnote {E-mail nmetwally@gmail.com}\\
		$^{1}$Math. Dept., Faculty of Science, Al-Azhar University, Assiut 71524, Egypt.\\
		$^{2}$Math. Dept., Faculty of Science, Aswan University, Aswan 81528, Egypt.\\
$^{3}$ Department of Mathematics.,
	College of Science, University of Bahrain, \\ P. O. Box 32038
	Kingdom of Bahrain	}


	\maketitle
	{\large \centerline{{\bf Abstract}}}
We examined the possibility of  recovering the losses of entanglement
and the non-local  advantage by using the local symmetric operations.
The improvement efficiency  may be increased by applying the symmetric operations on both qubits.
 The recovering process of both phenomenon is exhibited clearly when only one qubit is accelerated and the symmetric operations
is applied on both qubits. It is shown that, for large acceleration, the non-local coherent
advantage may be re-birthed by using these symmetric operations.

	\vspace{1 mm}
	
	\section{Introduction}\label{cohere1}
	Quantum coherence is one of the  most essential principles of quantum physics  that  plays a crucial role in many promising fields such as quantum biology \cite{cohe1,cohe2} and quantum thermodynamics \cite{cohe3,cohe4}.
Some  theoretical framework  concerning the  measure of  coherence of  quantum states  are given in \cite{cohe5,cohe6,cohe7,cohe8}.
Specifically,  Baumgratz et al.  \cite{cohe5} introduced  some criteria that must be satisfied by    any  good measure  of coherence.

In this contribution, we  consider  the $ l_{1} $-norm and the relative entropy of coherence as a measure of quantum coherence, where these measures satisfy the criteria suggested by  Baumgratz et al  \cite{cohe5}.
 For a combined system $AB$, we quantify  the non-local  advantage  $\mathcal{N}_{La}$ on party $B$ by local measurements on party $A$ and
 classical communication between its two parties \cite{cohe9}. However, it is well known that the Unruh effect on the accelerated systems causes degradation of entanglement \cite{cohe10} and consequently, the coherence between the accelerated subsystems decreases.  Moreover, the decay rate of the coherence increases  if the accelerated systems are subject to extra noise.  Recently, researchers have been interested in investigating the behavior of accelerated systems in different types of noise.   For example, Metwally \cite{cohe11} showed that there is   a more robust use of the generic pure state than the self-transposed classes and the optimum  communication between the users for small values of the acceleration is investigated in \cite{cohe12}.\\

In our approach, to investigate the possibility of enhancing  the non-local  coherent advantage of the accelerated system, we use  the non-Hermitian local symmetric operator, ($\mathcal{PT}$)  \cite{cohe13}, where it has been shown, that this operator can be used to minimize the losses of entanglement \cite{cohe14}. Also,  Guo et al \cite{cohe15} used this symmetric operator  to suppress decoherence and enhance the parameter estimation precision.  So, in this contribution, we assume  that a combined  system, initially prepared in a maximum entangled state of Bell type, is accelerated: either one or both subsystems are accelerated. Due to the acceleration the decoherence phenomena appears and consequently the entanglement degraded. We estimate the loss of   the non-local coherent advantage and  by means of the negativity we quantify the entanglement. The possibility of protecting  and minimizing the decay of these two phenomena are achieved by using the symmetric operator, where different possibilities are considered.

The paper is organized as follows. In Sec. \ref{cohere2}, we  describe briefly the acceleration process of the initial system, where it is assumed that either one or both subsystems are accelerated. The amount of the survival amount of entanglement is quantified  using the negativity measure.  The mathematical form of the non-local coherent advantage is introduced in Sec. \ref{cohere2.2}, where, we  investigate the decay behavior of this phenomena due to the acceleration process.  The improving process of the entanglement and the non-local coherent advantage is discussed in Sec. \ref{cohere3}, where  different scenarios are introduced. In Sec. \ref{cohere4}, we investigate numerically, the possibility of improving the entanglement of the accelerated systems. Sec. \ref{cohere5}, is devoted to discuss the behavior of the non-local coherent advantage in the presence of the symmetric operators, where it is shown that by controlling on the operator strength, one can protect the loss of this phenomena. Finally, our results are summarized in Sec. \ref{cohere6}.

\section{The  system and its evaluation}\label{cohere2}
\subsection{Acceleration Process}\label{cohere2.1}
Assume that, we have  a quantum system  initially prepared in a maximum entangled state of Bell type as,
\begin{eqnarray}
|\psi_{0}\rangle=\frac{1}{\sqrt{2}}(|0_{A}0_{B}\rangle+|1_{A}1_{B}\rangle).
\end{eqnarray}
It is assumed that, either one or both subsystems are accelerated. The acceleration process is performed with the  computational basis, $ |0_{R}\rangle $ and $ |1_{R}\rangle $, in the Minkowski space $(t, z)$ of the qubit state  is   transformed into the
Rindler space  $(\tau, x)$  using the transformations  \cite{cohe17,cohe18}.
\begin{equation}\label{trans}
\tau=r~tanh\left(\frac{t}{z}\right), \quad x=\sqrt{t^2-z^2},
\end{equation}
where  $-\infty<\tau<\infty$, $-\infty<x<\infty$  and $r$ is the
acceleration of the moving particle.
Note that,  the  Minkowsik coordinates $(t,z)$  and   Rindler coordinates $(\tau, x)$ are used   to describe Dirac field, in the inertial and  non-inertial frames, respectively.
The relations (\ref{trans})
describe  two regions in Rindler's spaces: the first region $I $
for $t>|z|$  and the second region $II$ for $t<-|z|$ \cite{PhysRevLett.91.180404}. Therefore the basis $ |0_{R}\rangle $ and $ |1_{R}\rangle $, can be transformed as,
\begin{eqnarray}
\begin{aligned}
|0_{R}\rangle & =\cos(r)|0_{I}\rangle|0_{II}\rangle+\sin(r)|1_{I}\rangle|1_{II}\rangle,\\
|1_{R}\rangle &=|1_{I}\rangle|0_{II}\rangle.
\end{aligned}	
\end{eqnarray}
 If only the first qubit is accelerated, then the final accelerated state in the first region
$I$  is given by,
\begin{eqnarray}
\rho(0)_{A,I_{acc}}=\left(
\begin{array}{cccc}
\frac{\cos ^2(r)}{2} & 0 & 0 & \frac{\cos (r)}{2} \\
0 & \frac{\sin ^2(r)}{2} & 0 & 0 \\
0 & 0 & 0 & 0 \\
\frac{\cos (r)}{2} & 0 & 0 & \frac{1}{2} \\
\end{array}
\right).
\end{eqnarray}
Similarly, if the two subsystems are accelerated, then the final accelerated state in the first region $I$ is given by
\begin{eqnarray}
\rho(0)_{A_{acc},I_{acc}}=\left(
\begin{array}{cccc}
\frac{\cos ^4(r)}{2} & 0 & 0 & \frac{\cos ^2(r)}{2} \\
0 & \frac{1}{8} \sin ^2(2 r) & 0 & 0 \\
0 & 0 & \frac{1}{8} \sin ^2(2 r) & 0 \\
\frac{\cos ^2(r)}{2} & 0 & 0 & \frac{1}{2} \mu \\
\end{array}
\right),
\end{eqnarray}
where $ \mu=(\sin ^4(r)+1) $.

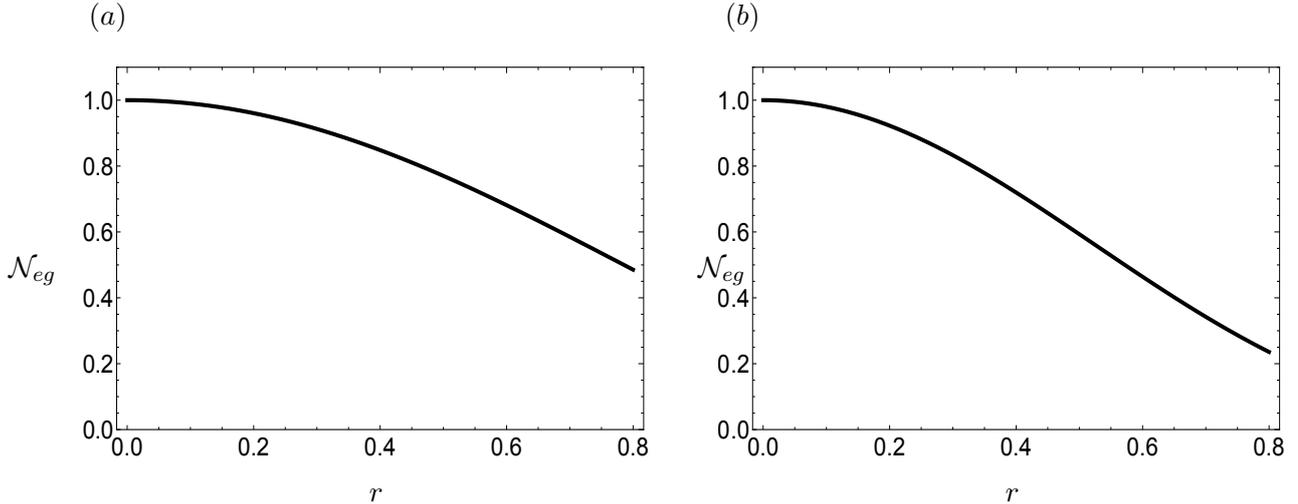
\begin{figure}[!h]
	\begin{center}
\includegraphics[width=0.45\textwidth, height=150px]{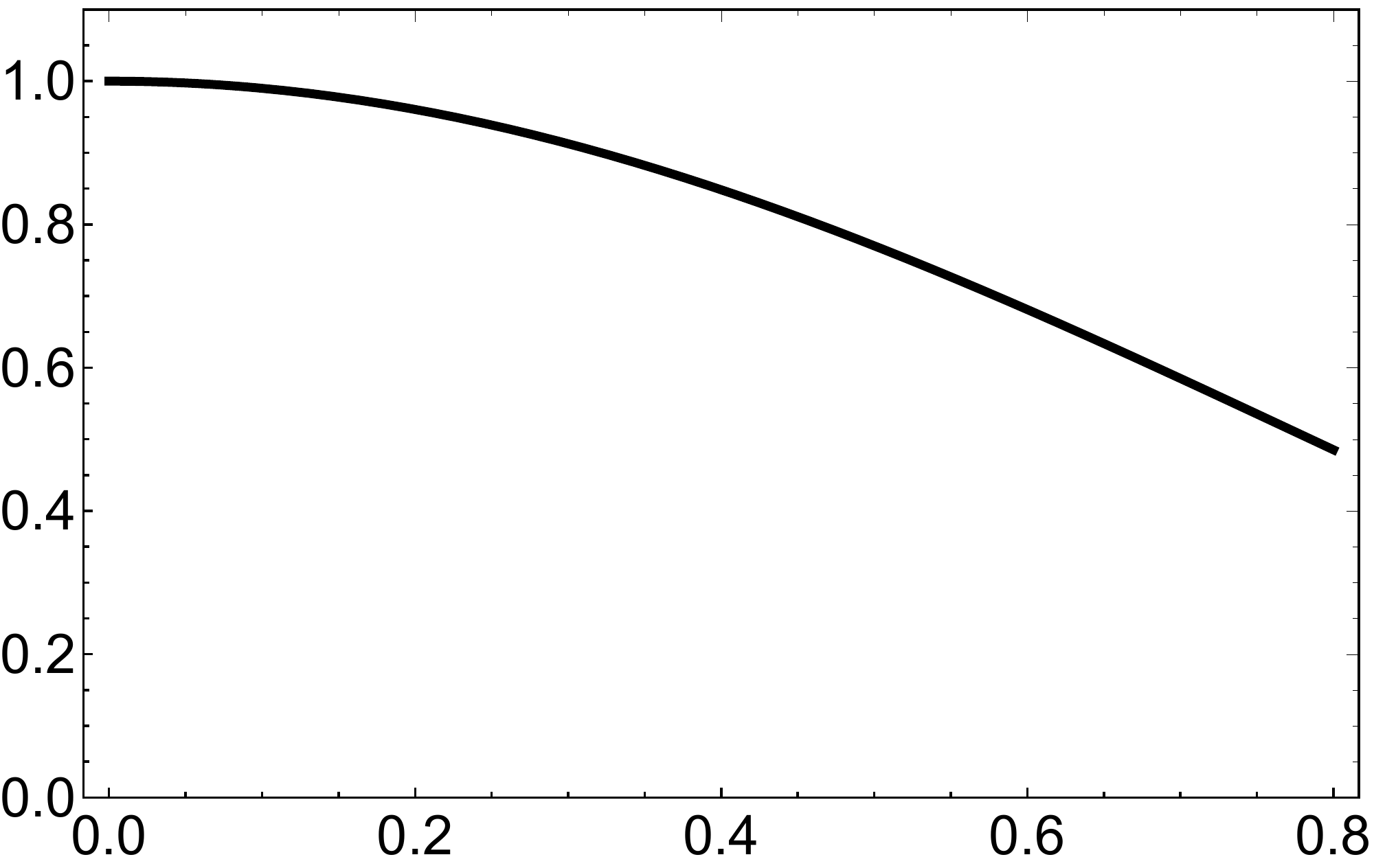}\put(-210,165){($ a $)}\put(-240,70){$\mathcal{N}_{eg}$}
		\put(-105,-15){$r$}~~~~~~
\includegraphics[width=0.45\textwidth, height=150px]{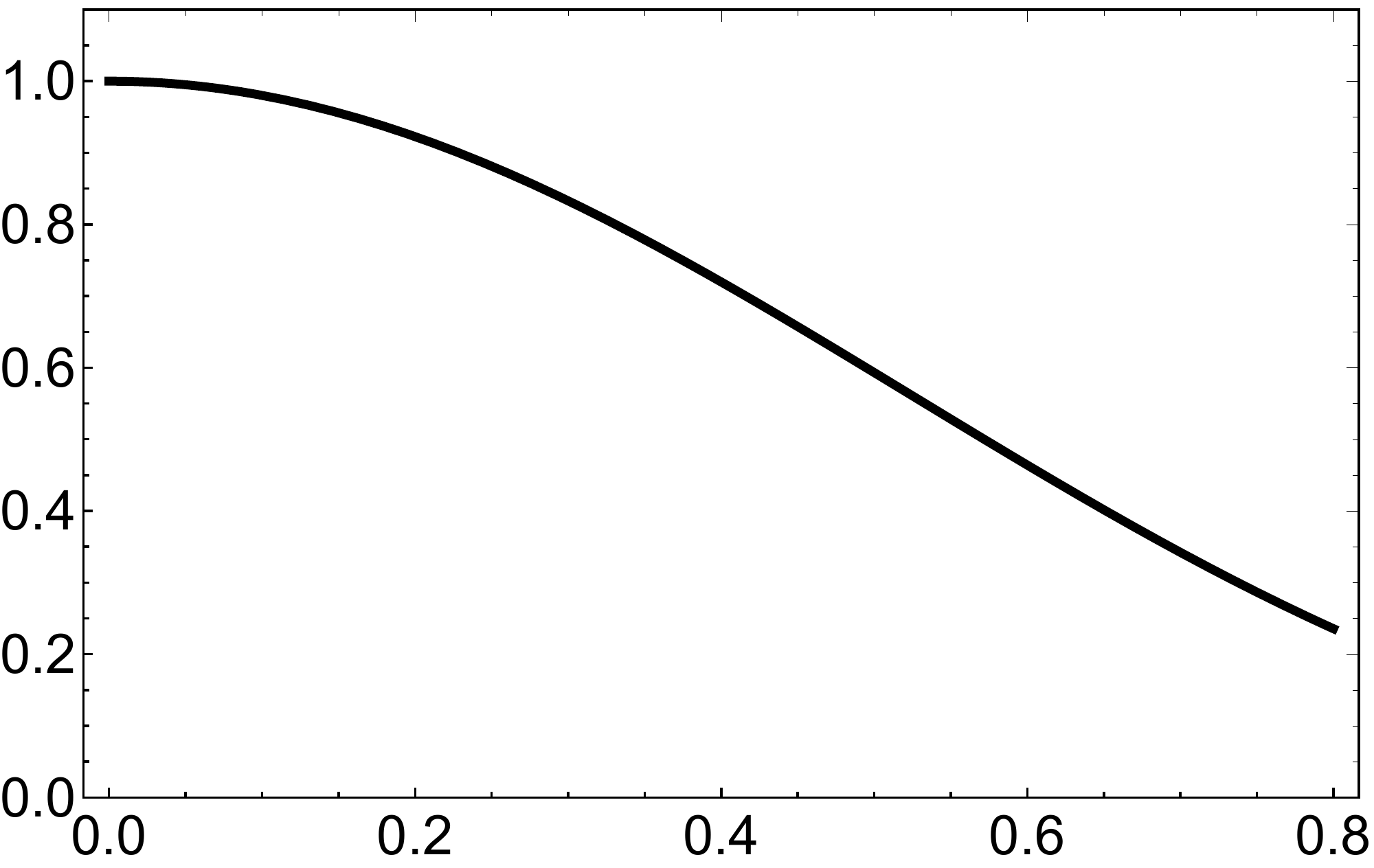}\put(-210,165){($ b $)}\put(-220,70){$\mathcal{N}_{eg}$}
		\put(-105,-15){$r$}
	\end{center}
	\caption{\label{coo1} In the absence of the $\mathcal{PT}$-local, the behavior of the negativity, $\mathcal{N}_{eg}$ of the accelerated  system    is described in Fig(a),(b), when Alice'qubit or both qubits are accelerated, respectively. }
\end{figure}
To quantify the amount of entanglement between the two subsystems, we use one of the most common quantitative  measures of entanglement, namely, the negativity \cite{cohe23,cohe21}.
 Mathematically it is defined by,
\begin{equation}
\mathcal{N}_{eg}=max[0,-2\min\{\lambda_{\mu}\}],
\end{equation}
where $\{{\lambda_{\mu}}\}, ~\mu=1..4$ represent the eigenvalues of $ \rho_{ab}^{T_{2}}$ and $ T_{2} $ refers to the partial transposition for the second subsystem. For a maximally-entangled  state the negativity is maximum, i.e.,  $\mathcal{N}_{eg} = 1$, while for the product  states,  $\mathcal{N}_{eg} = 0$ \cite{cohe21,cohe22}.
 Fig.(1),  shows the behavior of the survival amount of entanglement by means of negativity: at zero acceleration $(r=0)$, the negativity is maximum, $\mathcal{N}_{eg}=1$. As the acceleration increases, the negativity decreases to reach its minimum values at $r=0.8$.
It is  noted  from Figs.(1a) and (1b) that,  the decay rate of negativity  relatively increases, when both qubits are accelerated.

\subsection {The  non-local  advantage, $\mathcal{N}_{La}$}\label{cohere2.2}
We start by recalling one of the  favourite measures of coherence, namely,  the one norm coherence   $C_{l_{1}}^{na1}$.  For any density operator $\rho_{ij}$ it is defined on the basis  $ {|i\rangle}_{i=1}^{d} $ as,
\begin{eqnarray}
\begin{aligned}
C_{l_{1}}^{na1}&=\sum_{i\neq j}|\langle i|\rho_{ij}|j \rangle|.
\end{aligned}
\end{eqnarray}
Let us assume that the two users Alice and Bob share two qubit- state, $\rho_{AB}$. Let, t Alice who hold the first qubit $(A)$, performs a local measurements  $ \prod_{i}^{a}=\frac{1}{2}(I_{2}+(-1)^{a}\sigma_{i}) $ on  her qubit $A$ and informs Bob of her randomly selected observable $ \sigma_{i} $  and the measurement results $ a\in\{0,1\} $, where $ \sigma_{i}$ being one of the Pauli operators  $ \sigma_{x,y,z}$ and $ I_{2} $ is the 2-dimensional identity operator. By averaging over the three possible measurements of Alice and the corresponding eigenbases choosen by Bob,
 Mondal et al \cite{cohe19} derived the criterion for achieving the non-local advantage  of  quantum coherence ( NAQC)
\begin{eqnarray}\label{coh2}
\tilde{C}_{\alpha}^{na}=\frac{1}{2}\sum_{\substack{i,j,a\\i\neq j}}p_{a|\prod_{i}^{a}}C_{\alpha}^{\sigma_{j}}(\rho_{B|\prod_{i}^{a}})>C_{\alpha}^{m},
\end{eqnarray}
where $ C_{\alpha}^{\sigma_{j}}(.) $($ \alpha=l_{1} $ ) represents the quantum coherence with respect to the reference basis spanned by the eigenstates of $ \sigma_{j} $, and the two critical values are given by $ C_{l_{1}}^{m} =\sqrt{6}$.
 The probability for Alice's measurement is $p_{a|\prod_{i}^{a}}=Tr[(\prod_{i}^{a}\otimes I_{2})\rho_{AB}]$, and $\rho_{B|\prod_{i}^{a}} $ is the corresponding postmeasurement state defined as,
\begin{equation}
\rho_{B|\prod_{i}^{a}}=\frac{Tr_{A}[(\prod_{i}^{a}\otimes I_{2})\rho_{AB}] }{p_{a|\prod_{i}^{a}}}.
\end{equation}
  Based on the criterion of equation \ref{coh2}, Hu et al\cite{cohe20} proposed to characterize quantitatively the degree of  $\mathcal{N}_{La}$
of   bipartite state  as,
\begin{equation}
\mathcal{N}_{La}(\rho_{AB})=\max\{0,\frac{\tilde{C}_{\alpha}^{na}(\rho_{AB})-C_{\alpha}^{m}}{\tilde{C}_{\alpha,\max}^{na}-C_{\alpha}^{m}}\},
\end{equation}
where $\tilde{C}_{\alpha,\max}^{na}=\max_{\rho_{AB}\in D(C^{d\times d})}\tilde{C}_{\alpha}^{na}(\rho_{AB})$ and  for the
two-qubit states we have  $ \tilde{C}_{\alpha,\max}^{na}=3 $.\\

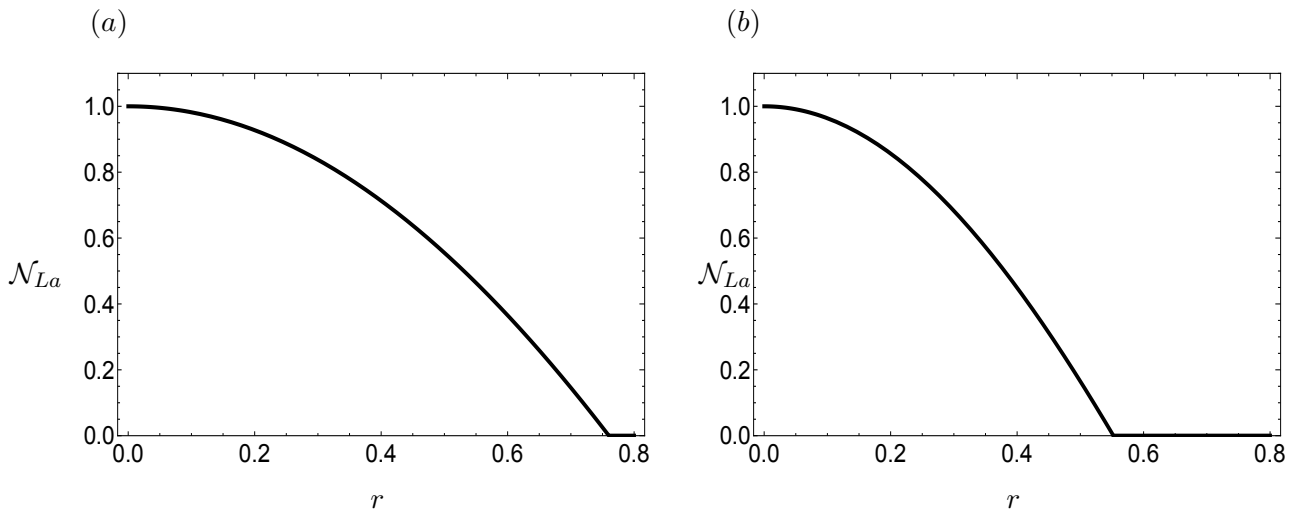
\begin{figure}[!h]
	\begin{center}
		\includegraphics[width=0.45\textwidth, height=150px]{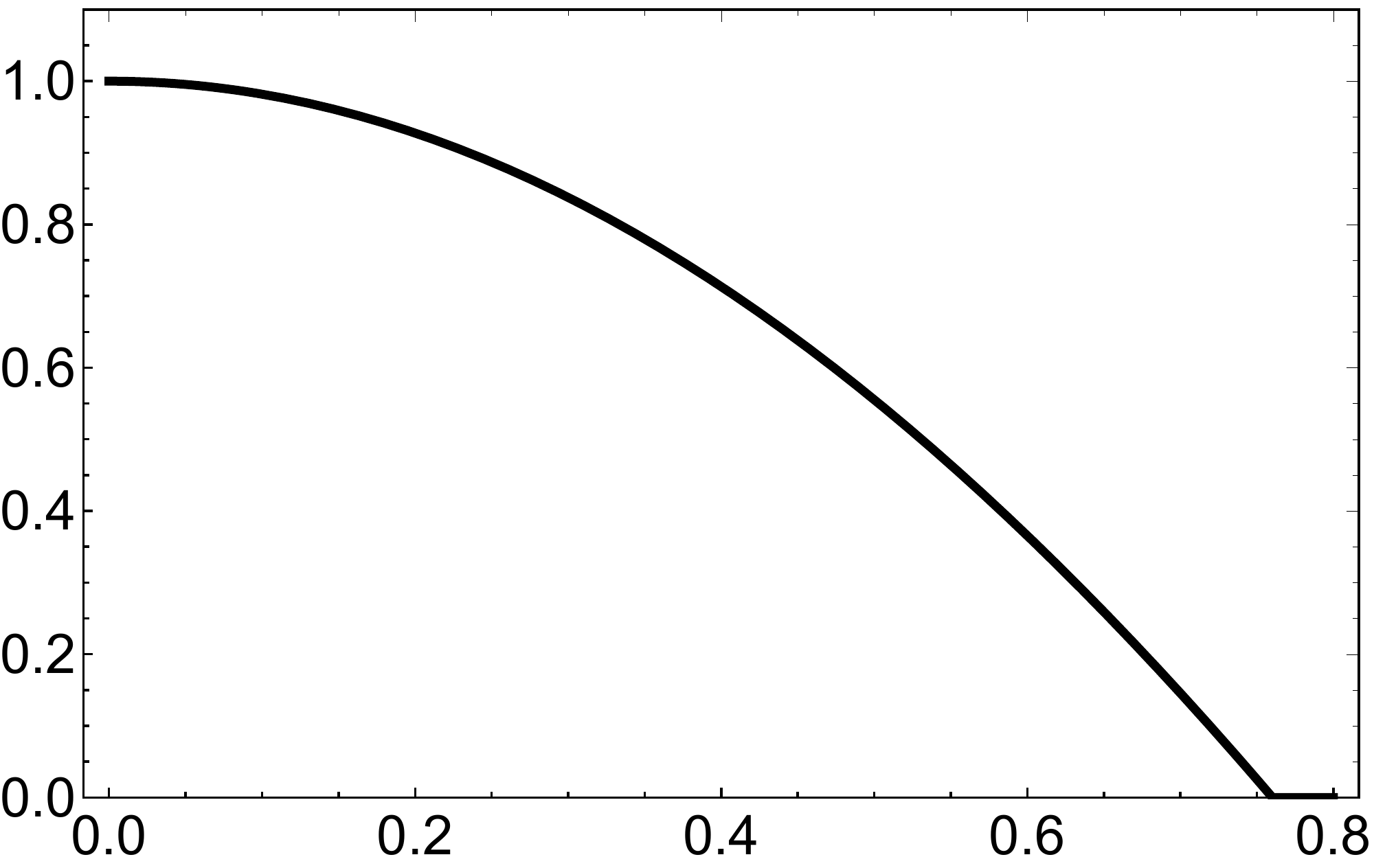}\put(-210,165){($ a $)}\put(-240,70){$\mathcal{N}_{La}$}
		\put(-105,-15){$r$}~~~~~~
		\includegraphics[width=0.45\textwidth, height=150px]{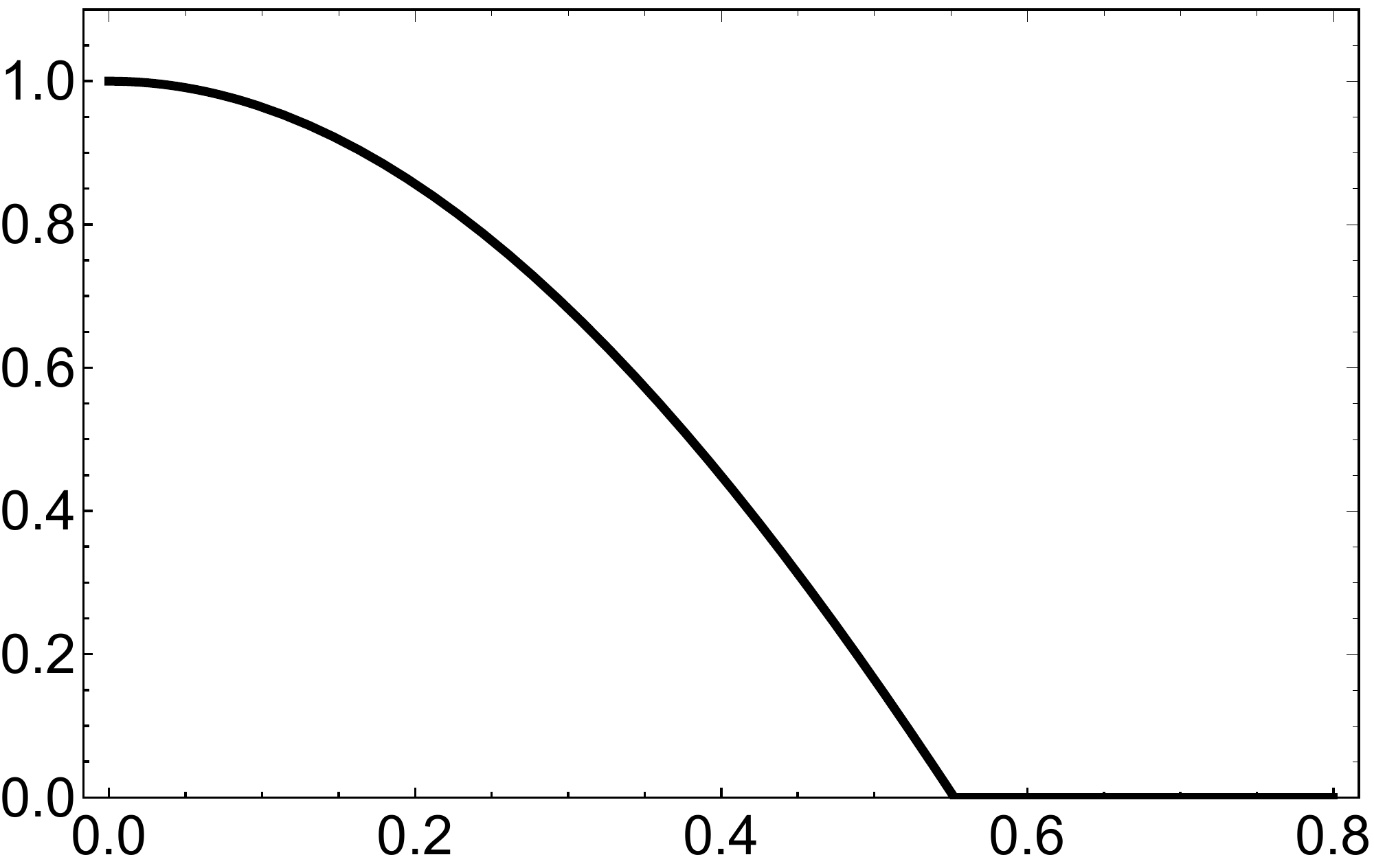}\put(-210,165){($ b $)}\put(-220,70){$\mathcal{N}_{La}$}
		\put(-105,-15){$r$}
	\end{center}
	\caption{\label{coo2} The behavior of the non-local advantage , $\mathcal{N}_{La}$ of the accelerated  system (a) only qubit system is accelerated and (b)the two  qubits are accelerated, respectively. }
\end{figure}

In Fig.(\ref{coo2}), we plot the non-local  advantage $\mathcal{N}_{La} $ for the accelerated system in the absence of the local $\mathcal {PT}$-symmetric operator. As seen  $\mathcal{N}_{La} $, decreases as the acceleration increases. However, when only one qubit  is accelerated,(see Fig.(\ref{coo2}a)) $\mathcal{N}_{La} $  decreases gradually in Fig.(\ref{coo2}b),  it decays faster if both qubits are accelerated  and   vanishes  at small values of acceleration, $r>0.55$ and  survives   for  large acceleration if only one qubit is accelerated.

\section{The evaluation of the system }\label{cohere3}

In this section, we use the  $\mathcal{PT}$-symmetric operator \cite{cohe13} to recover   the losses of quantum entanglement and the non local advantage , as well as, the possibility of restraining  their decay, $\mathcal{N}_{La}$. This  symmetric operator, $\mathcal{PT}$  is defined as,
	\begin{eqnarray}
	\mathcal{H}_{\mathcal{PT}}=\left(
	\begin{array}{cc}
	i \sin (\alpha ) & 1 \\
	1 & -i \sin (\alpha ) \\
	\end{array}
	\right),
	\end{eqnarray}
where $ \alpha $ is the real number characterizes the non-Hermitically  if  $ \rvert \alpha  \lvert <\frac{\pi}{2} $, while $\alpha=0$ corresponds to  $ H_{\mathcal{PT}} $ Hermitian.	The time-evolving operator of $  \mathcal{H}_{\mathcal{PT} }$ according to non-Hermitian quantum theory \cite{cohe13,cohe14} is written as,
		\begin{eqnarray}\label{coh1}
U_{\mathcal{PT}}=\exp(-i\mathcal{H}_{\mathcal{PT}}t)=\sec (\alpha ) \left(
\begin{array}{cc}
\cos (\alpha -\alpha_{1}) & -i \sin (\alpha_{1}) \\
-i \sin (\alpha_{1}) & \cos (\alpha +\alpha_{1}) \\
\end{array}
\right),
	\end{eqnarray}
	where $\alpha_{1}=t \cos (\alpha ) $.  Now, we consider the following cases:
\begin{enumerate}

\item{\underline {\it Only one qubit is accelerated}}

Here we assume that, the users Alice and Bob share the state(3), where only one qubit is accelerated. If the symmetric operator is applied on  the first qubit only, then the output state is defined as:

\begin{eqnarray}
\rho(t)&=&\frac{(U_{\mathcal{PT}}\otimes I_{B})\varrho_{A,I_{acc}}(U_{\mathcal{PT}}\otimes I_{B})^{\dagger}}{Tr[(U_{\mathcal{PT}}\otimes I_{B})\varrho_{A,I_{acc}}(U_{\mathcal{PT}}\otimes I_{B})^{\dagger}]}
\end{eqnarray}
where  the final density operator $ \rho(t) $ on the computational basis  is defined by the elements,
\begin{equation}
\begin{aligned}[c]
\rho_{11}&=\frac{\delta^2}{2Z} \cos ^2(r) ,\\
\rho_{12}&=\frac{i \nu\delta}{2Z}  \cos (r) ,\\
\rho_{13}&=\frac{i \nu\delta }{2Z} \cos ^2(r) ,\\
\rho_{14}&=\frac{\beta \delta}{2Z} \sec (\alpha ) \cos (r) ,\\
\rho_{22}&=\frac{1}{2Z} (\nu^{2}+\sin ^2(r) \delta^2),\\
\rho_{23}&=\frac{\nu^{2}}{2Z}  \cos (r),
\end{aligned}
\qquad
\begin{aligned}[c]
\rho_{24}&=-\frac{ i \nu}{2Z} (\sec (\alpha ) \beta-\sin ^2(r) \delta),\\
\rho_{33}&=\frac{\nu^{2}}{2Z}  \cos ^2(r),\\
\rho_{34}&=-\frac{i\beta}{2Z}  \sec ^2(\alpha ) \sin (\alpha_{1}) \cos (r) ,\\
\rho_{44}&=\frac{1}{2Z} \sec ^2(\alpha ) (\beta ^2+\sin ^2(\alpha_{1}) \sin ^2(r)),\\
Z&=\sec ^2(\alpha )-\tan ^2(\alpha ) \cos (2 \alpha_{1} ),\\
\rho_{21}&=\rho_{12}^{\ast},~~\rho_{31}=\rho_{13}^{\ast},~~\rho_{32}=\rho_{23},~~\rho_{41}=\rho_{14},~~
\rho_{42}=\rho_{24}^{\ast},\\
\rho_{43}&=\rho_{34}^{\ast},~~\delta=\tan (\alpha ) \sin (\alpha_{1})+\cos (\alpha_{1}),\\
\beta&=\cos (\alpha +\alpha_{1})~~\nu=\sec (\alpha ) \sin (\alpha_{1}).
\end{aligned}
\end{equation}
The second possibility is  applying the symmetric-operator ( \ref{coh1}) on both qubits. In this case, the final output state  is defined as,
\begin{eqnarray}
\rho(t)&=&\frac{(U_{\mathcal{PT}}\otimes U_{\mathcal{PT}})\varrho_{A,I_{acc}}(U_{\mathcal{PT}}\otimes U_{\mathcal{PT}})^{\dagger}}{Tr[(U_{\mathcal{PT}}\otimes I_{B})\varrho_{A,I_{acc}}(U_{\mathcal{PT}}\otimes I_{B})^{\dagger}]}
\end{eqnarray}
where its elements are given by,
\begin{eqnarray}
\rho_{11}&=&\frac{1}{2Z} (\xi^{2}\sec ^2(\alpha ) +\xi^{2}\sin ^2(\alpha )  \cos ^2(r)+\xi^{2}\tan ^2(\alpha )  (\sin ^2(r)-2 \cos (r))+\cos ^2(\alpha_{2}) \cos ^2(r)),
\nonumber\\
\rho_{12}&=&\frac{\xi}{Z}\tan (\alpha )  \sin ^2(\frac{r}{2}) (\cos (\alpha_{2})+i (2 \sec ^2(\alpha )-1) \sin (\alpha_{2})),
\nonumber\\
\rho_{13}&=&\frac{\xi}{4Z} \tan (\alpha )  (e^{-i \alpha_{2}} (-2 \cos (r)+\cos (2 r)+1)-4 i \sec ^2(\alpha ) \sin (\alpha_{2})\omega),
\nonumber\\
\rho_{14}&=&\frac{1}{2Z} (-2\xi^{2} \tan ^2(\alpha ) +\cos (r) (\xi^{2} (\sin ^2(\alpha )+\sec ^2(\alpha ))+\cos ^2(\alpha_{2}))-i\xi (\sec (\alpha ) \cos (\alpha_{2}) \sin ^2(r))),
\nonumber\\
\rho_{21}&=&\frac{1}{2Z} \tan (\alpha ) \xi \sin ^2(\frac{r}{2}) (2 \cos (\alpha_{2})+i \xi\sec (\alpha ) (\cos (2 \alpha )-3) ),
\nonumber\\
\rho_{22}&=&\frac{1}{2Z} \sin ^2(r) (\xi^{2}\sin ^2(\alpha )+\cos ^2(\alpha_{2}))+2 \tan ^2(\alpha ) \xi^{2} \sin ^4(\frac{r}{2}),
\nonumber\\
\rho_{23}&=&\frac{\xi}{2Z} \sec (\alpha )  (-2 \tan ^2(\alpha ) \sin (\alpha_{2})\omega+i \cos (\alpha_{2}) \sin ^2(r)),
\nonumber\\
\rho_{24}&=&\frac{1}{4Z} \tan (\alpha ) \xi \sin ^2(\frac{r}{2}) (-4 \cos (\alpha_{2}) (\cos (r)+2)-4 i \sin (\alpha_{2}) (\cos (r)-2 \tan ^2(\alpha ))),
\nonumber\\
\rho_{31}&=&\frac{\xi}{4Z} \tan (\alpha )  (4 i \xi\sec (\alpha ) \omega+e^{i \alpha_{2}} (-2 \cos (r)+\cos (2 r)+1)),
\nonumber\\
\rho_{33}&=&\frac{\xi^{2}}{2Z} \sec ^2(\alpha )\sin ^2(\frac{r}{2}) (\cos (2 \alpha )\omega+\cos (r)+3),
\nonumber\\
\rho_{34}&=&\frac{i\xi}{Z} \tan (\alpha )  \sin ^2(\frac{r}{2}) ((2 \sec ^2(\alpha )-1) \sin (\alpha_{2})+i \cos (\alpha_{2})),
\nonumber\\
\rho_{43}&=&\frac{i\xi}{Z} \tan (\alpha )  \sin ^2(\frac{r}{2}) (-2 \xi\sec (\alpha ) +\sin (\alpha_{2})+i \cos (\alpha_{2})),
\nonumber\\
\rho_{44}&=&\frac{1}{4Z} (2 \cos ^2(\alpha_{2})+\xi^{2} (4 \sec ^2(\alpha )+ (-4 \tan ^2(\alpha ) \cos (r)+\cos (2 r)-5)+2\sin ^{2}(\alpha_{2}))),
\nonumber\\
\rho_{32}&=&\rho_{23},~~\rho_{41}=\rho_{14}^{\ast},~~\rho_{42}=\rho_{24}^{\ast},~~Z=8 \tan ^2(\alpha ) \xi^{2} \sin ^2(\frac{r}{2})+1,
\nonumber\\
\alpha_{2}&=&t cos^2(\alpha ),~~\xi=\sec (\alpha ) \sin (\alpha_{2}),~~\omega= (\cos (r)-1).
\end{eqnarray}

\item {\underline{\it Both qubits are accelerated:}}
In the second case, we consider that both subsystems are
accelerated, namely the users share the state (4). If the symmetric operator  ( \ref{coh1}) is applied on one qubit only, then the final output state is defined as $4\times 4$ matrix, where its elements are given by,
 \begin{eqnarray}
 \begin{aligned}
 \rho_{11}&=\frac{1}{8Z} (\nu^{2} \sin ^2(2 r)+4 \cos ^4(r) \delta^2),~\quad
  \rho_{12}=\frac{i \delta\nu}{2Z}  \cos ^2(r) ,\\
  \rho_{13}&=-\frac{ i \nu}{8Z} (\sec (\alpha ) \sin ^2(2 r) \beta-4 \cos ^4(r) \delta),~\quad
   \rho_{14}= \frac{ \beta \delta}{2Z} \sec (\alpha ) \cos ^2(r),\\
   \rho_{22}&= \frac{1}{8Z} \sec ^2(\alpha ) (\sin ^2(2 r) \cos ^2(\alpha -\alpha_{1})+4 \sin ^2(\alpha_{1}) \mu),~\quad
    \rho_{23}=  \frac{\nu^{2}}{2Z}  \cos ^2(r),\\
  \rho_{24}&= -\frac{i \nu}{8Z}  (4 \sec (\alpha ) \mu \beta-\sin ^2(2 r) \delta),~\quad
  \rho_{33}=\frac{1}{8Z} \sec ^2(\alpha ) (\sin ^2(2 r) \beta^{2}+4 \sin ^2(\alpha_{1}) \cos ^4(r)),\\
     \rho_{34}&=  -\frac{i\beta}{2Z}  \sec ^2(\alpha ) \sin (\alpha_{1}) \cos ^2(r) ,~\quad
    \rho_{44}=\frac{1}{8Z} \sec ^2(\alpha ) (4 \mu \beta^{2}+\sin ^2(\alpha_{1}) \sin ^2(2 r)),\\
     \rho_{21}&=   \rho_{12},~~\rho_{31}=  \rho_{13}^{\ast},~~ \rho_{32}=  \rho_{23},~~ \rho_{41}=   \rho_{14},~~\rho_{42}=  \rho_{24}^{\ast},~~
    \rho_{43}= \rho_{34}^{\ast},\\
Z&= \sec ^2(\alpha )-\tan (\alpha ) (\tan (\alpha ) \cos (2 \alpha_{1})+\sin (2 \alpha_{1}) \sin ^2(r))  ,~~\alpha_{1}=t \cos (\alpha ).
\end{aligned}
\end{eqnarray}
The final state of the initial accelerated system   given by,
\begin{eqnarray}
\begin{aligned}
\rho(t)&=\frac{(U_{\mathcal{PT}}\otimes U_{\mathcal{PT}})\varrho_{A_{acc},I_{acc}}(U_{\mathcal{PT}}\otimes U_{\mathcal{PT}})^{\dagger}}{Tr[(U_{\mathcal{PT}}\otimes U_{\mathcal{PT}})\varrho_{A_{acc},I_{acc}}(U_{\mathcal{PT}}\otimes U_{\mathcal{PT}})^{\dagger}]}
\end{aligned},
\end{eqnarray}
In the computational basis the elements of the density operator (19) are  defined by
\begin{eqnarray}
\begin{aligned}
\rho_{11}&=\frac{1}{Z}(\sec ^4(\alpha ) \gamma+\frac{\cos ^4(r)}{2}),~\quad
\rho_{12}=\frac{i\gamma}{Z}( \tan (\alpha ) \sec ^3(\alpha ) ),\\
\rho_{14}&=\frac{1}{2Z} (-2 \tan ^2(\alpha ) \xi^{2}+\text{b1} \cos ^2(r)-2i \xi(\sec (\alpha ) \cos (\alpha_{2}) \sin ^2(r))),\\
\rho_{22}&=\frac{1}{8Z} \sec ^2(\alpha ) \sin ^2(r) (-4 \tan ^2(\alpha ) \cos (2 \alpha_{2})+4 \sec ^2(\alpha )+2 \cos (2 \alpha ) \cos ^2(r)+\cos (2 r)-3),\\
\rho_{23}&=\frac{\gamma}{Z}(\tan ^2(\alpha ) \sec ^2(\alpha ) ),~\quad
\rho_{24}=\frac{\lambda\xi}{Z}(\tan (\alpha )  \sin ^2(r) ),\\
\rho_{33}&=\frac{1}{8Z} \sec ^2(\alpha ) \sin ^2(r) (-4 \tan ^2(\alpha ) \cos (2 \alpha_{2})+4 \sec ^2(\alpha )+2 \cos (2 \alpha ) \cos ^2(r)+\cos (2 r)-3),\\
\rho_{34}&=\frac{\lambda}{Z}(\tan (\alpha ) \xi \sin ^2(r) ),\\
\rho_{41}&=\frac{1}{2Z} (\cos ^2(r)-2 \sec ^2(\alpha ) \sin (\alpha_{2}) \sin ^2(r) (\tan ^2(\alpha ) \sin (\alpha_{2})-i \cos (\alpha_{2}))),\\
\rho_{44}&=\frac{1}{2Z} (\sin ^2(\alpha_{2}) (\tan ^4(\alpha ) \mu+\cos (2 \alpha ) \sec ^4(\alpha ) \cos ^4(r))+\mu\cos ^2(\alpha_{2}) ),\\
\rho_{21}&=\rho_{12}^{\ast},~~\rho_{13}=\rho_{12},~~\rho_{31}=\rho_{12}^{\ast},~~\rho_{32}=\rho_{23},~~\rho_{42}=\rho_{24}^{\ast},~~
\rho_{43}=\rho_{34}^{\ast}.
\end{aligned}
\end{eqnarray}
where,

$Z=4 \tan ^2(\alpha ) \sec ^2(\alpha ) \gamma+1,~~\gamma=\sin ^2(\alpha_{2}) \sin ^2(r)~\quad
b1=\sin ^2(\alpha_{2}) (\tan ^4(\alpha )+\sec ^4(\alpha ))+\cos ^2(\alpha_{2}),~~\alpha_{2}=t \cos ^2(\alpha ),\quad
\lambda=(-\cos (\alpha_{2})+i \tan ^2(\alpha ) \sin (\alpha_{2}))$.

\end{enumerate}

\section{Recovering the entanglement}\label{cohere4}

Here, we  discuss the behavior of entanglement by means of the negativity, where the effect of the symmetric operator with different strengths on the negativity is investigated. It is well  known that, due to the acceleration process, the  decay rate of negativity increases when both qubits are accelerated.

\begin{figure}[!h]
	\begin{center}
		\includegraphics[width=0.3\textwidth, height=125px]{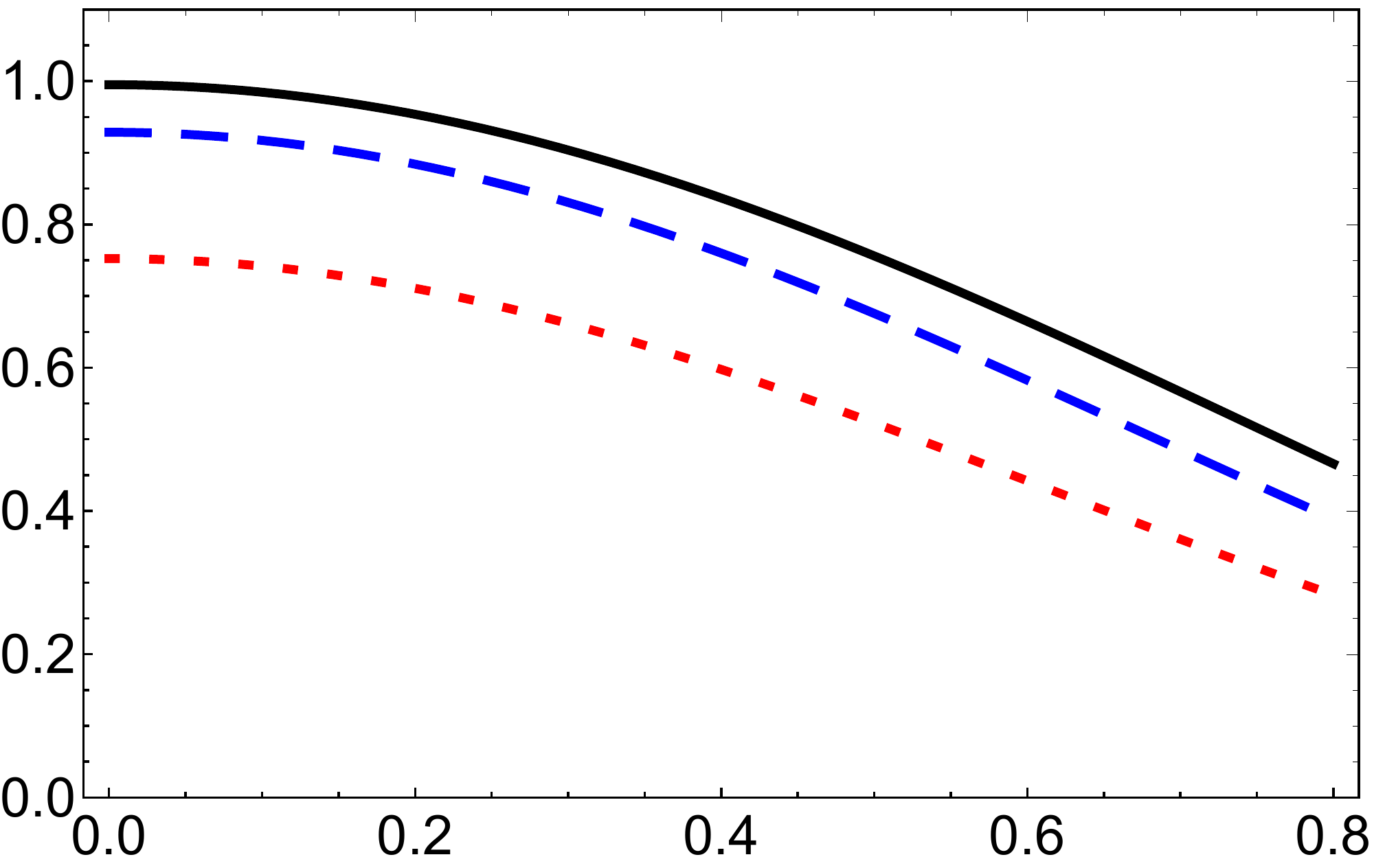}
\put(-160,100){($ a $)}\put(-160,60){$\mathcal{N}_{eg}$}
	\put(-70,-15){$r$}~~~\quad\quad
		\includegraphics[width=0.3\textwidth, height=125px]{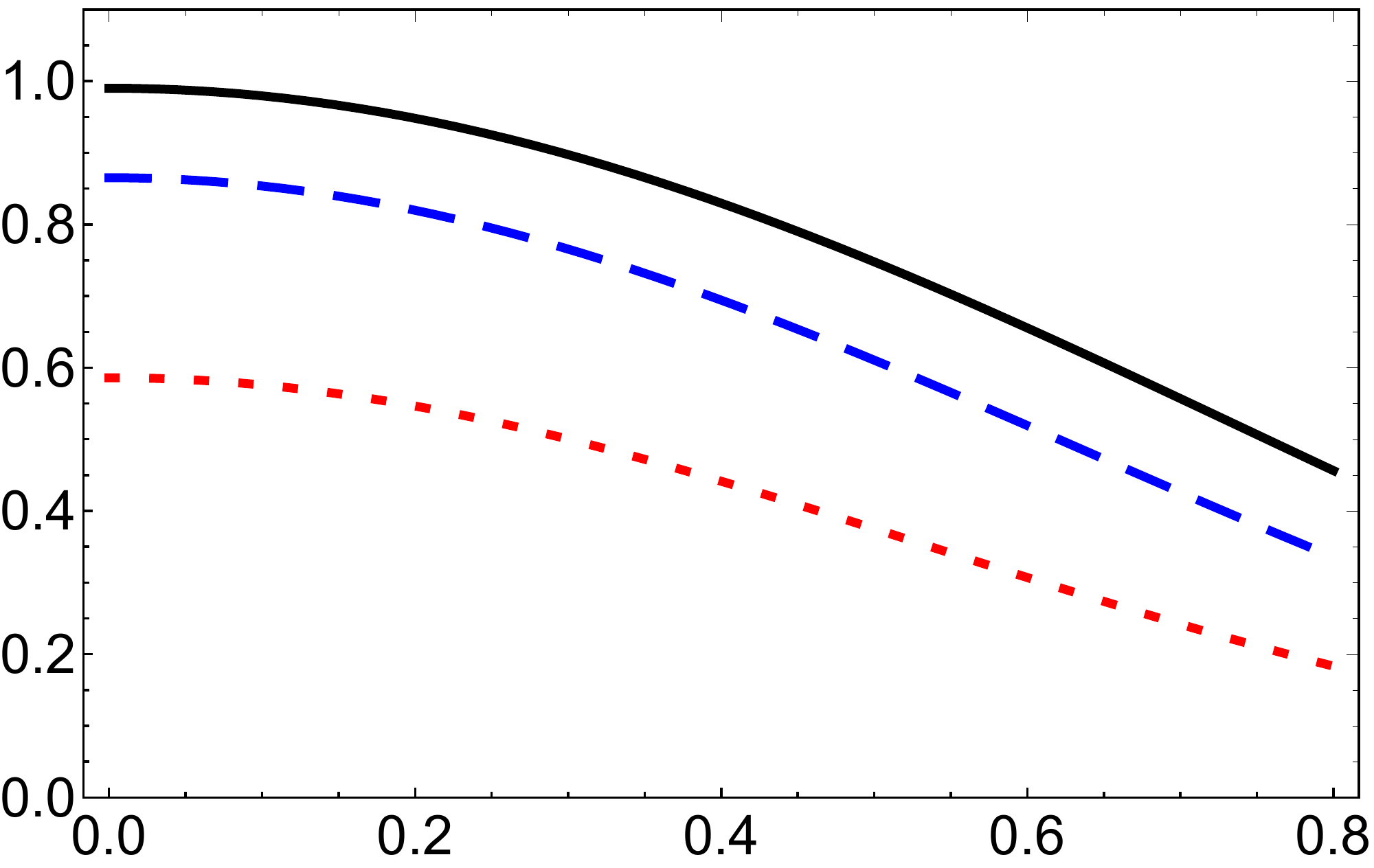}
\put(-160,100){($ b $)}\put(-160,60){$\mathcal{N}_{eg}$}
	\put(-70,-15){$r$}~~~\quad\quad
			\includegraphics[width=0.3\textwidth, height=125px]{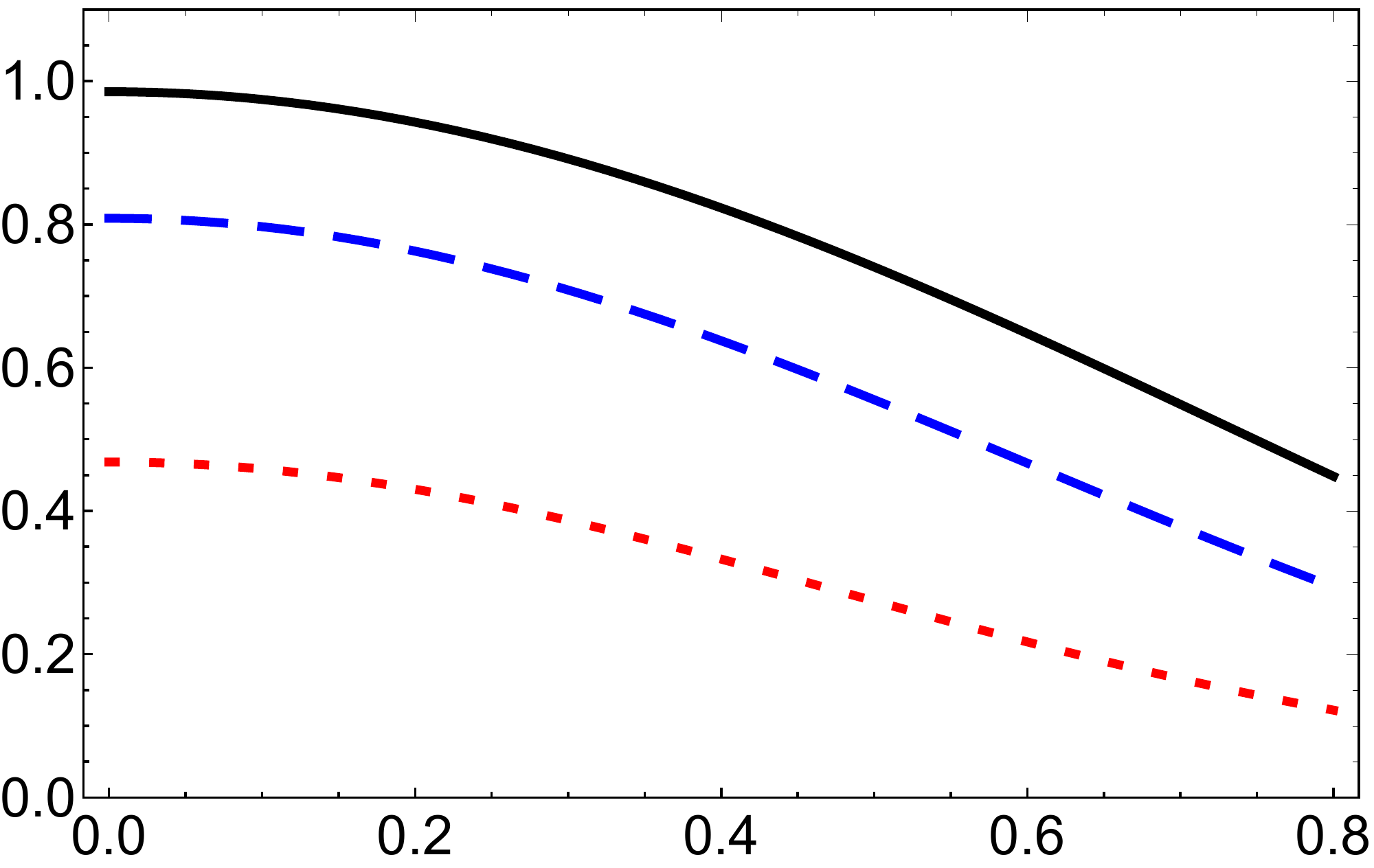}
\put(-160,100){($ c $)}\put(-160,60){$\mathcal{N}_{eg}$}
	\put(-70,-15){$r$}~~~\quad\quad
	\end{center}
	\caption{\label{coo3}The behavior of the negativity  $\mathcal{N}_{eg}$ of the accelerated system at different values of interaction time  and
for different initial settings of the $\mathcal{PT}$. It is assumed that the symmetric operator is applied only on the accelerated first qubit (a) $\alpha=\frac{\pi}{6}$, (b) $\alpha=\frac{\pi}{4}$ and (c) $\alpha=\frac{\pi}{3}$. The solid, dash and dot curves represent $\mathcal{N}_{eg}$ at $t=0.1,0.4$ and $0.9$, respectively. }
\end{figure}
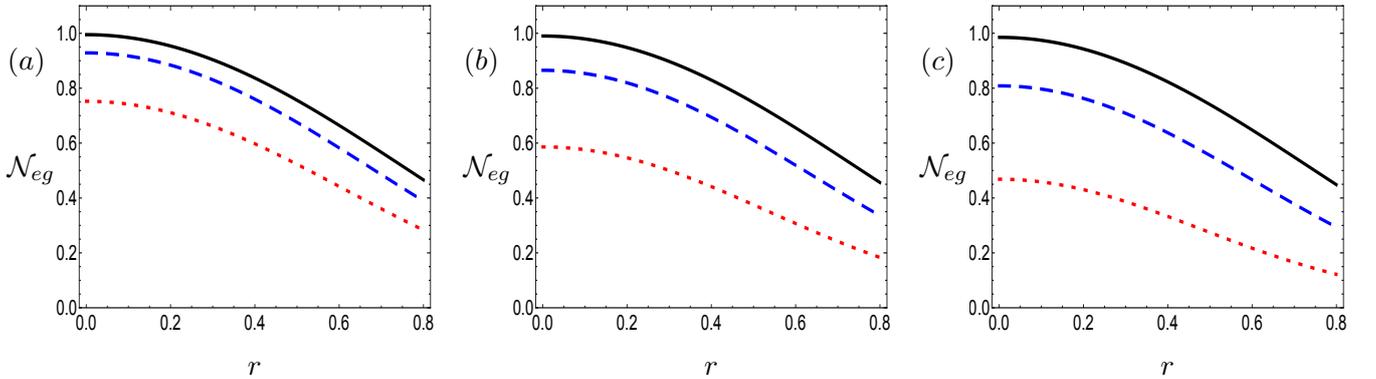

In Fig.(\ref{coo3}), it is shown  the effect of the $\mathcal{PT}$ symmetric operation  on the behavior of the survival amount of the entanglement by means of the negativity, $\mathcal{N}_{ng}$.  The negativity decreases as $r$ increases at different values of interaction time.  As, it is displayed in Fig.(\ref{coo3}b) and (\ref{coo3}c), the decay rate decreases as $\alpha$ decreases, where the  maximum values  of the  negativity are improved as one increases the interaction time. Moreover,  the entanglement doesn't vanish even  for  large acceleration $r$.

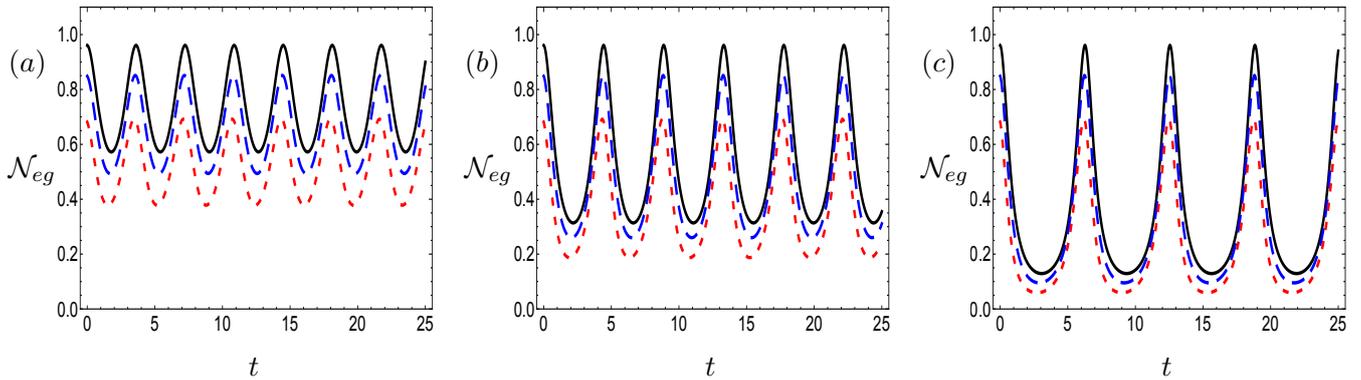
\begin{figure}
\begin{center}
	\includegraphics[width=0.3\textwidth, height=125px]{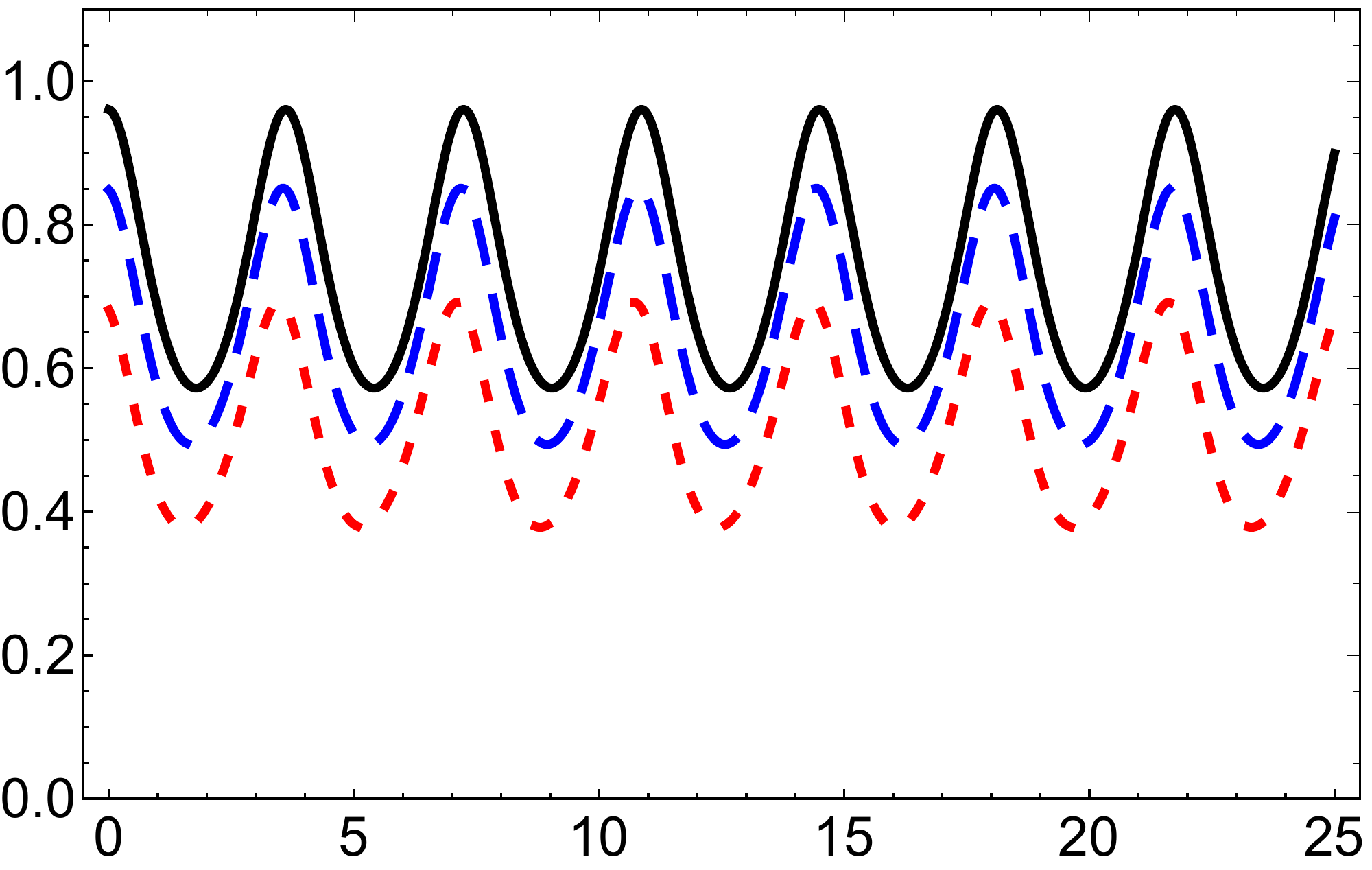}
\put(-160,100){($ a $)}\put(-160,60){$\mathcal{N}_{eg}$}
	\put(-70,-15){$t$}~~~\quad\quad
	\includegraphics[width=0.3\textwidth, height=125px]{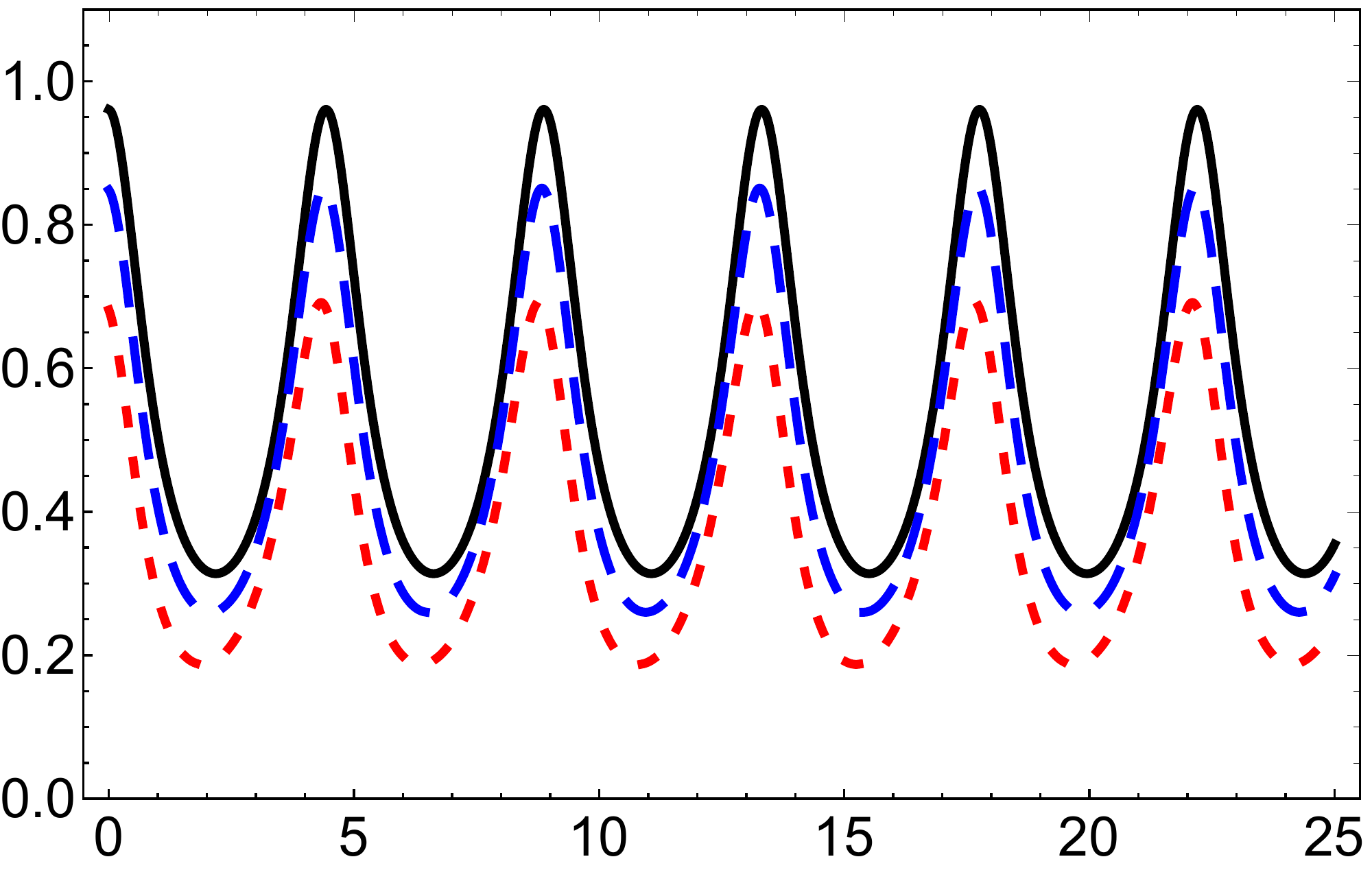}
\put(-160,100){($ b $)}\put(-160,60){$\mathcal{N}_{eg}$}
	\put(-70,-15){$t$}~~~\quad\quad
	\includegraphics[width=0.3\textwidth, height=125px]{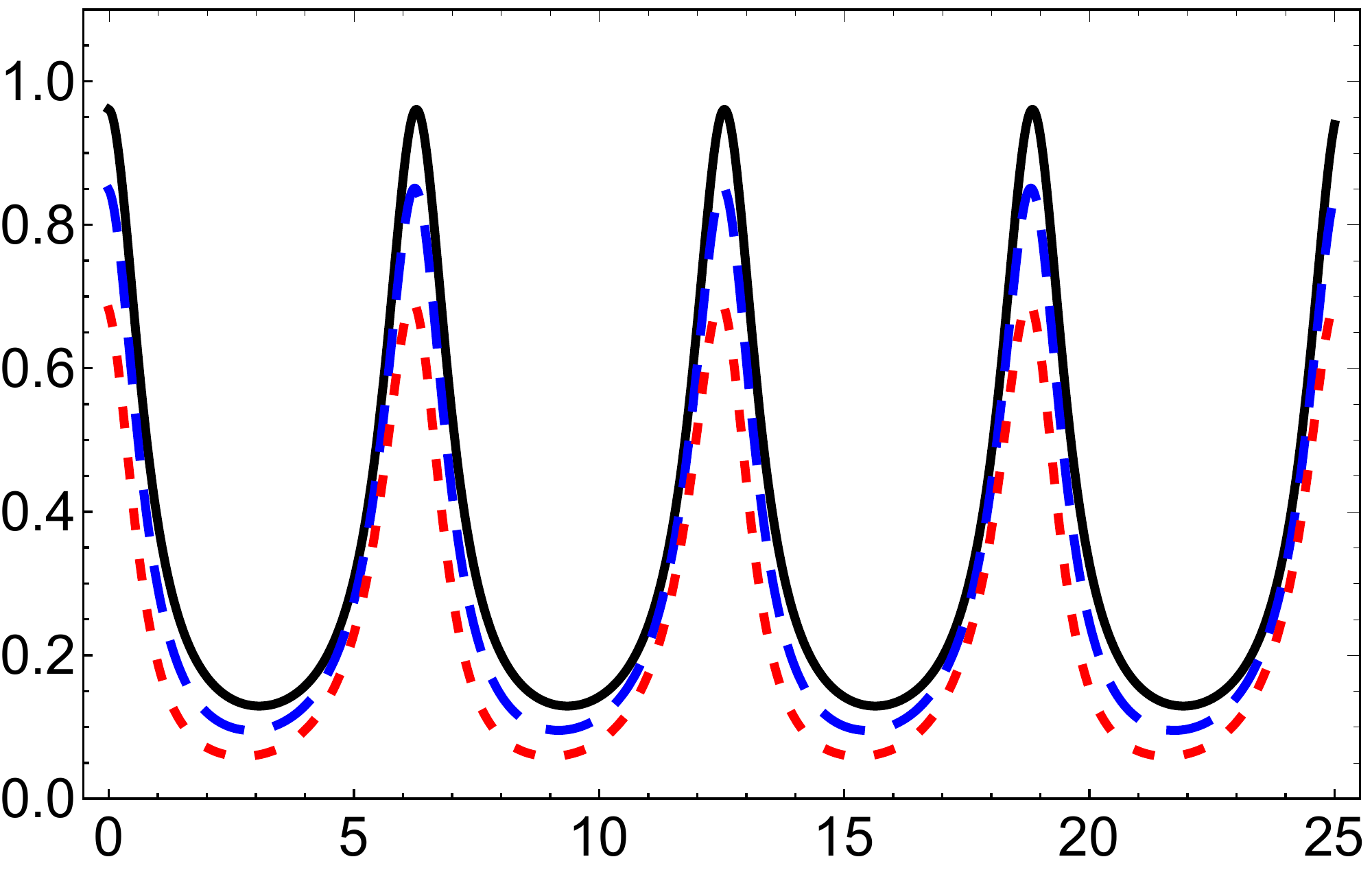}
\put(-160,100){($ c $)}\put(-160,60){$\mathcal{N}_{eg}$}
	\put(-70,-15){$t$}~~~\quad\quad
\end{center}
\caption{\label{coo4}The behavior of the negativity  $\mathcal{N}_{eg}$ in the presence of the local $\mathcal{PT}$-symmetric operator on the accelerated qubit  $A$. The solid, dash and the dot lines represent   $\mathcal{N}_{eg}$ at $r=0.2,0.4$ and $0.6$, respectively, where  the local symmetric operator is described by (a) $\alpha =\frac{\pi}{6}$, (b) $\alpha =\frac{\pi}{4}$  and (c) $\alpha =\frac{\pi}{3}$.}
\end{figure}
The behavior of the survival amount of entanglement, $\mathcal{N}_{eg}$  is shown in Fig.(\ref{coo4}), where it displays the  instantaneous effect of the symmetric operator $\mathcal{PT}$ on the negativity. For small value of the operator strength, which is characterised  by the angle $\alpha$, the negativity decays gradually  to reach its minimum values. However, the decay rate increases as the acceleration increases.  As one increases the operator strength, the behavior  of $\mathcal{N}_{eg}$ changes dramatically, as it is  displayed in Fig.(\ref{coo4}b), where  the entanglement   fluctuates  between its maximum and minimum values and  the amplitudes of these oscillations increase and consequently,the  minimum values  of the negativity decrease. So, the minimum values of the negativity decreases and its the maximum values don't exceed the initial ones. Moreover the minimum  values depend on the initial  values of accelerations and the operator's strength.

\begin{figure}[!h]
\begin{center}
		\includegraphics[width=0.3\textwidth, height=125px]{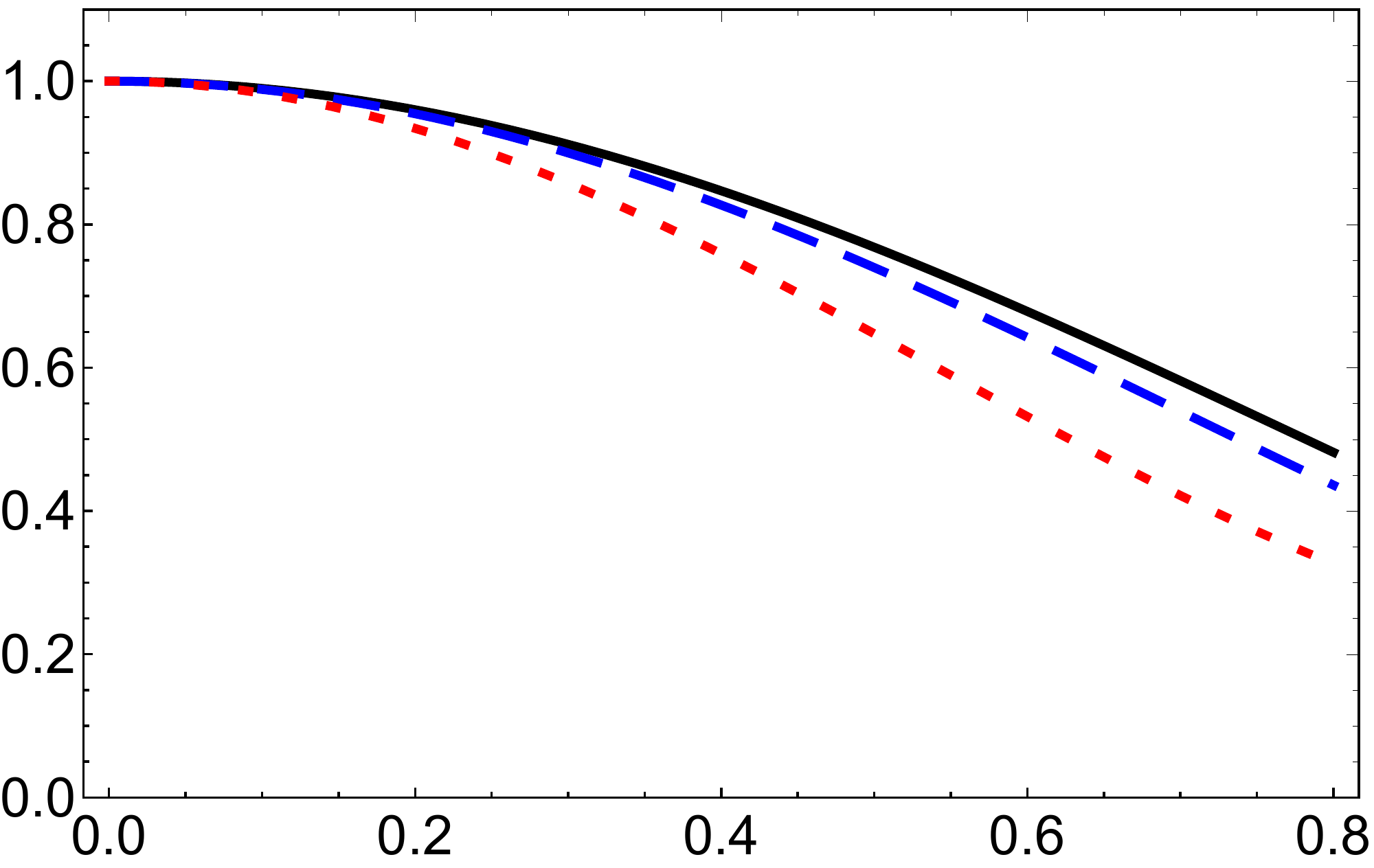}
	\put(-160,100){($ a $)}\put(-160,60){$\mathcal{N}_{eg}$}
	\put(-70,-15){$r$}~~~\quad\quad
		\includegraphics[width=0.3\textwidth, height=125px]{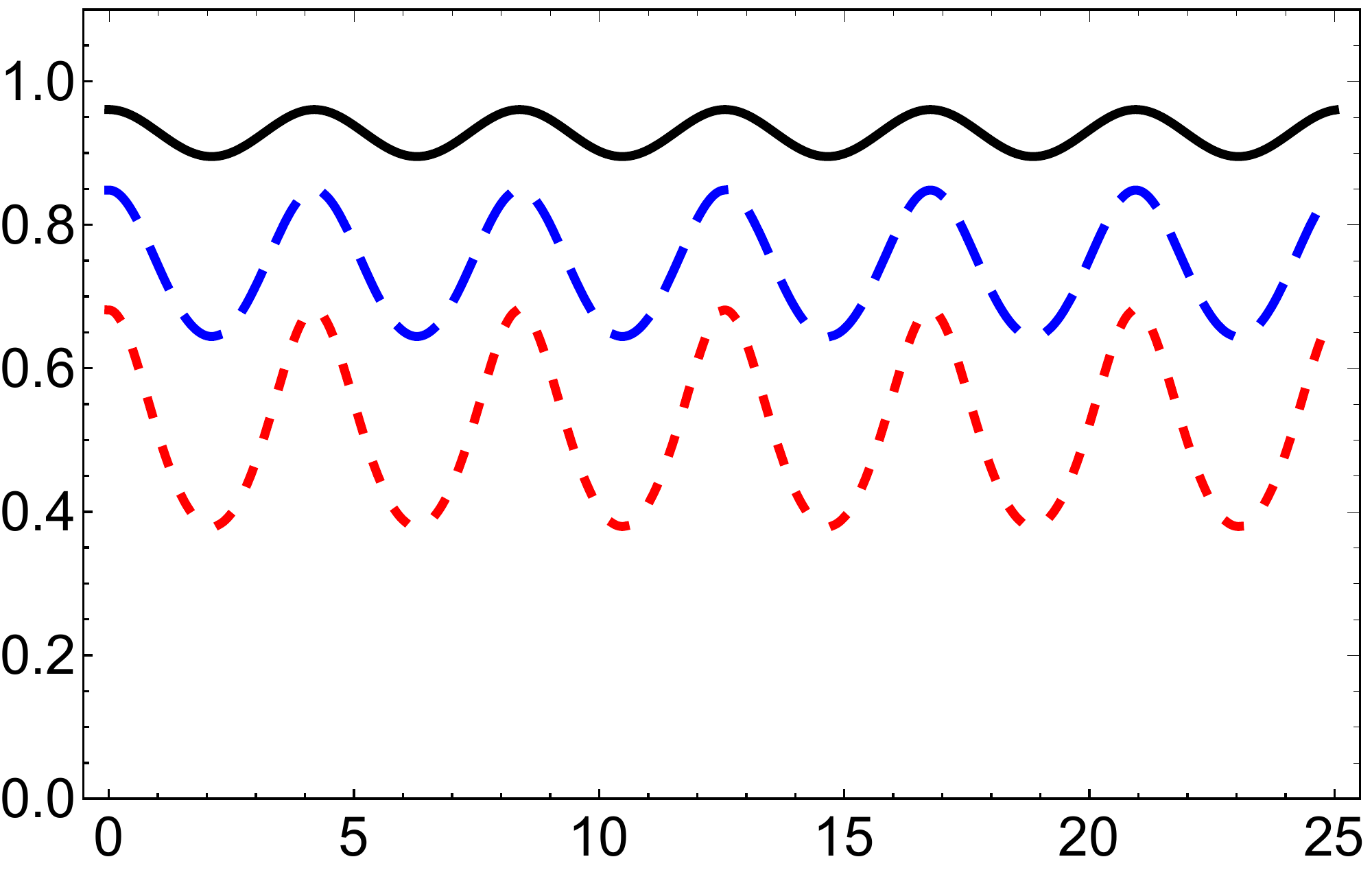}
	\put(-160,100){($ b $)}\put(-160,60){$\mathcal{N}_{eg}$}
	\put(-70,-15){$t$}~~~\quad\quad
\end{center}
\caption{\label{coo5}(a) The same as Fig.(\ref{coo3}a), where the negativity is a function of the acceleration, i.e.,  $\mathcal{N}_{eg}(r)$ (b) The same as Fig(\ref{coo4}a) where the negativity is a function of the interaction time, i.e., $\mathcal{N}_{eg}(t)$. It is assumed that,  the local $\mathcal{PT}$-symmetric operation on both qubits, $\alpha=\pi/6$ and only the first qubit is accelerated. }
\end{figure}
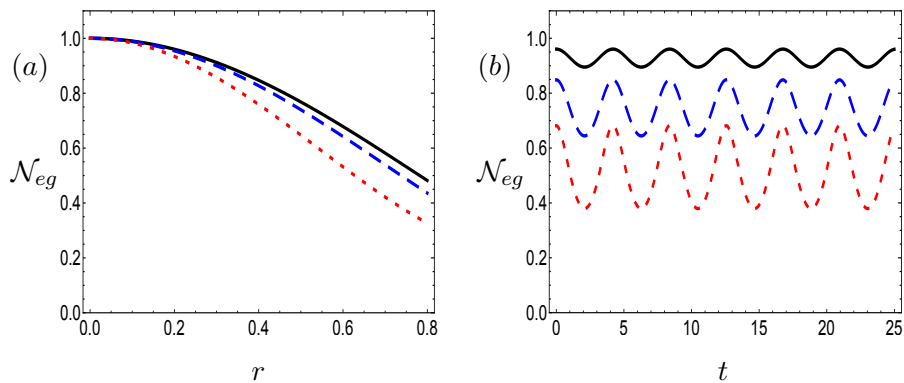

 Fig(\ref{coo5}), shows the possibility of improving the minimum values of the entanglement, when the symmetric operator is applied on both qubits. The behavior of the negativity  as a function of the acceleration, $\mathcal{N}_{eg}(r)$ is displayed in Fig.(\ref{coo5}a), where  the decay rate  decreases and the minimum values are much better than those displayed in Fig(\ref{coo3}a), with the symmetric operator is applied on only one qubit. On the other hand,  Fig.(\ref{coo5}b),  shows the behavior of the negativity as a function of the interaction time, $\mathcal{N}_{eg}(t)$.  The depicted behavior is similar to that displayed  in Fig.(\ref{coo4}a), for  the same set of parameters. However, by applying the $\mathcal{PT}$ on both qubits, the number of fluctuations decreases, where  their amplitudes decrease, and consequently the minimum values are improved. Moreover, at small values of the interaction time, the maximum values are much better than those displayed at larger $t$.

\begin{figure}[!h]
\begin{center}
	\includegraphics[width=0.3\textwidth, height=125px]{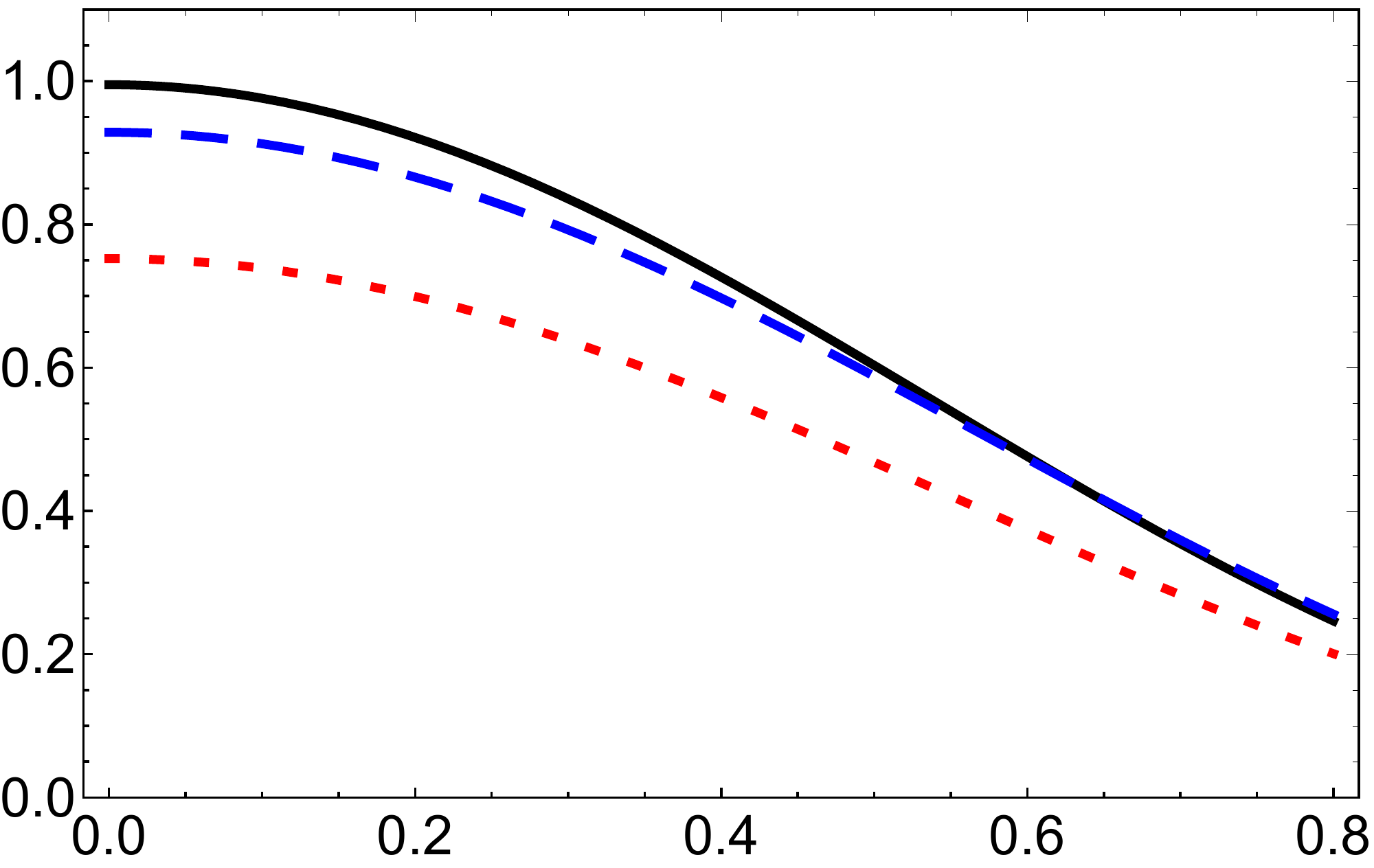}
	\put(-160,100){($ a $)}\put(-160,60){$\mathcal{N}_{eg}$}
	\put(-70,-15){$r$}~~~\quad\quad\quad\quad\quad
	\includegraphics[width=0.3\textwidth, height=125px]{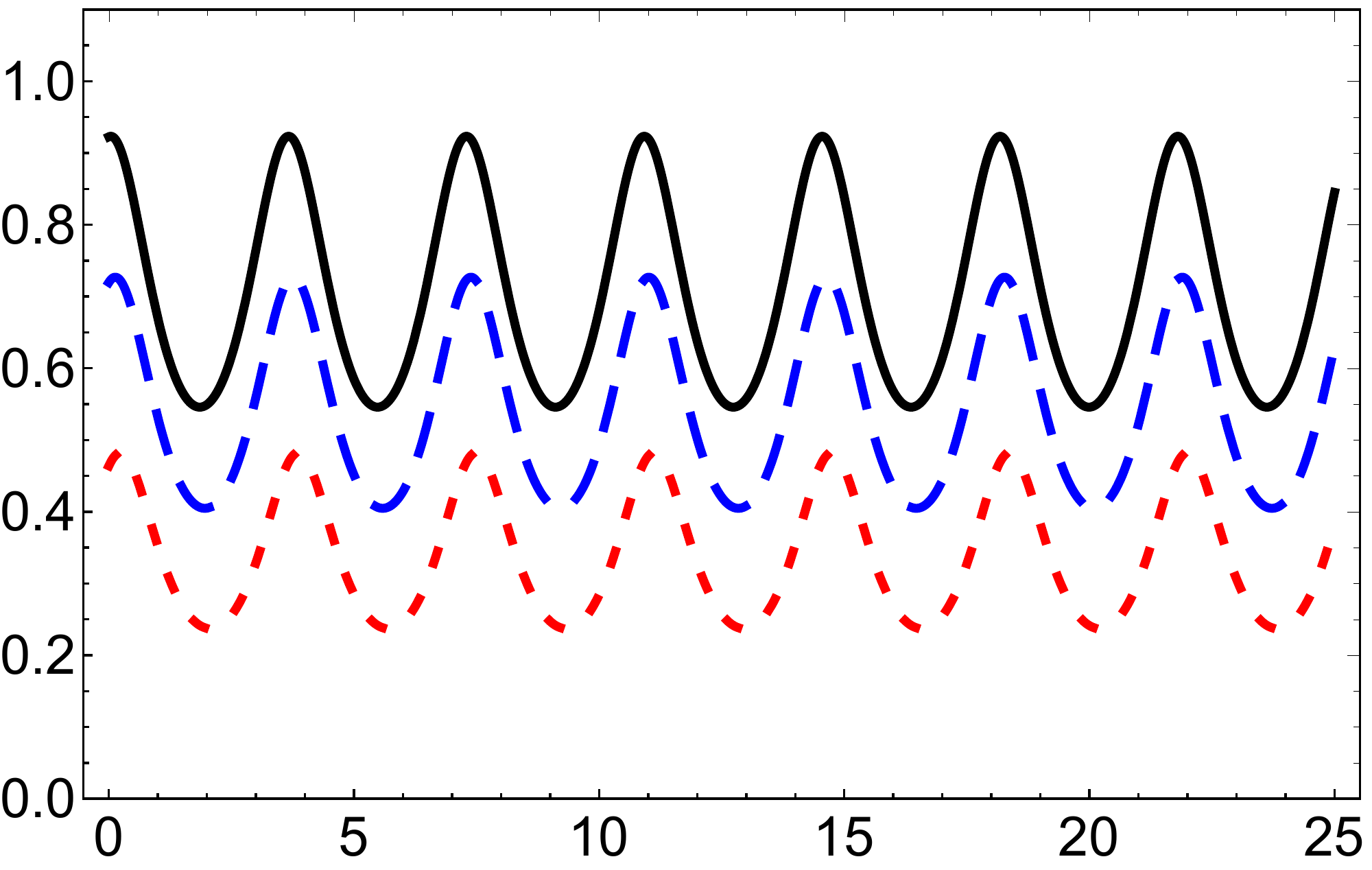}
	\put(-160,100){($ b $)}\put(-160,60){$\mathcal{N}_{eg}$}
	\put(-70,-15){$t$}~~~\quad\quad\\
		\includegraphics[width=0.3\textwidth, height=125px]{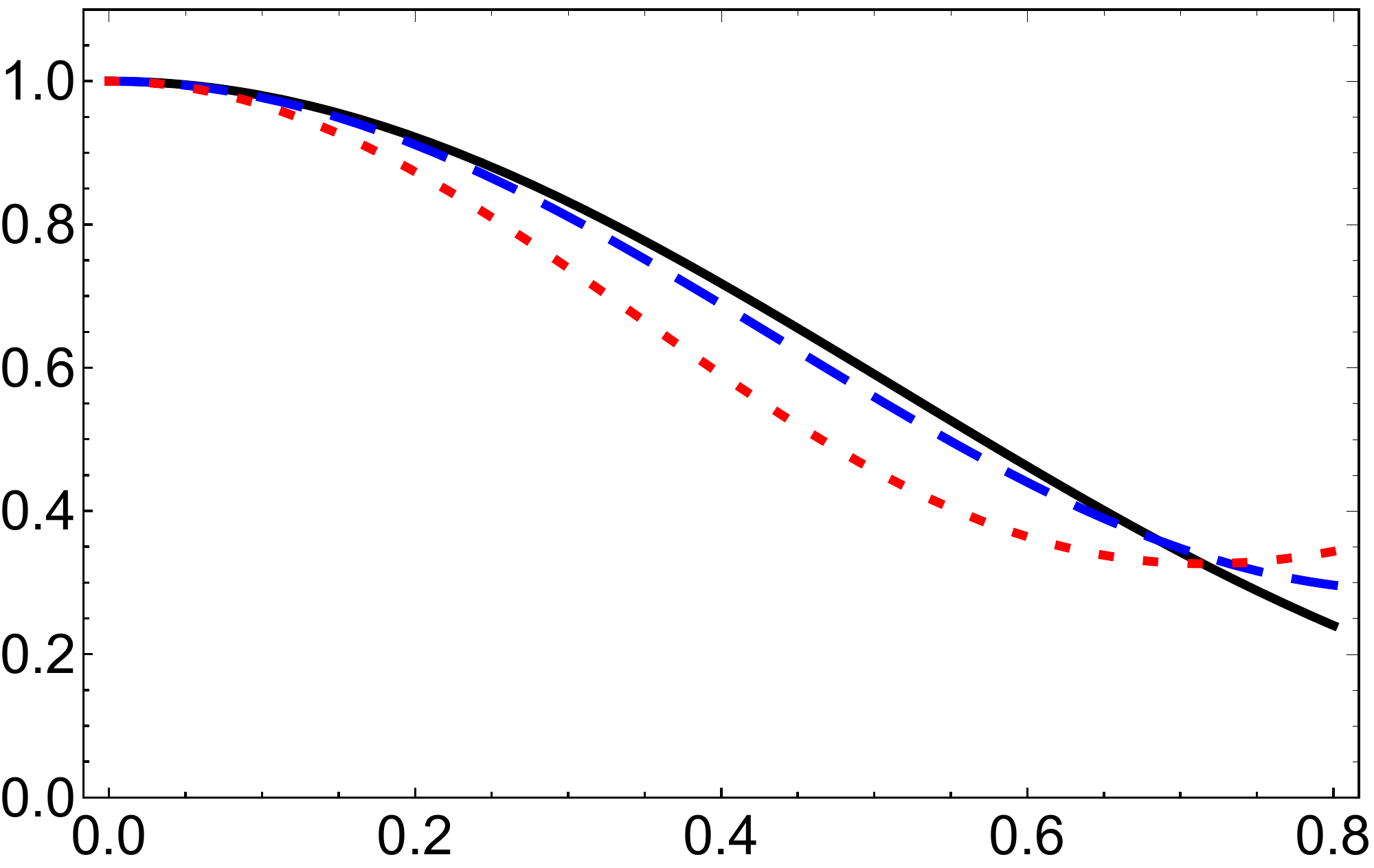}
	\put(-160,100){($ c $)}\put(-160,60){$\mathcal{N}_{eg}$}
	\put(-70,-15){$r$}~~~\quad\quad\quad\quad\quad
		\includegraphics[width=0.3\textwidth, height=125px]{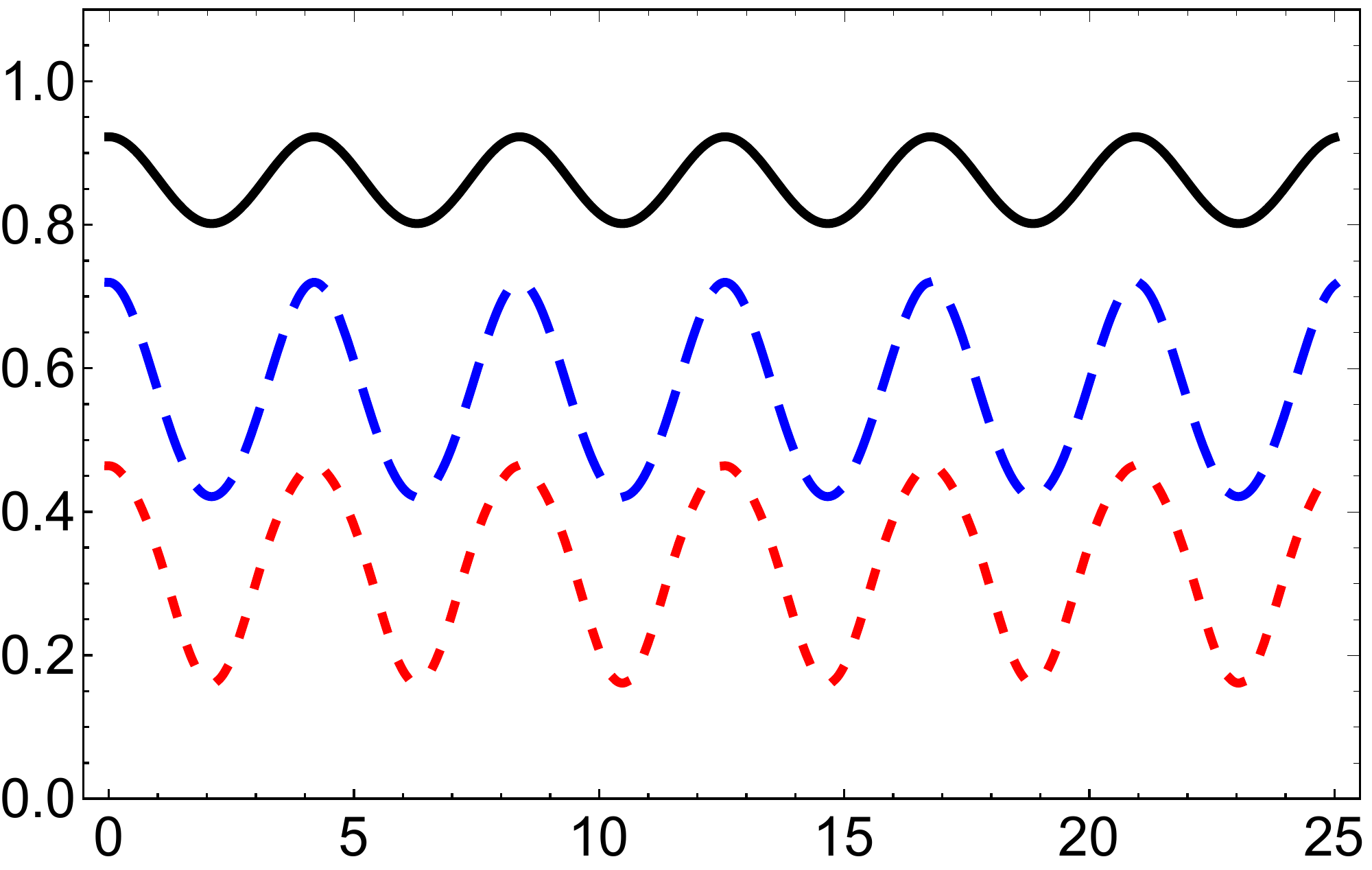}
	\put(-160,100){($ d $)}\put(-160,60){$\mathcal{N}_{eg}$}
	\put(-70,-15){$t$}~~~\quad\quad
\caption{\label{coo6}The behavior of the negativity $\mathcal{N}_{eg}$  of the accelerated system, where it assumed that both qubits are accelerated. In Fig.(a),(b) it is assumed that the symmetric-operator is applied on only  one qbit and (c), (d) it is applied on both qubits. The used parameters are the same as that used in Figs.(\ref{coo3}a), (\ref{coo4}a), respectively.}
\end{center}
\end{figure}
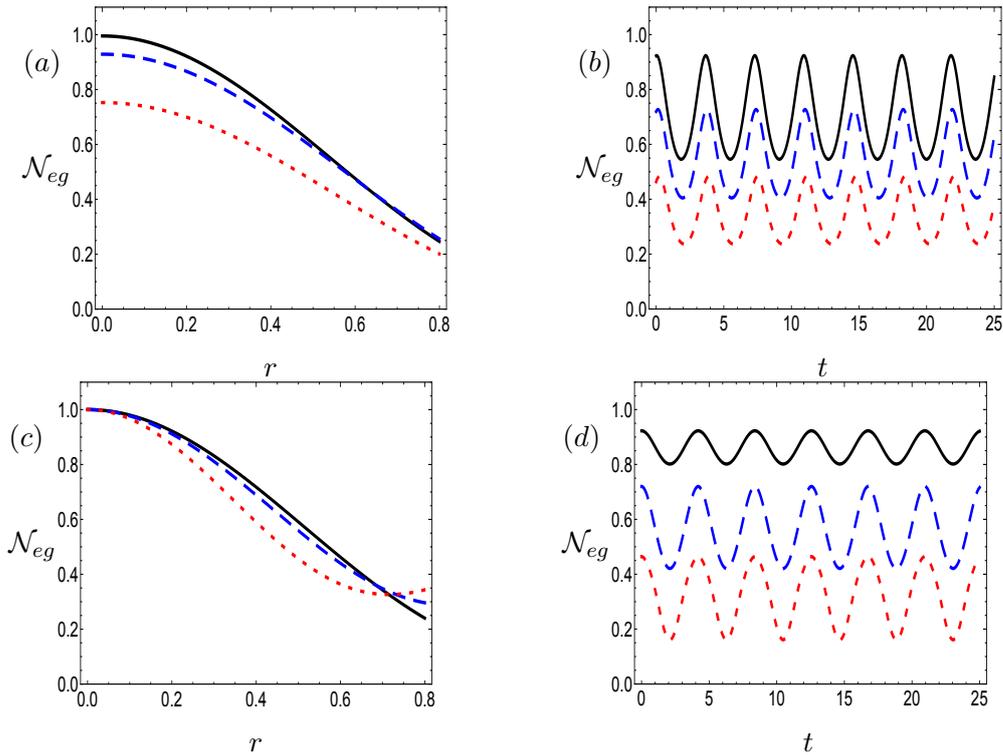

In reality, the two qubits may be  accelerated. In this case, the behavior  of the negativity  depicted in Fig.(\ref{coo6}) is similar to those predicted for cases of accelerating only one qubit in Fig.(\ref{coo3}-\ref{coo4}).  However, due to the acceleration process of both qubits, the decay rate is much larger than that displayed in Fig.(\ref{coo3}a) at large $r$.  This phenomena is clearly displayed by comparing Fig.(\ref{coo3}a) and Fig.(\ref{coo5}a), where the maximum values are much larger, when only one qubit is accelerated.  The possibility of improving and restraining the losses of the negativity can be achieved by applying the symmetric-operator on both qubits. As it is displayed in Fig.(\ref{coo5}b),the negativity decays gradually and the maximum values are much better than that displayed in Fig.(\ref{coo6}a).
 As it is displayed from Fig.(\ref{coo6}b) and (\ref{coo6}d), at small acceleration, the oscillations' amplitude decreases when the symmetric-operator is applied on  both qubits and consequently, the minimum values of the negativity are improved.

 From the previous results of the negativity, it is clear that the effect  of the symmetric-operator on the accelerated systems is much better when it is applied on both qubits during the interaction. However,  the instant effect of the symmetric operator  improve the minimum values of entanglement even at large acceleration.  On the other hand, the possibility of improving   the  entanglement  and minimizing its losses is achieved, if only one qubit is accelerated and the symmetric-operator is applied  on both qubits.

In this context, it is important to mention that, our results consistences with that obtained by Chen et.al\cite{cohe14}, where the entanglement of a two qubits system  can be restored when one of them undergoes the $\mathcal{PT}$ symmetric operation and the entanglement can exceed its initial values. However, although  our approach is different from that suggested by Chen et.al\cite{cohe14}, where we applied the  $\mathcal{PT}$ on accelerated systems, our conclusion is similar, where we showed that local symmetric operation  improves the entanglement of the shared state between the two  parties and minimize  its losses.

\section{Recovering the losses of $\mathcal{N}_{La}$}\label{cohere5}
In this section, we investigate  numerically the possibility of recovering the losses of the non-local coherent quantum advantage, where different  cases will be considered: one or both   qubits   are  accelerated, and the symmetric operator  is applied either on a single or both qubits, with  different initial settings of the symmetric operators.

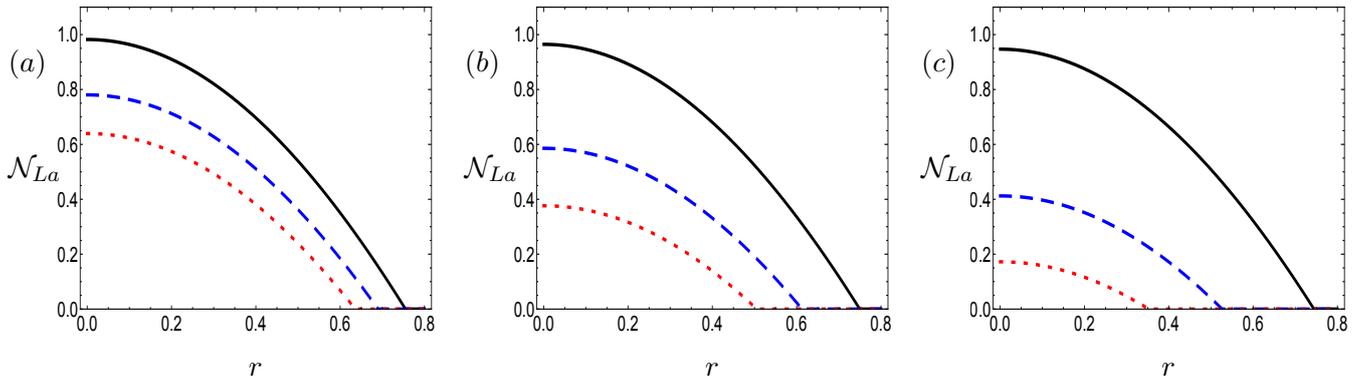
\begin{figure}[t!]
	\begin{center}
		\includegraphics[width=0.3\textwidth, height=125px]{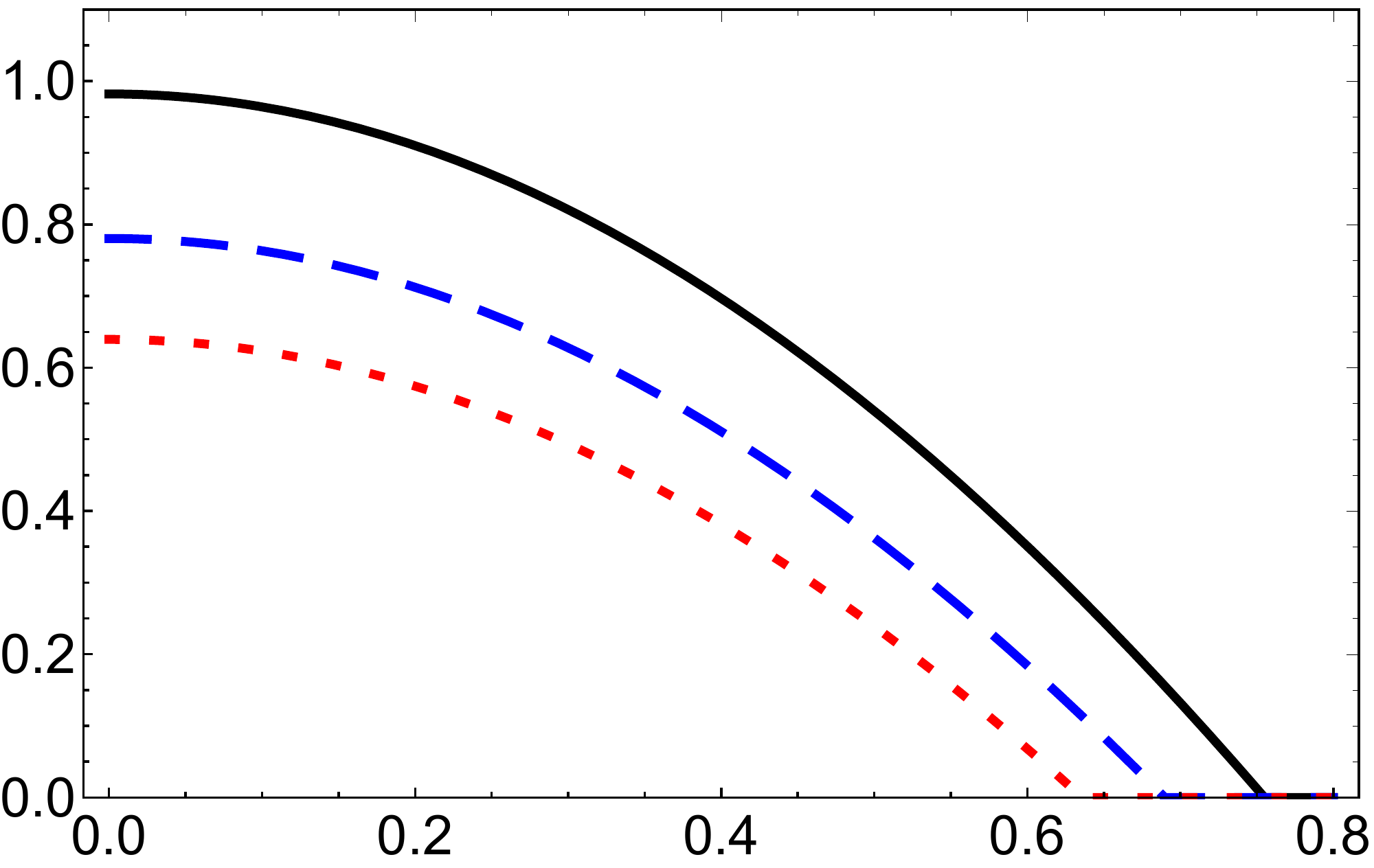}
\put(-160,100){($ a $)}\put(-160,60){$\mathcal{N}_{La}$}
	\put(-70,-15){$r$}~~~\quad\quad
		\includegraphics[width=0.3\textwidth, height=125px]{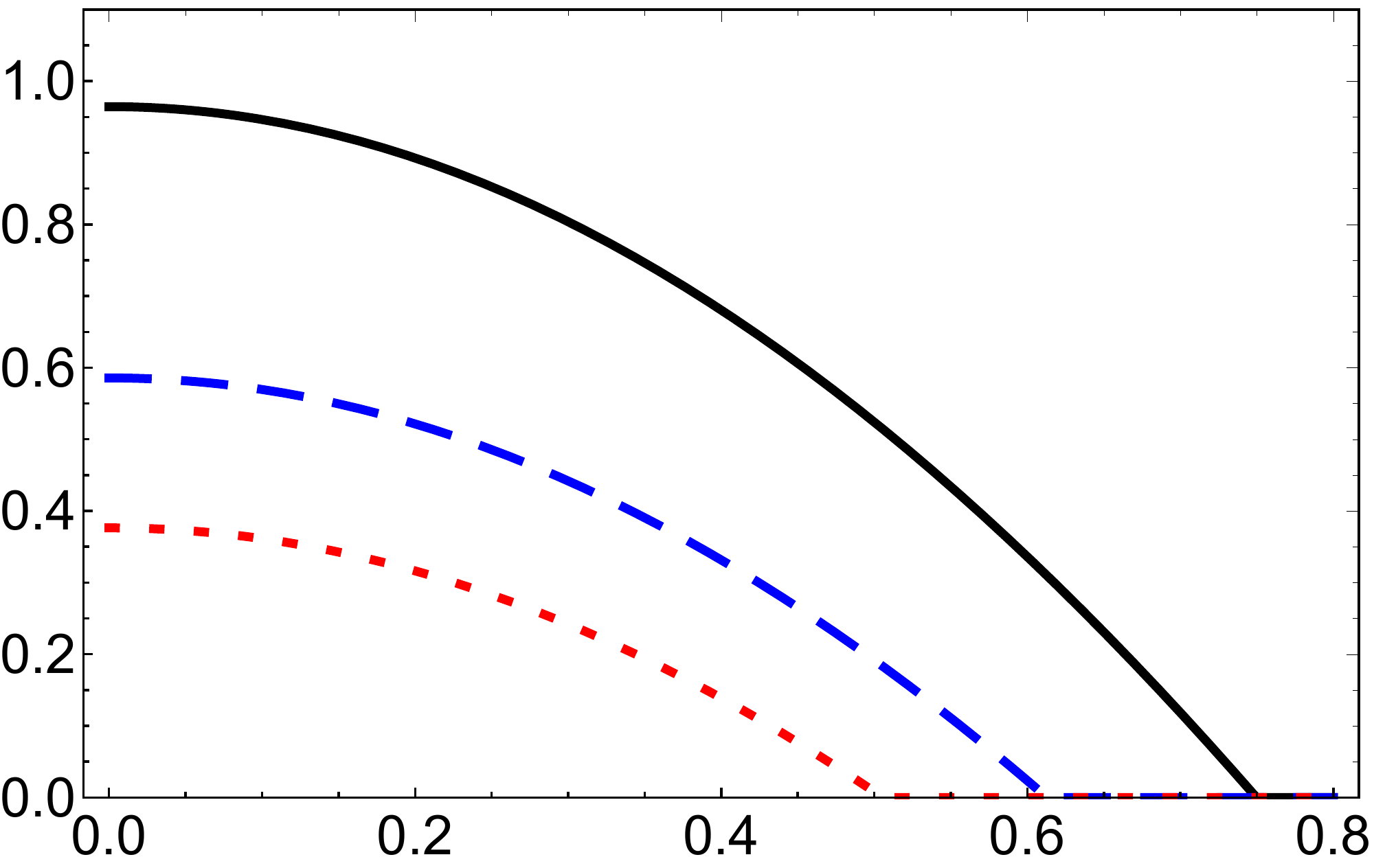}
\put(-160,100){($ b $)}\put(-160,60){$\mathcal{N}_{La}$}
	\put(-70,-15){$r$}~~~\quad\quad
\includegraphics[width=0.3\textwidth, height=125px]{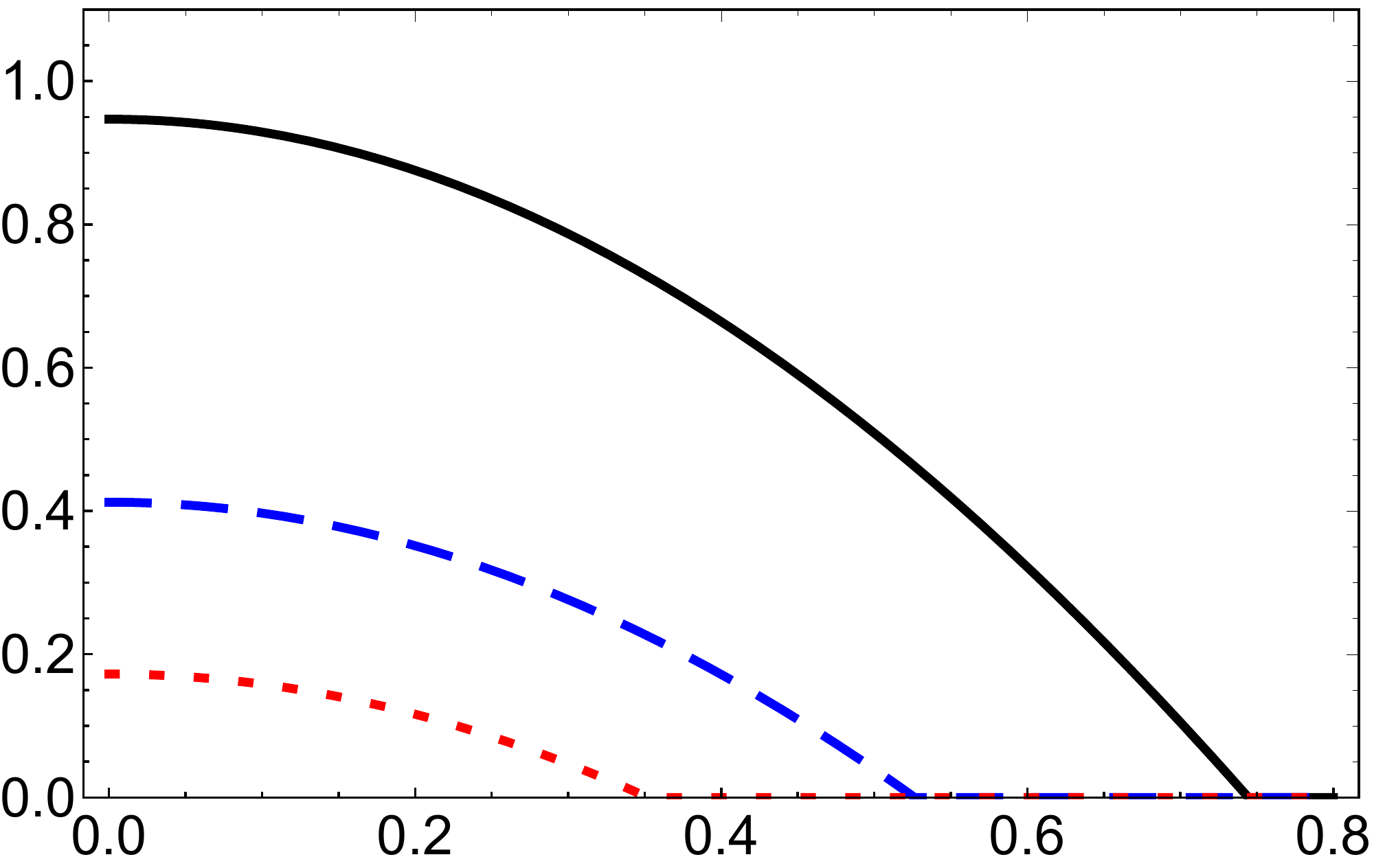}
\put(-160,100){($ c $)}\put(-160,60){$\mathcal{N}_{La}$}
	\put(-70,-15){$r$}~~~\quad\quad
	\end{center}
	\caption{\label{coo7}The  behavior of  the non-local coherent advantage $\mathcal{N}_{La}$ , when only one qubit is accelerated at different interaction time, where the solid, dash and dot curves are evaluated at $t=0.1,0.4$ and $0.9$, respectively, and  the $\mathcal{PT}$ symmetric is described (a) $\alpha=\frac{\pi}{6}$, (b) $\alpha=\frac{\pi}{4}$ and (c) $\alpha=\frac{\pi}{3}$.}
\end{figure}

\subsection{Only one qubit is accelerated}\label{cohere5.1}

Fig.(\ref{coo7}),  shows the behavior of $\mathcal{N}_{La}$ at different interaction time, where it is assumed that the $\mathcal{PT}-operator $ is described by  different angles. The general behavior is similar to that shown in Fig.(\ref{coo2}), where the non-local coherent advantage decreases as the acceleration increases.   However, at small values of the interaction time $t=0.1$,   $\mathcal{N}_{La}$ decreases gradually as the acceleration increases. The  maximum values of the non-local coherent advantage  decreases as one increases the interaction time. Moreover, as it is displayed in Fig.(\ref{coo2}a)  $\mathcal{N}_{La}$ vanishes at small values of accelerations.  Fig.(\ref{coo7}b), show that when we increase the angle $\alpha$, which describes the strength of  $\mathcal{PT}$, where we set $\alpha=\pi/4$,  $\mathcal{N}_{La}$ decreases and the maximum values are much smaller  than those displayed in Fig.(\ref{coo7}a), where  $\alpha=\frac{\pi}{6}$. Also, the non-local advantage survives at large values of acceleration, even at large interaction time. These phenomena is clearly exhibited  in Fig.(\ref{coo7}c), where we set $\alpha=\frac{\pi}{3}$. In this case, the maximum  values  of $\mathcal{N}_{La}$ are much smaller  than those displayed at small values of the operator strength $\alpha$.

\begin{figure}[!h]
\begin{center}
	\includegraphics[width=0.3\textwidth, height=125px]{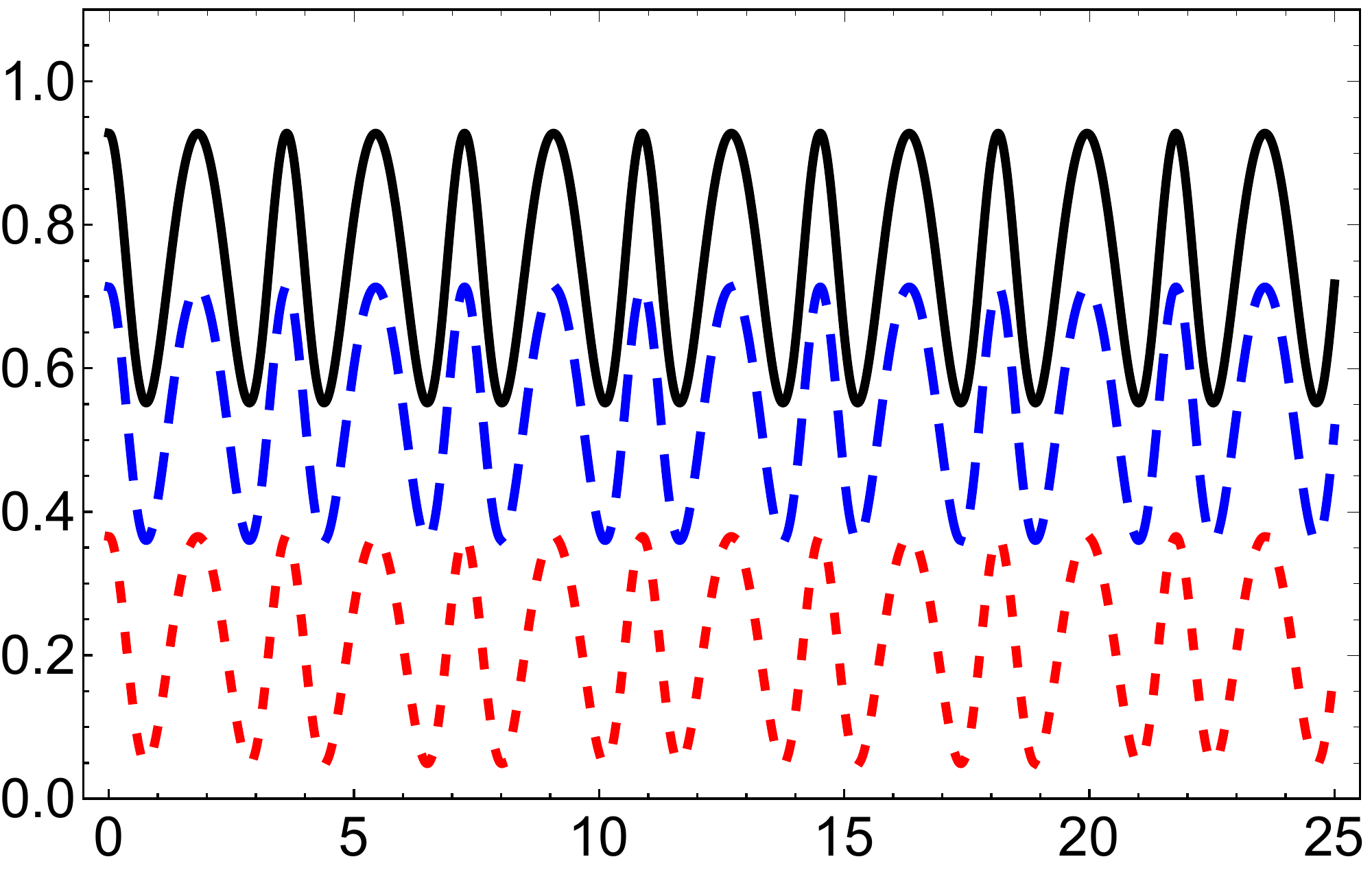}\put(-160,100){($ a $)}\put(-160,60){$\mathcal{N}_{La}$}
	\put(-70,-15){$t$}~~~\quad\quad
	\includegraphics[width=0.3\textwidth, height=125px]{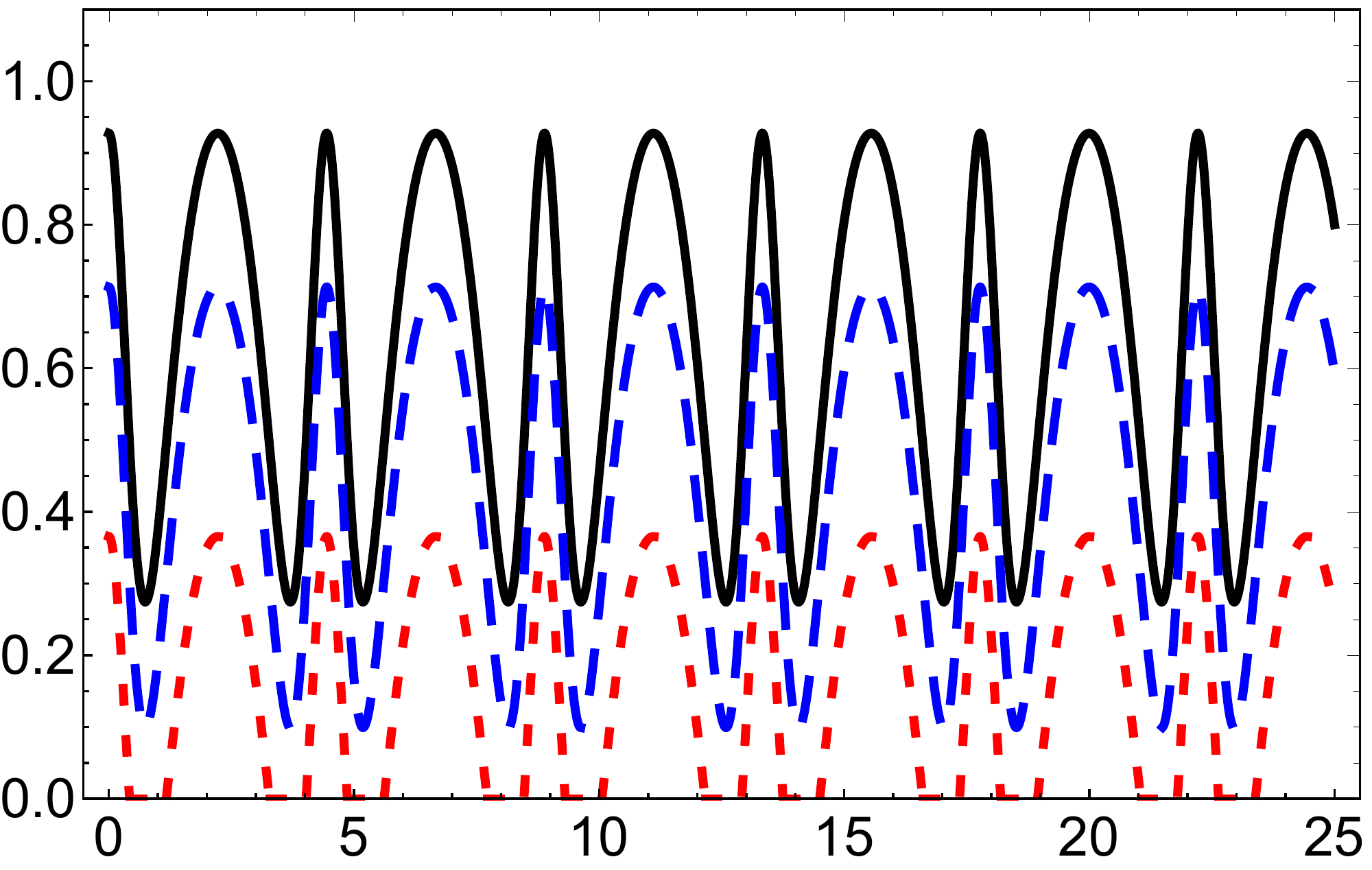}\put(-160,100){($ b $)}\put(-160,60){$\mathcal{N}_{La}$}
	\put(-70,-15){$t$}~~~\quad\quad
	\includegraphics[width=0.3\textwidth, height=125px]{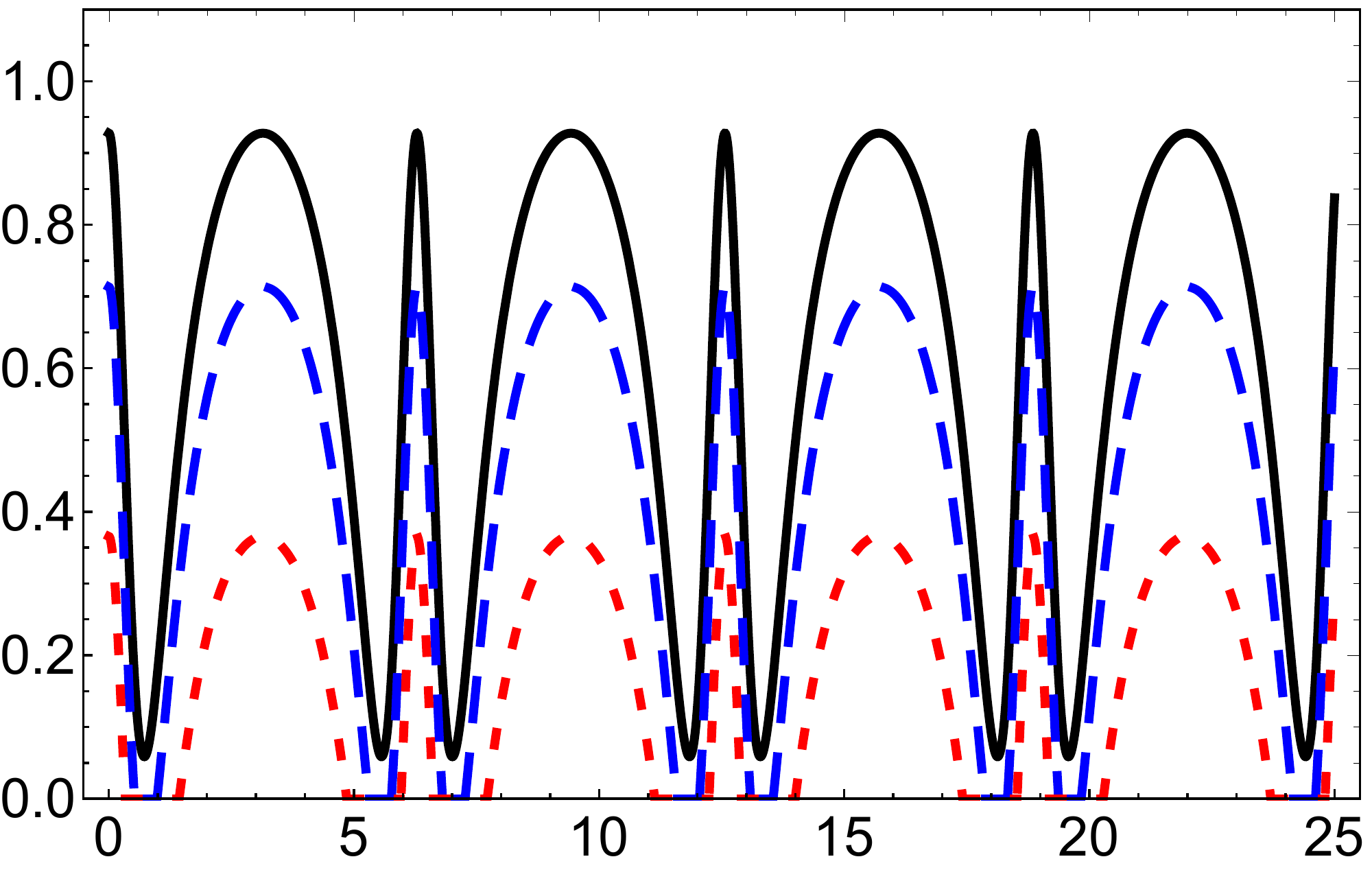}\put(-160,100){($ c $)}\put(-160,60){$\mathcal{N}_{La}$}
	\put(-70,-15){$t$}
\end{center}
\caption{\label{coo8}The behavior of the non-local coherent advantage, $\mathcal{N}_{La}$ in the presence of the local $\mathcal{PT}$-symmetric operator  on the accelerated qubit  $A$. The solid, dash and the dot lines represents the   $\mathcal{N}_{La}$ at  $r=0.2$, $r=0.4$ and $0.6$, respectively, where  the local symmetric operator is described by (a) $\alpha =\frac{\pi}{6}$, (b) $\alpha =\frac{\pi}{4}$  and (c) $\alpha =\frac{\pi}{3}$.}
\end{figure}
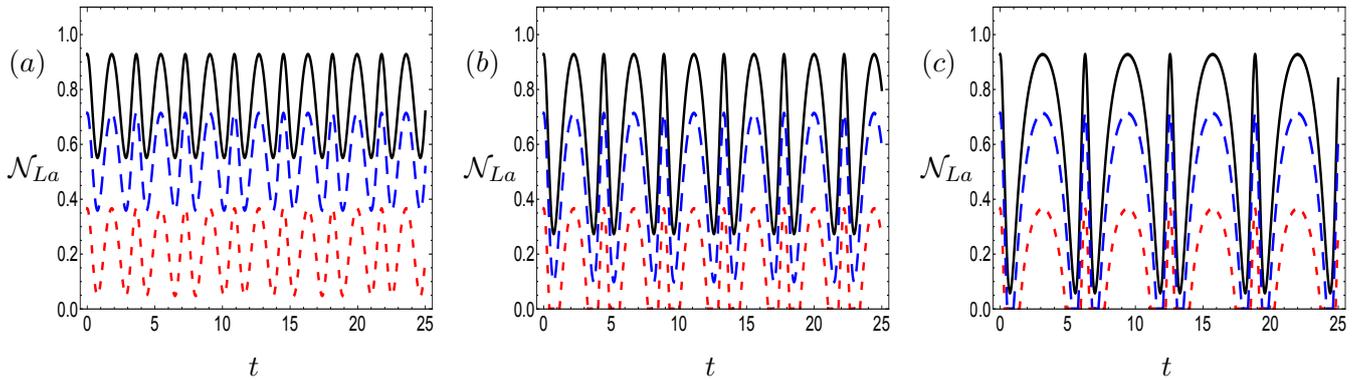

The effect of the local symmetric operator $\mathcal{PT}$ on the behavior of the non-local coherent advantage, $\mathcal{N}_{La}(t)$ is shown in Fig.(\ref{coo8})  with  different initial acceleration, where it is assumed that   the $\mathcal{PT}$ is applied only on one  qubit. In  Fig.(\ref{coo8}a) we consider that, the symmetric operator is described by a small angle $\alpha=\pi/6$. The general behavior  shows that, the non-local coherent advantage  oscillates periodically between its maximum and minimum values.  The minimum values  depend on the initial acceleration, where as one decreases the acceleration, the minimum values are much larger than those displayed at small acceleration.

 The effect of larger values of the initial angle settings of the symmetric operator  is displayed in Fig.(\ref{coo8}b), where we set $\alpha=\pi/4$. The periodic oscillations   of the non-local coherent advantage are  displaced with larger values of interaction time, where  for  large acceleration, it vanishes faster.  Moreover, as the interaction time increases further, it rebirths to reach its maximum value, which doesn't exceed its initial value.

  From Fig.(\ref{coo8}),  we observe that  at small values of the operator's strength  ($\alpha=\pi/6$),  the behavior of the non-local coherent advantage changes   dramatically, where it oscillates  fast and the amplitudes of the oscillations are smaller. This means that, the minimum values of the   $\mathcal{N}_{La}$ are much larger than those displayed in Figs.(\ref{coo8}b),(\ref{coo8}c). Moreover, the vanishing   phenomena  of $\mathcal{N}_{La}$   disappears and the long-lived  phenomena of the non-local coherent advantage is predicted.

\begin{figure}[!h]
	\begin{center}
		\includegraphics[width=0.3\textwidth, height=125px]{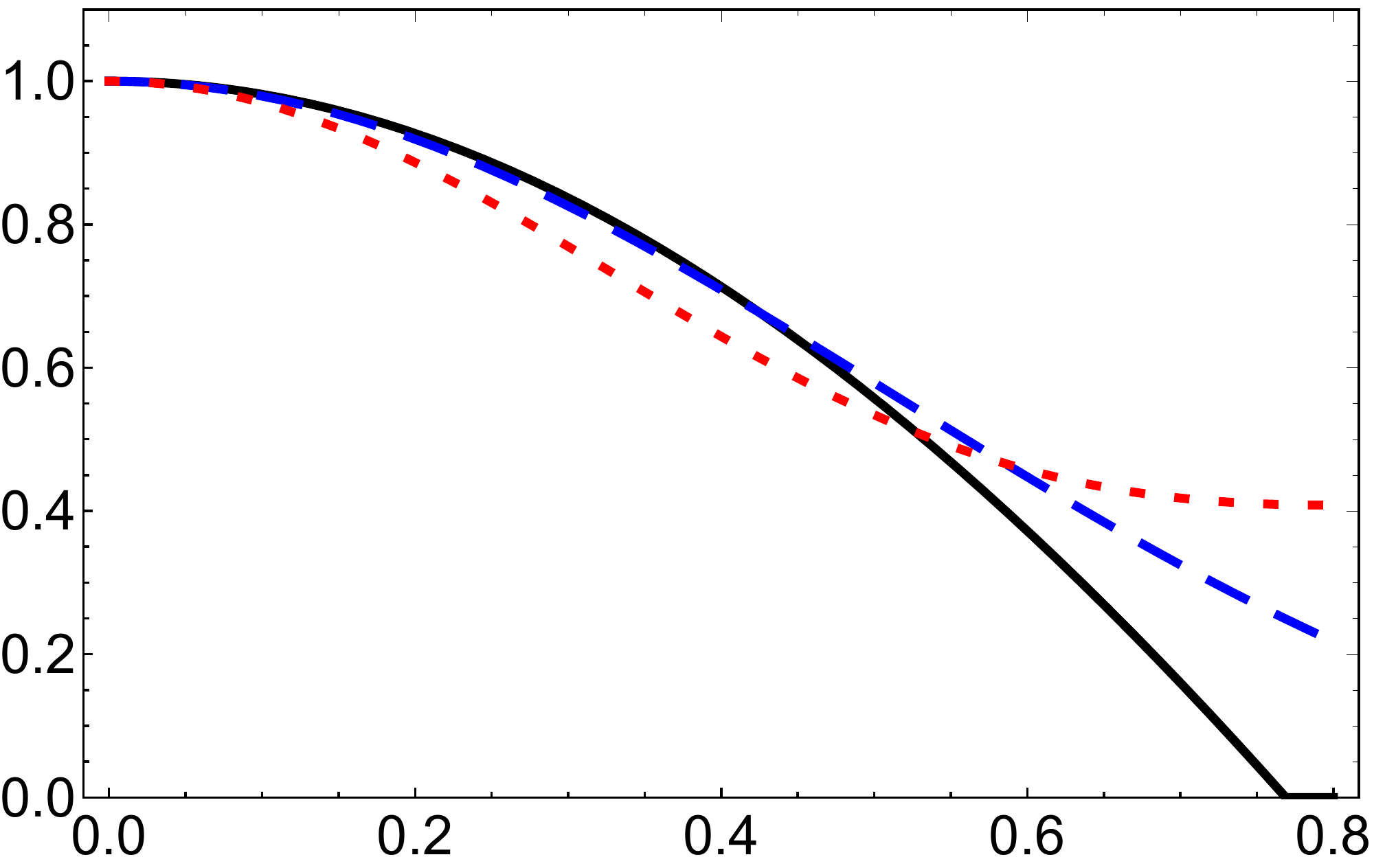}
\put(-160,100){($ a $)}\put(-160,60){$\mathcal{N}_{La}$}
	\put(-70,-15){$r$}~~~\quad\quad
		\includegraphics[width=0.3\textwidth, height=125px]{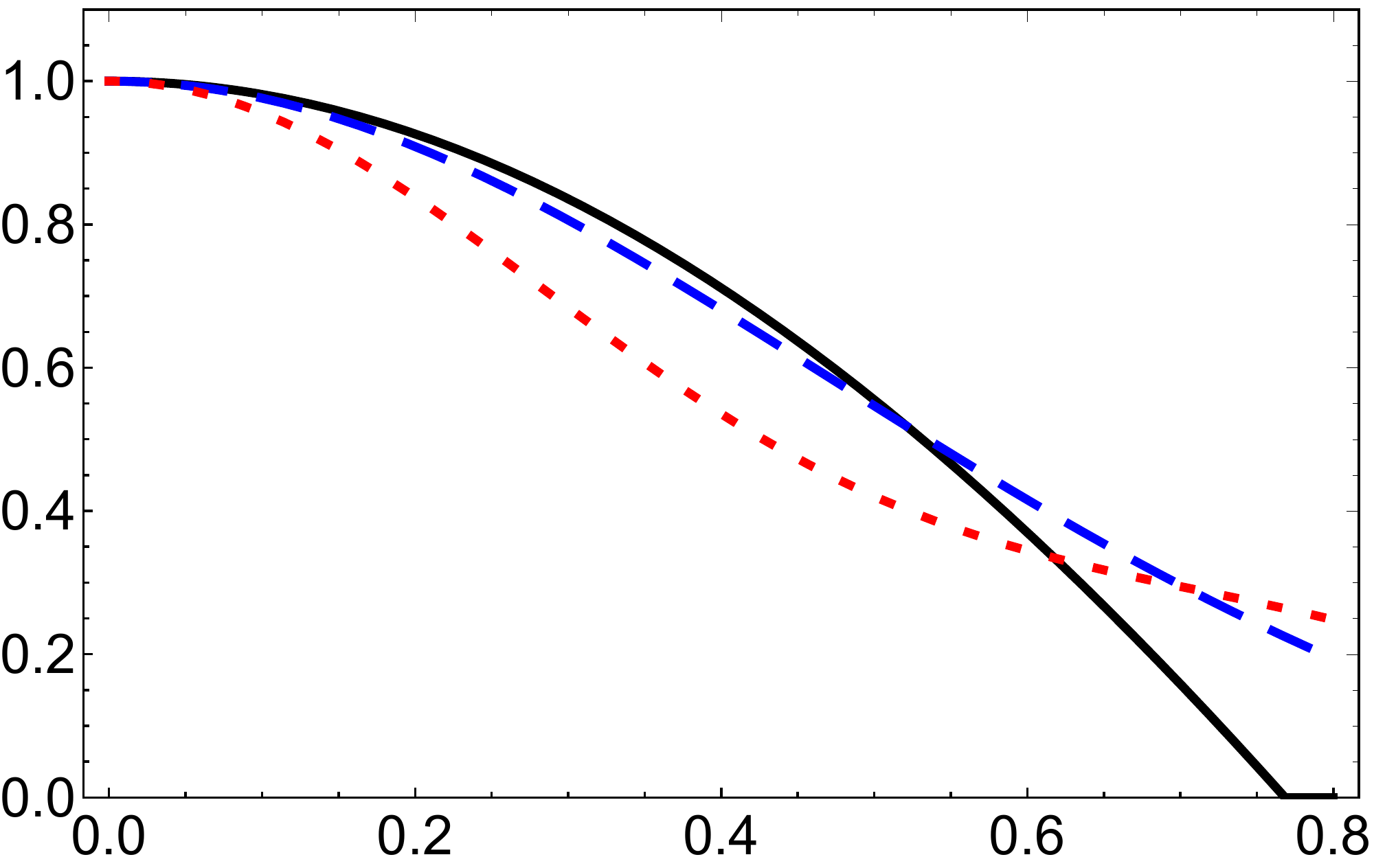}
\put(-160,100){($ b $)}\put(-160,60){$\mathcal{N}_{La}$}
	\put(-70,-15){$r$}~~~\quad\quad
			\includegraphics[width=0.3\textwidth, height=125px]{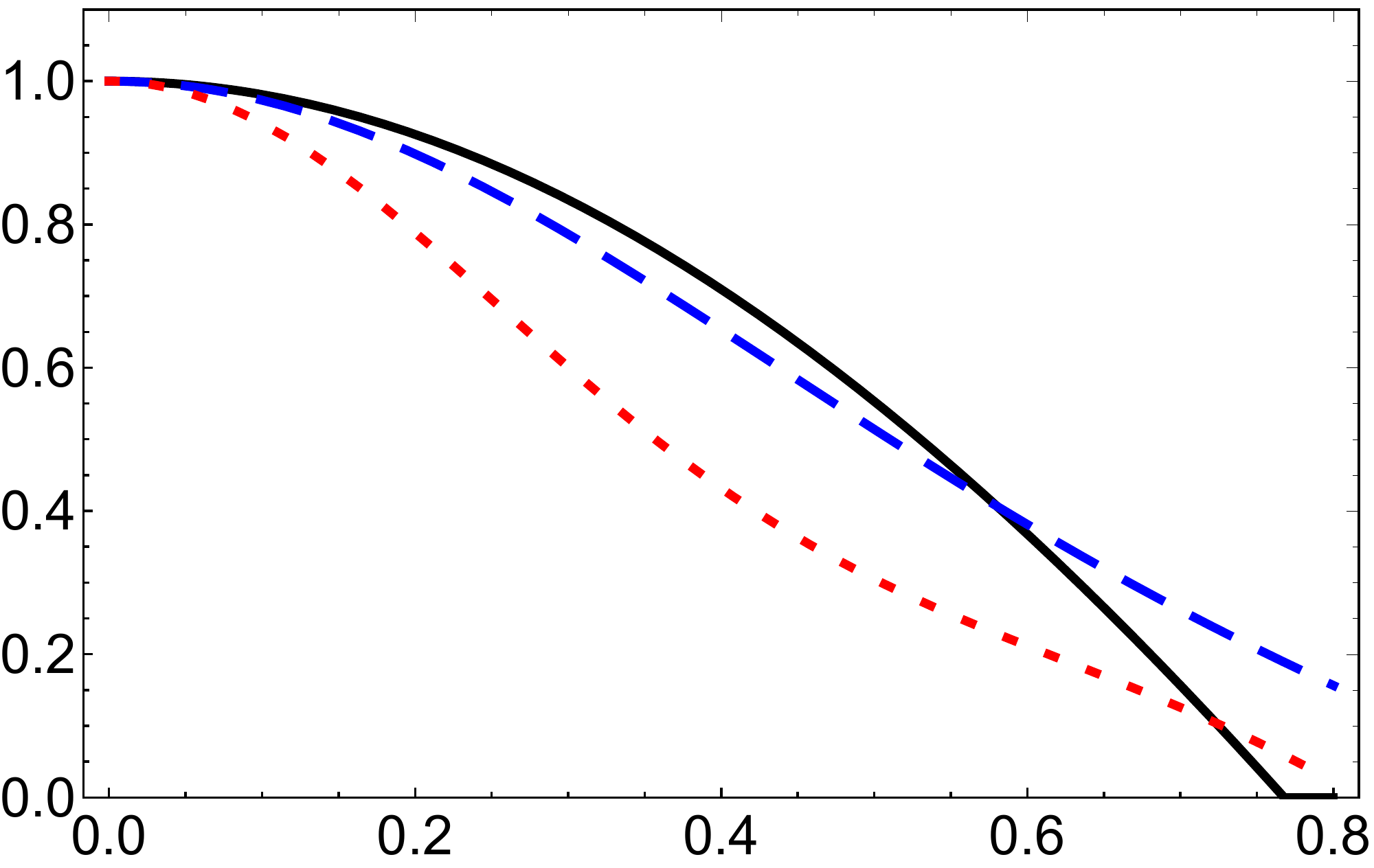}
\put(-160,100){($ c $)}\put(-160,60){$\mathcal{N}_{La}$}
	\put(-70,-15){$r$}~~~\quad\quad
	\end{center}
	\caption{\label{coo9}The same as Fig.(\ref{coo7}), but it is assumed that the non local symmetric operator $\mathcal{PT}$ is applied on both qubits and only the first qubit is accelerated.}
\end{figure}
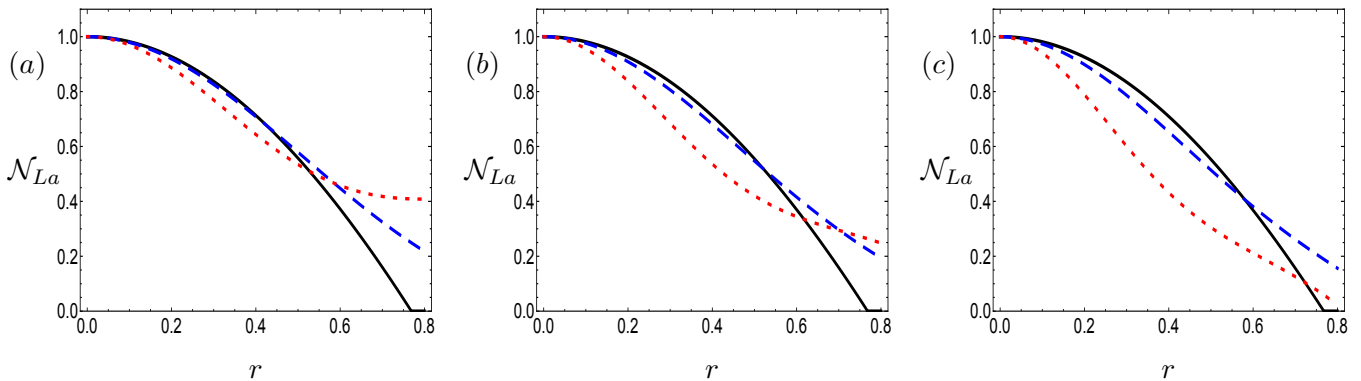

In Fig.(\ref{coo9}), it is assumed that the symmetric operator $\mathcal{PT}$  is  applied  on both qubits, where only one qubit is accelerated.  The general behavior shows that,   $\mathcal{N}_{La}$  decays as the acceleration increases.
However, at small interaction time, the decay strength induced from the acceleration is stronger than the improvement strength of the symmetric operator.  The efficiency of  symmetric-operator increases as one increases the interaction time, where  as one increases the angle settings of the operator, the non-local coherent advantage  decreases. Moreover, at $r=0$, the non-local advantage is maximum for all interaction times.

From Figs.(\ref{coo7}) and (\ref{coo9}), one may conclude that, the powerful of the local symmetric operator, $\mathcal{PT}$ is  clearly exhibited, where one can keep the non-local coherent advantage survival and consequently the efficiency of the accelerated state increases for any value  of the acceleration. By applying the local symmetric operator on both qubits, the  vanishing phenomena of $\mathcal{N}_{La}$ disappears. Moreover, at zero acceleration, the effect of the  symmetric operator is large, where the non-local coherent advantage is maximum and independent of the interaction time.

\begin{figure}[!h]
		\begin{center}
			\includegraphics[width=0.3\textwidth, height=125px]{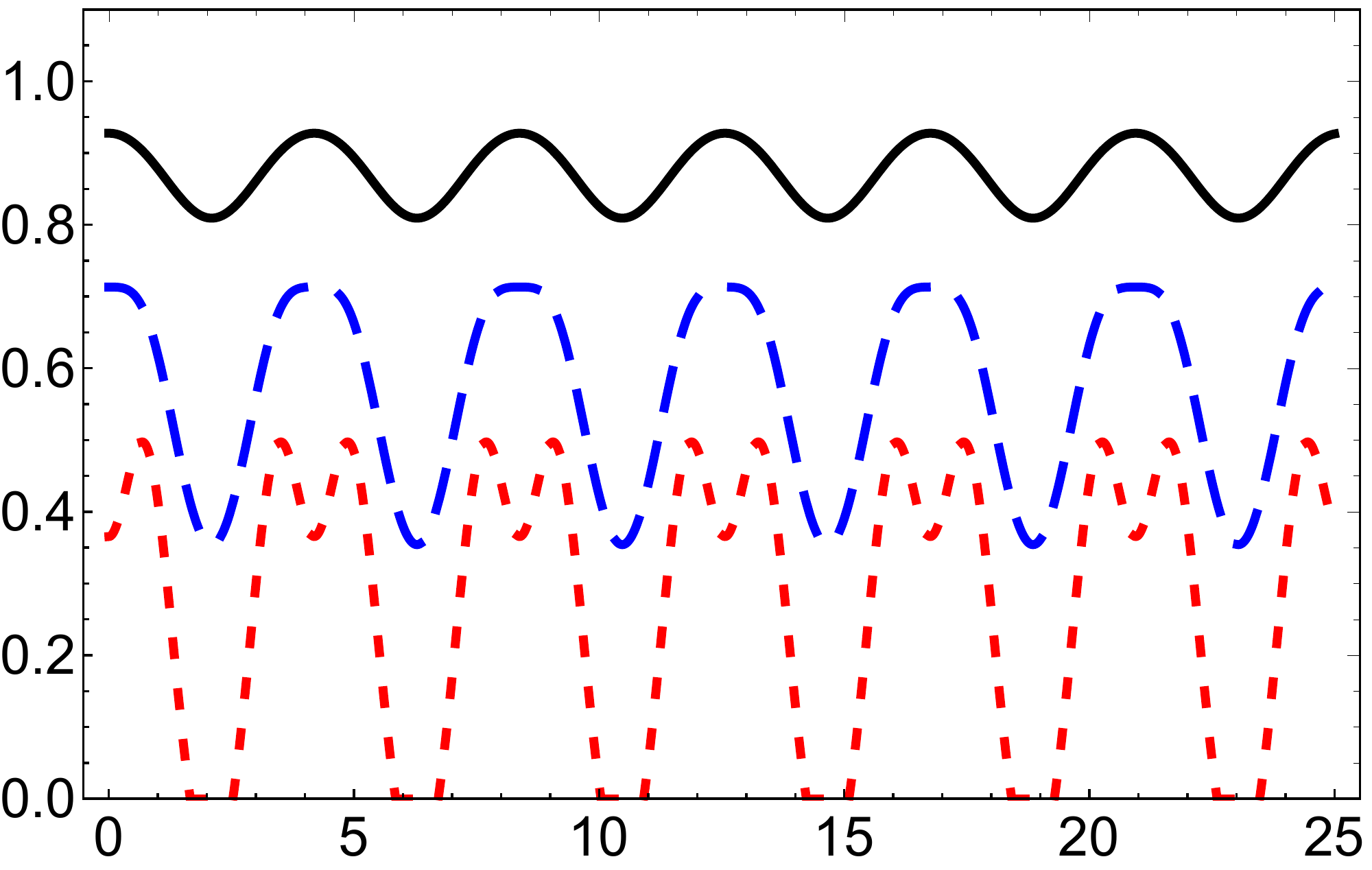}
\put(-160,100){($ a $)}\put(-160,60){$\mathcal{N}_{La}$}
	\put(-70,-15){$t$}~~~\quad\quad
		\includegraphics[width=0.3\textwidth, height=125px]{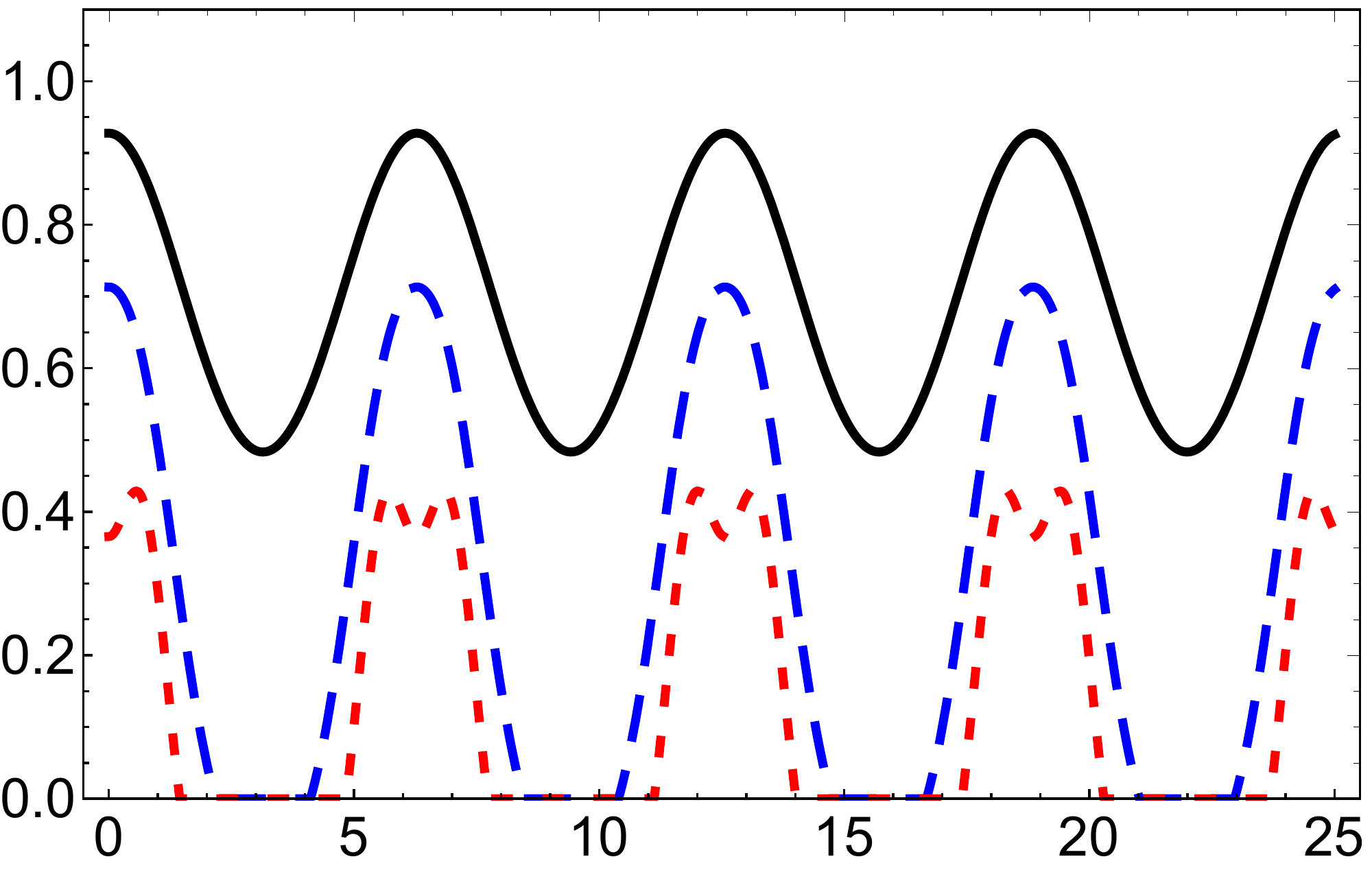}
\put(-160,100){($ b $)}\put(-160,60){$\mathcal{N}_{La}$}
	\put(-70,-15){$t$}~~~\quad\quad
			\includegraphics[width=0.3\textwidth, height=125px]{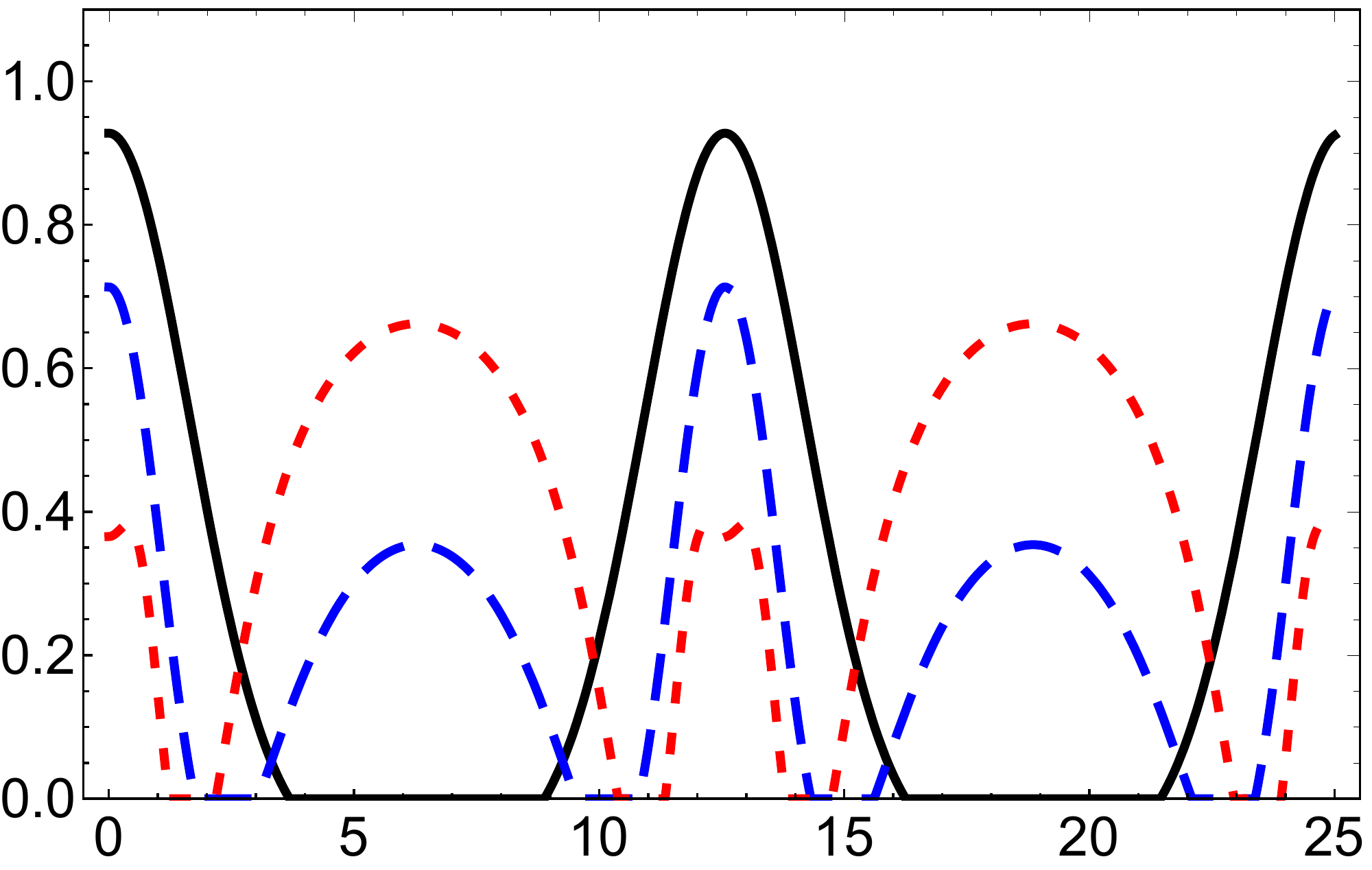}
\put(-160,100){($ c $)}\put(-160,60){$\mathcal{N}_{La}$}
	\put(-70,-15){$t$}~~~\quad\quad
\end{center}
		\caption{\label{coo10}The same as Fig.(\ref{coo8}), but it is assumed that, the non local symmetric operator $\mathcal{PT}$ is applied on both qubits and only the first qubit is accelerated.}
	\end{figure}
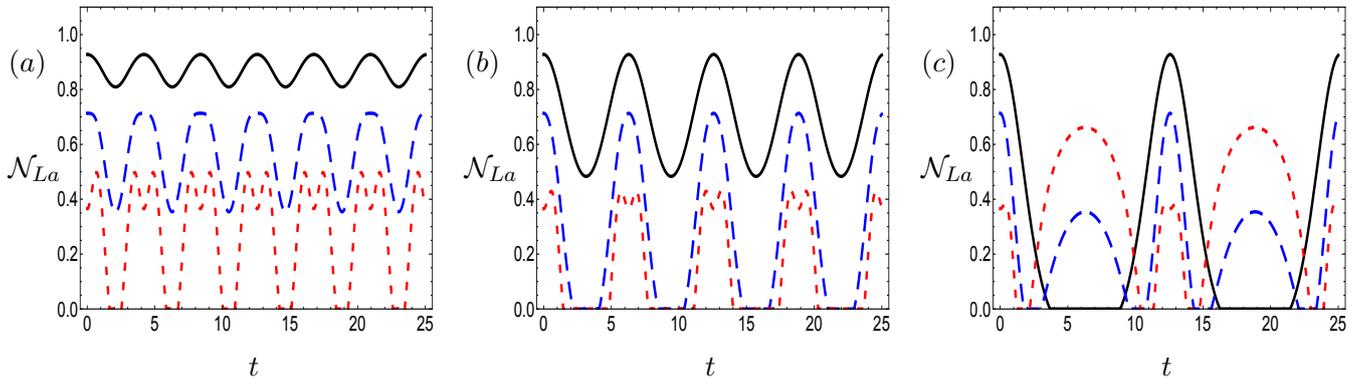

Fig.(\ref{coo10}) exhibits the improvements of the non-local advantage $\mathcal{N}_{La}$,  when the symmetric operator $\mathcal{PT}$ is applied on both  qubits. It is clear that, the behavior is similar to that displayed in Fig.(\ref{coo8}a), but  the number of oscillations is smaller  and  the amplitudes of these oscillations at small value of the interaction time, are smaller than those displayed in Fig.(\ref{coo8}). This behavior shows that, the non-local coherent advantage is improved. The disadvantage that predicted in this figure is  the vanishing  phenomena of $\mathcal{N}_{La}$ at large acceleration. Moreover, as it is displayed in Figs.(\ref{coo10}b) and (\ref{coo10}c), the vanishing time of the non-local coherent advantage increases as the strength of the  symmetric operator, $\mathcal{PT}$ increases, while,  the maximum values of $\mathcal{N}_{La}$ at large acceleration ($r=0.6$, $\alpha=\pi/3$)   exceeds their  initial values

From, Fig.(\ref{coo8}) and Fig.(\ref{coo10}) one may conclude that, the non-local advantage of the accelerated system  may be improved if the local symmetric operator is applied on both qubits  for  small acceleration. These improvements are depicted for   different aspects:  increasing the  rate of the non-local advantage during the interaction time, with  decreased   vanishing time., and  improvement  of the maximum values to exceeds their  initial values. However, the losses of the non-local advantage  are not only recovered but  may be increased  if the local symmetric operator is applied on both qubits.

\subsection{Both qubits are accelerated}\label{cohere5.2}
In this section, we investigate the effect of the symmetric-operator on the accelerated system, where it is assumed that both qubits are accelerated. Now, we discuss the possibility of recovering the losses of the non-local advantage due to the decoherence. Similarly to the previous section, we consider two cases,  one of them is assuming  that the symmetric operator applies only on one qubit, while for the second case, the symmetric operator is applied on both qubits.

\begin{figure}[!h]
	\begin{center}
		\includegraphics[width=0.3\textwidth, height=125px]{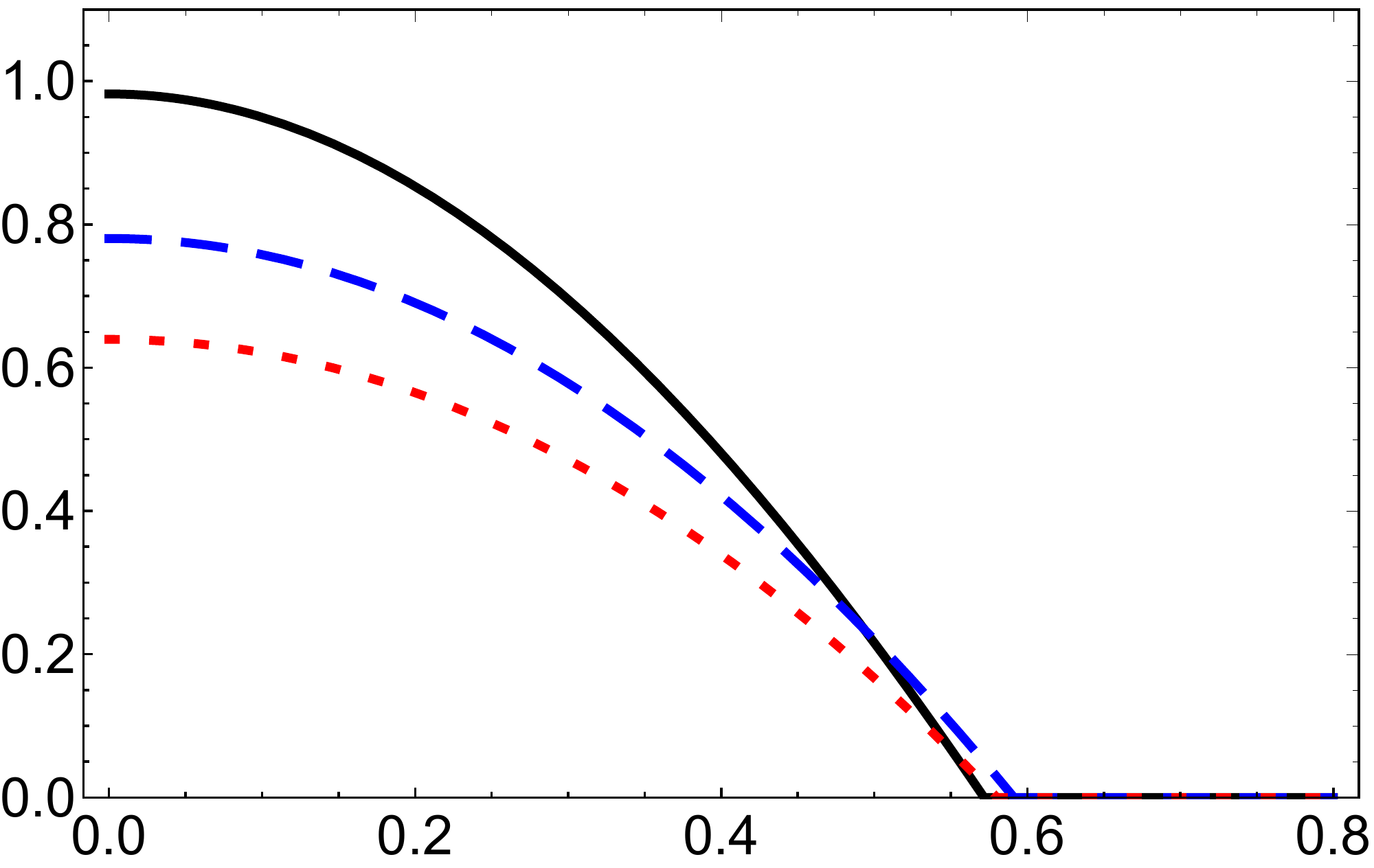}
\put(-160,100){($ a $)}\put(-160,60){$\mathcal{N}_{La}$}
	\put(-70,-15){$r$}~~~\quad\quad
		\includegraphics[width=0.3\textwidth, height=125px]{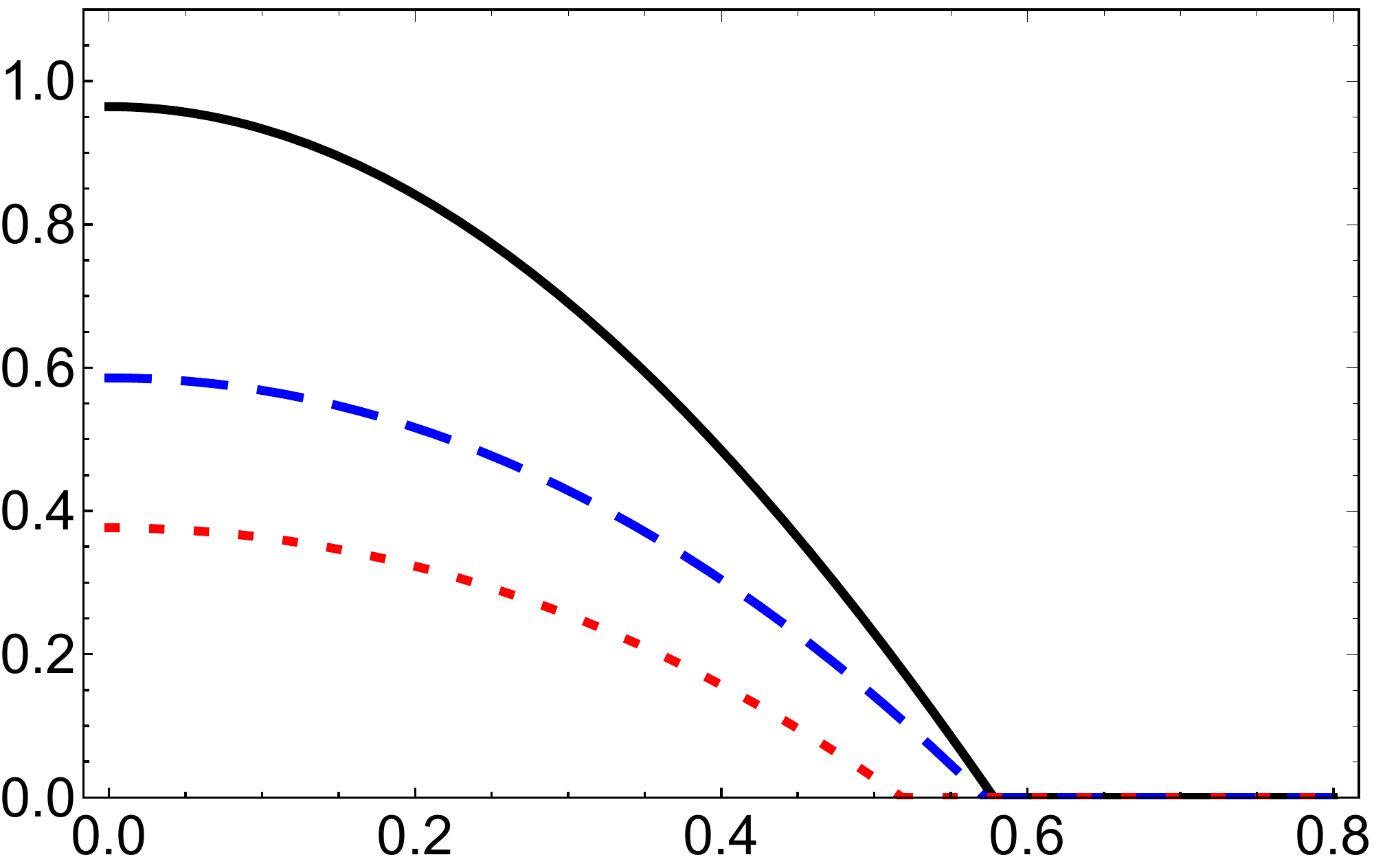}
\put(-160,100){($ b $)}\put(-160,60){$\mathcal{N}_{La}$}
	\put(-70,-15){$r$}~~~\quad\quad
			\includegraphics[width=0.3\textwidth, height=125px]{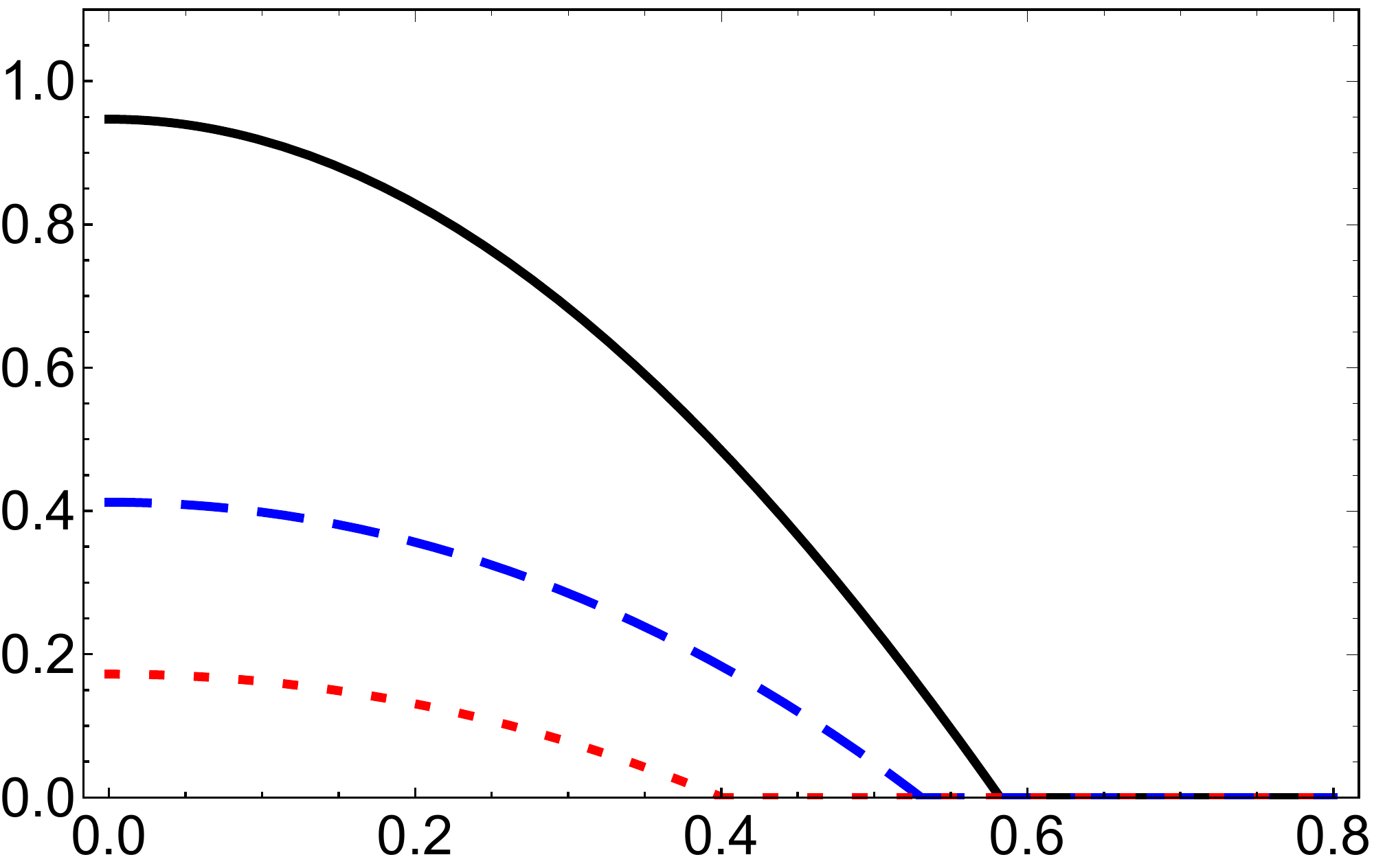}
\put(-160,100){($ c $)}\put(-160,60){$\mathcal{N}_{La}$}
	\put(-70,-15){$r$}~~~\quad\quad
	\end{center}
	\caption{\label{coo11}The  behavior of the non-local advantage, $\mathcal{N}_{La}$, when both qubit are accelerated and the local symmetric  operator is applied on  only one qubits. The solid, dash and dot are evaluated $t=0.1,0.4$ and 0.9., where the strength of the operator is (a) $\alpha=\frac{\pi}{6}$, (b) $\alpha=\frac{\pi}{4}$ and (c) $\alpha=\frac{\pi}{3}$.}
\end{figure}
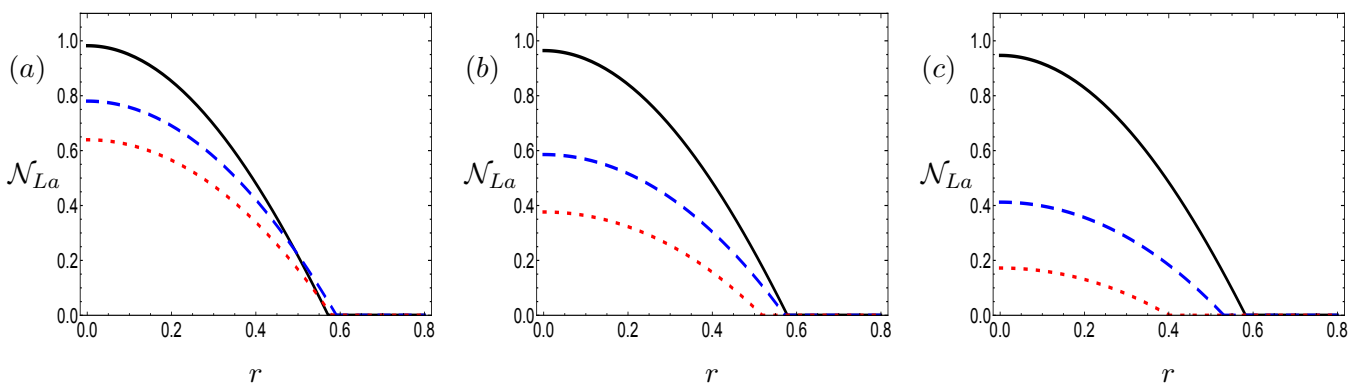

 Fig.(\ref{coo11}),  shows the possibility of improving the non-local coherent advantage $\mathcal{N}_{La}(r)$, when both qubits are accelerated at different interaction time. The behavior is similar to that displayed in Fig.(\ref{coo7}), where only one qubit is accelerated. However, the decay is much larger than that  displayed in Fig.(\ref{coo7}). The effect of the  symmetric operator $\mathcal{PT}$, is displayed at different values of the operator's strength. It is clear that, at small values of $\alpha$, the non-local coherent advantage, decays gradually  and vanishes at large  values of interaction time.  However, as one increases the operator' strength ($\alpha$), the maximum values decrease  and the decay rate increases. Moreover, the non-local  advantage vanishes at small  values of  the acceleration.  As it is displayed, in Fig.(\ref{coo11}c),  $\mathcal{N}_{La}(r)$ decays fast, with  maximum values are much smaller than those displayed at small values of $\alpha$.

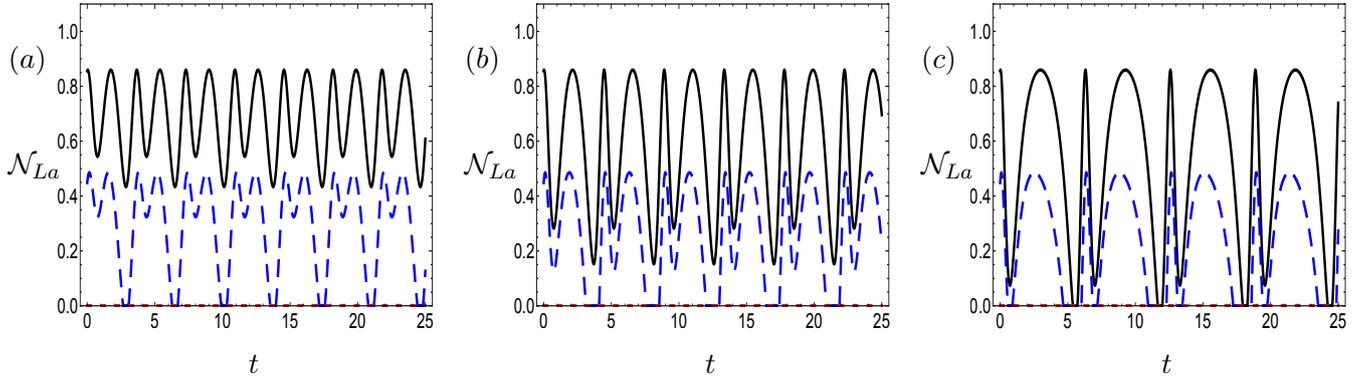
\begin{figure}[h!]
	\begin{center}
		\includegraphics[width=0.3\textwidth, height=125px]{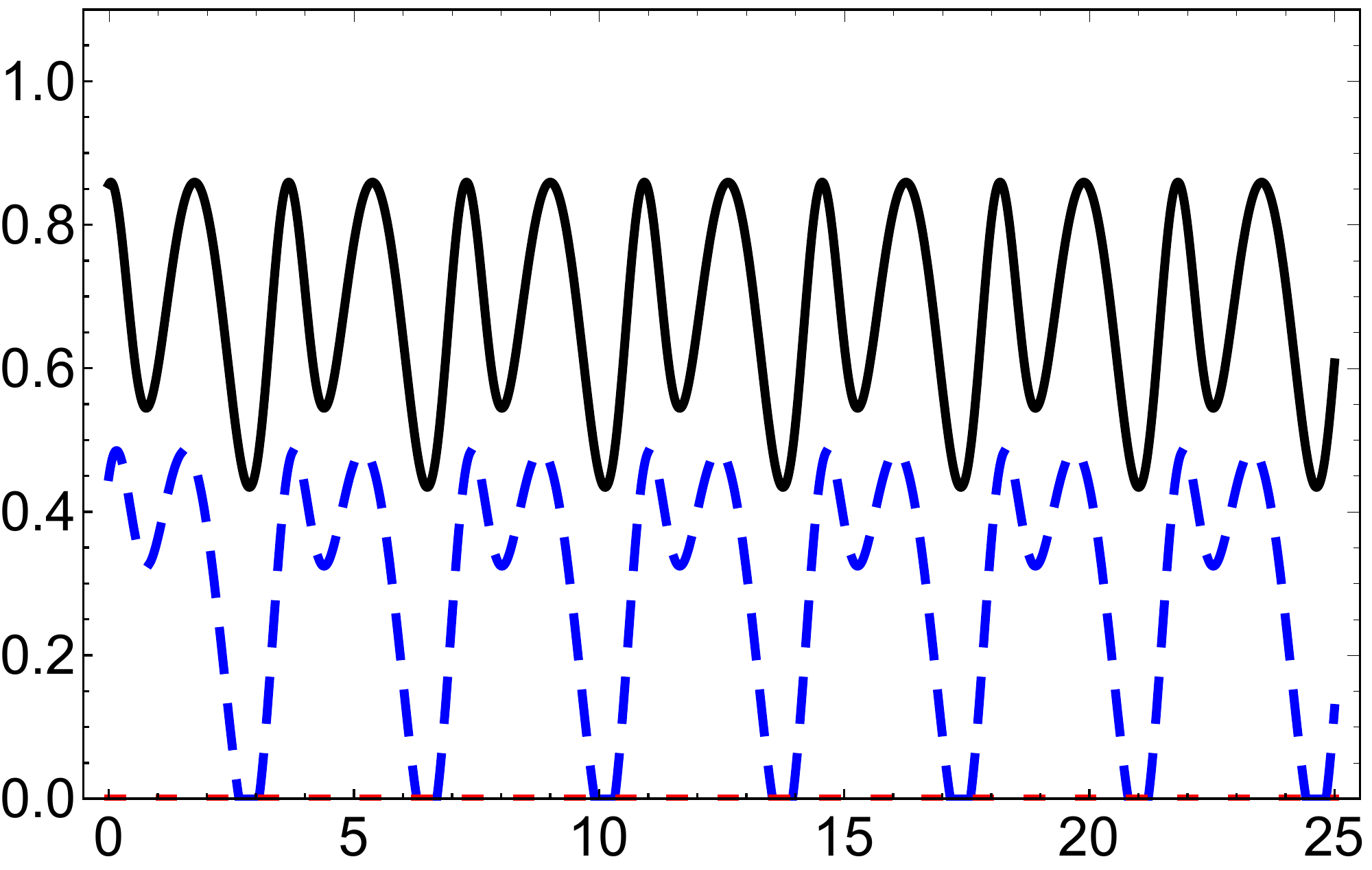}
\put(-160,100){($ a $)}\put(-160,60){$\mathcal{N}_{La}$}
	\put(-70,-15){$t$}~~~\quad\quad
		\includegraphics[width=0.3\textwidth, height=125px]{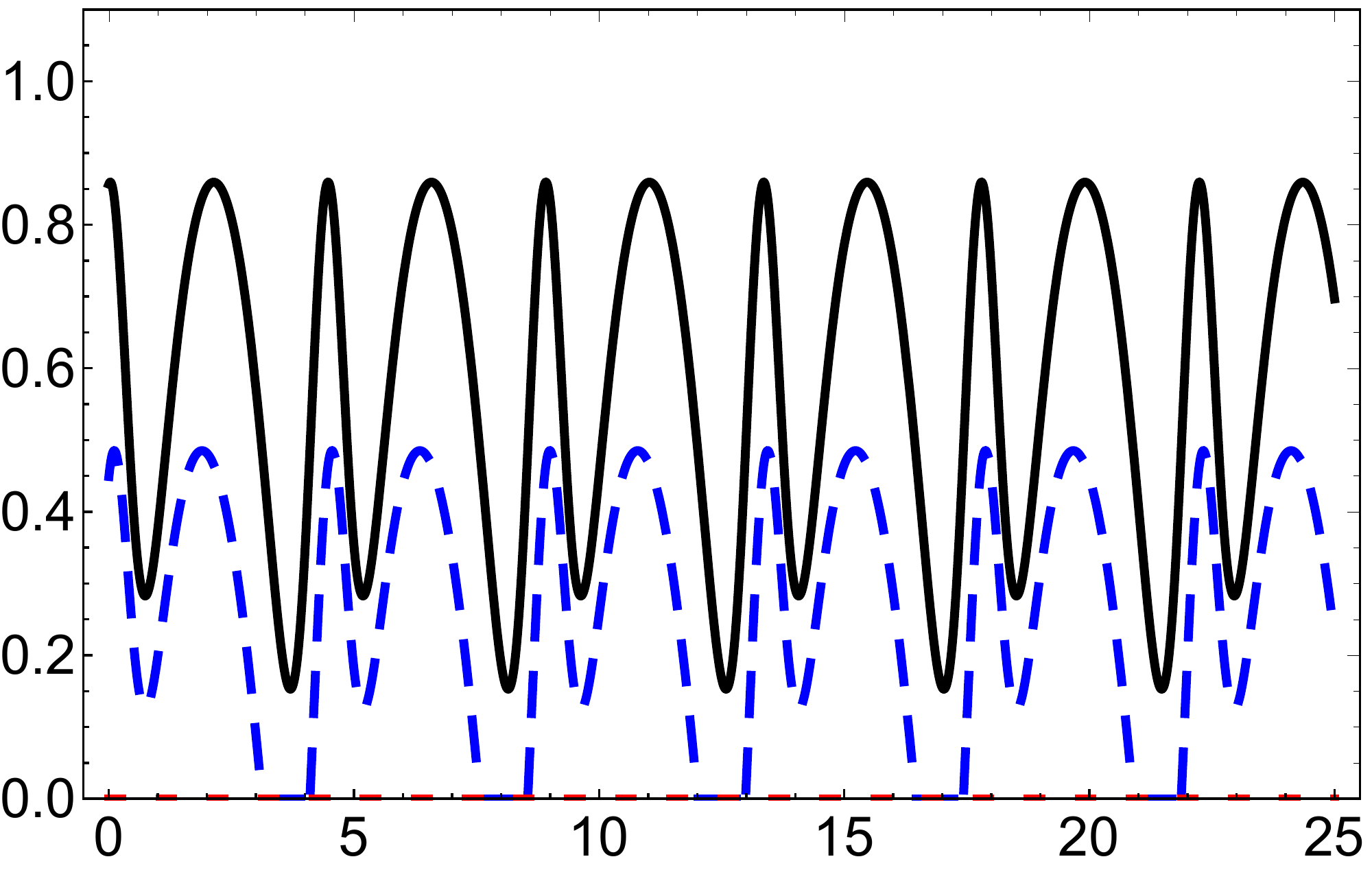}
\put(-160,100){($ b $)}\put(-160,60){$\mathcal{N}_{La}$}
	\put(-70,-15){$t$}~~~\quad\quad
		\includegraphics[width=0.3\textwidth, height=125px]{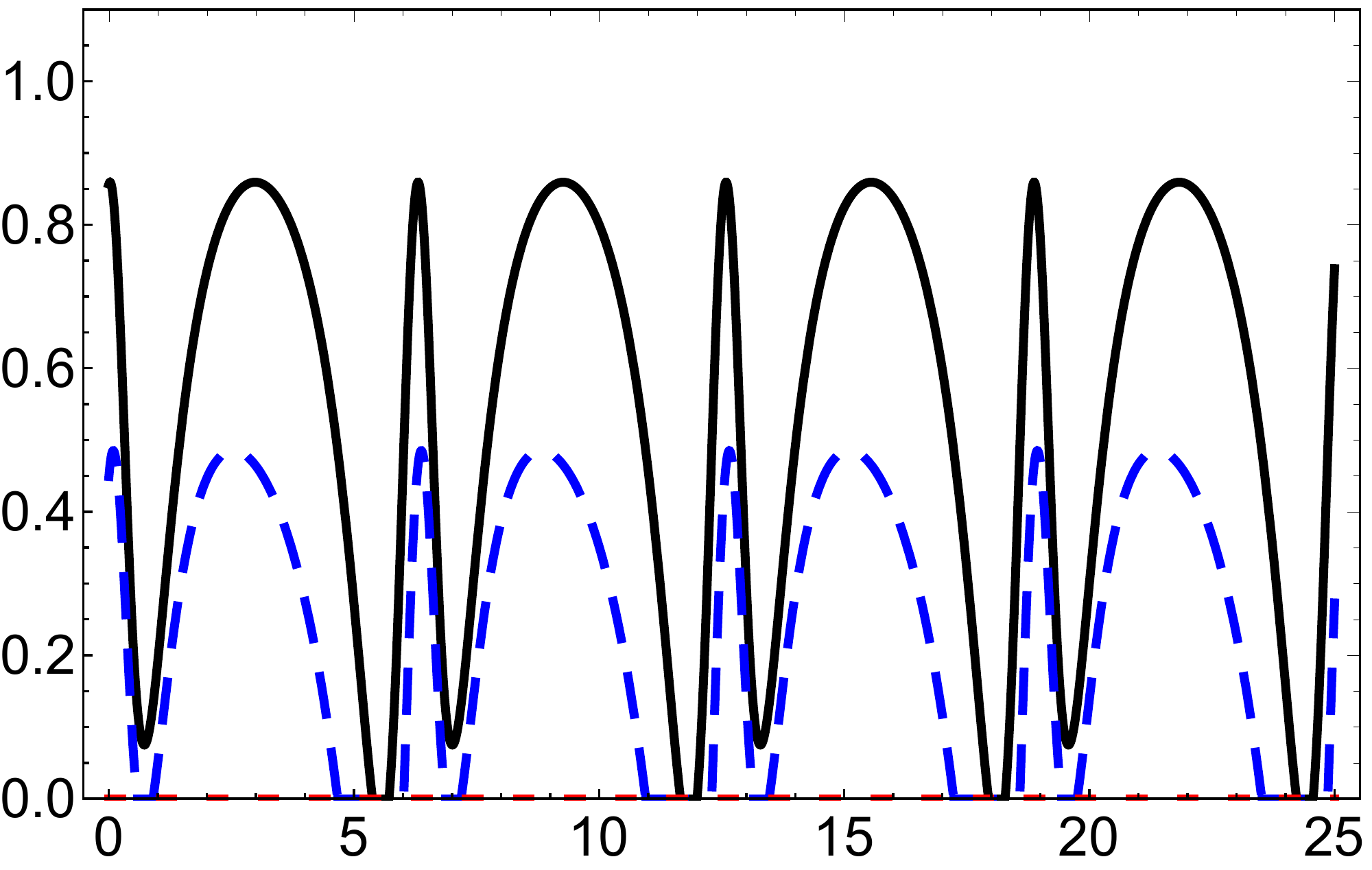}
\put(-160,100){($ c $)}\put(-160,60){$\mathcal{N}_{La}$}
	\put(-70,-15){$t$}~~~\quad\quad
	\end{center}
	\caption{\label{coo12}The behavior of the non-local advantage, $\mathcal{N}_{La}(t)$ when the symmetric operator is applied on both qubit. The sold, dash and dot curves  are evaluated at $r=0.2,0.4$ and $0.6$, respectively. The strength of the symmetric -operator is (a) $\alpha=\frac{\pi}{6}$, (b) $\alpha=\frac{\pi}{4}$ and (c) $\alpha=\frac{\pi}{3}$.}
\end{figure}

The behavior of $\mathcal{N}_{La}(t)$ is displayed in Fig.(\ref{coo12}), where  different values of the initial acceleration are considered. The fluctuations behavior of the non-local coherent advantage is displayed, where the number of  oscillations decreases as the symmetric operator' strength $\alpha$ increases. The periodic behavior shows that the amplitude of the oscillations decreases as  $\alpha$ decreases and consequently, the minimum values are much large at small values of the operator' strength. Moreover,  as it is shown  in  Figs.(\ref{coo12}), the non-local advantage vanishes completely with large acceleration. Further, the vanishing time of $\mathcal{N}_{La}(t)$  increases as the  $\mathcal{PT}$-symmetric operator strength increases.

\begin{figure}
	\begin{center}
		\includegraphics[width=0.3\textwidth, height=125px]{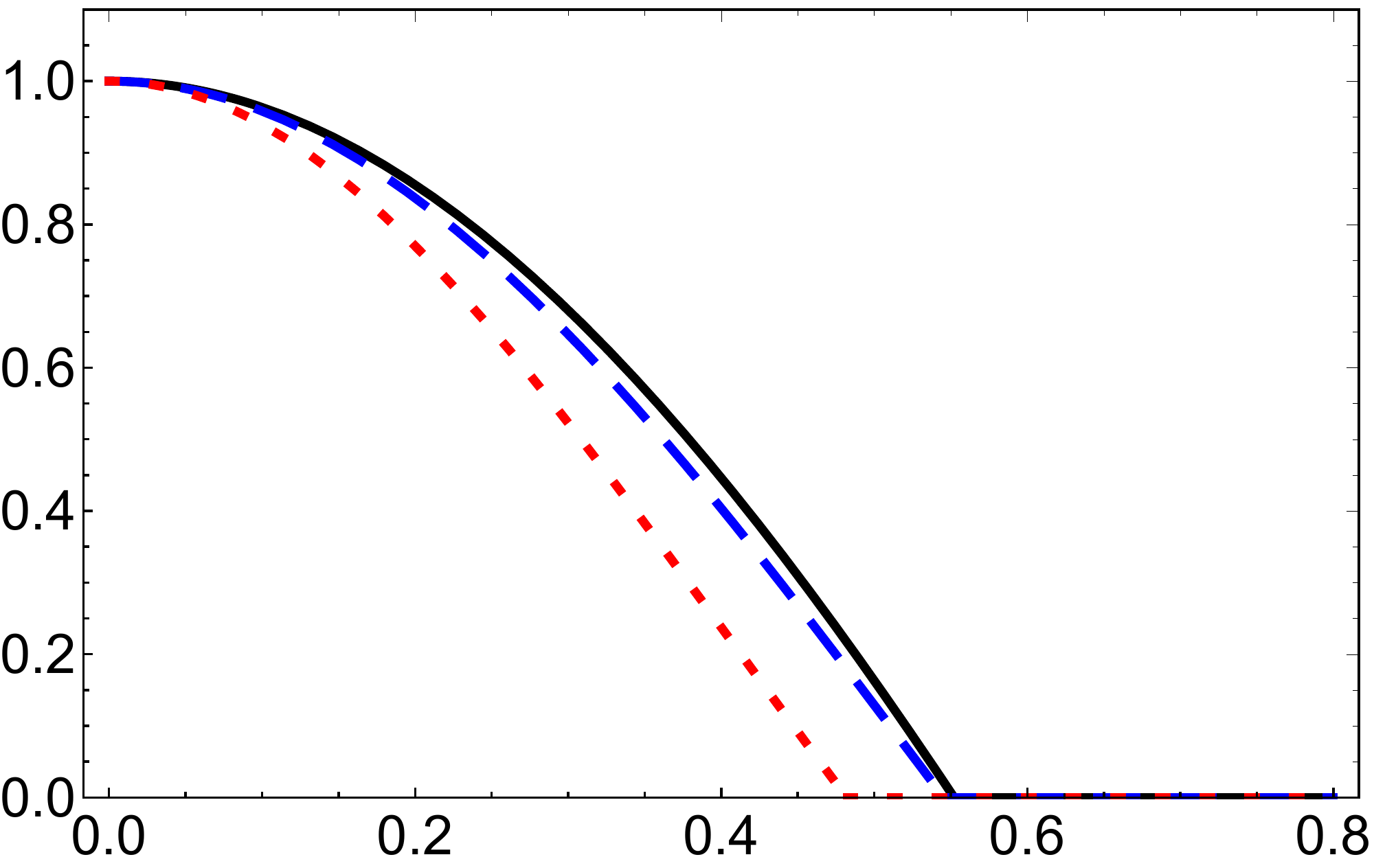}
\put(-160,100){($ a $)}\put(-160,60){$\mathcal{N}_{La}$}
	\put(-70,-15){$r$}~~~\quad\quad~~~~~~
	\includegraphics[width=0.3\textwidth, height=125px]{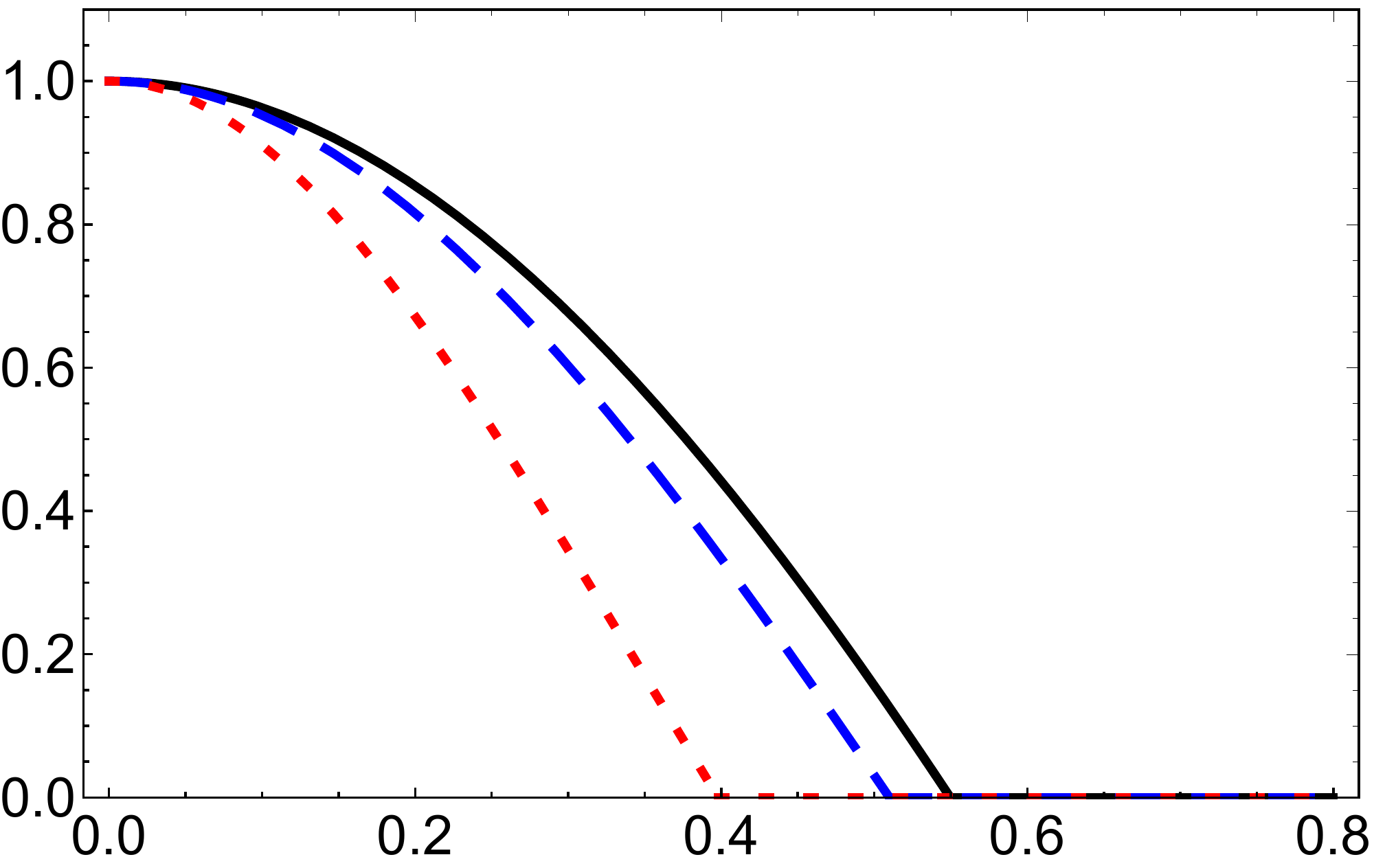}
\put(-160,100){($ b $)}\put(-160,60){$\mathcal{N}_{La}$}
	\put(-70,-15){$r$}~~~\quad\quad
		\includegraphics[width=0.3\textwidth, height=125px]{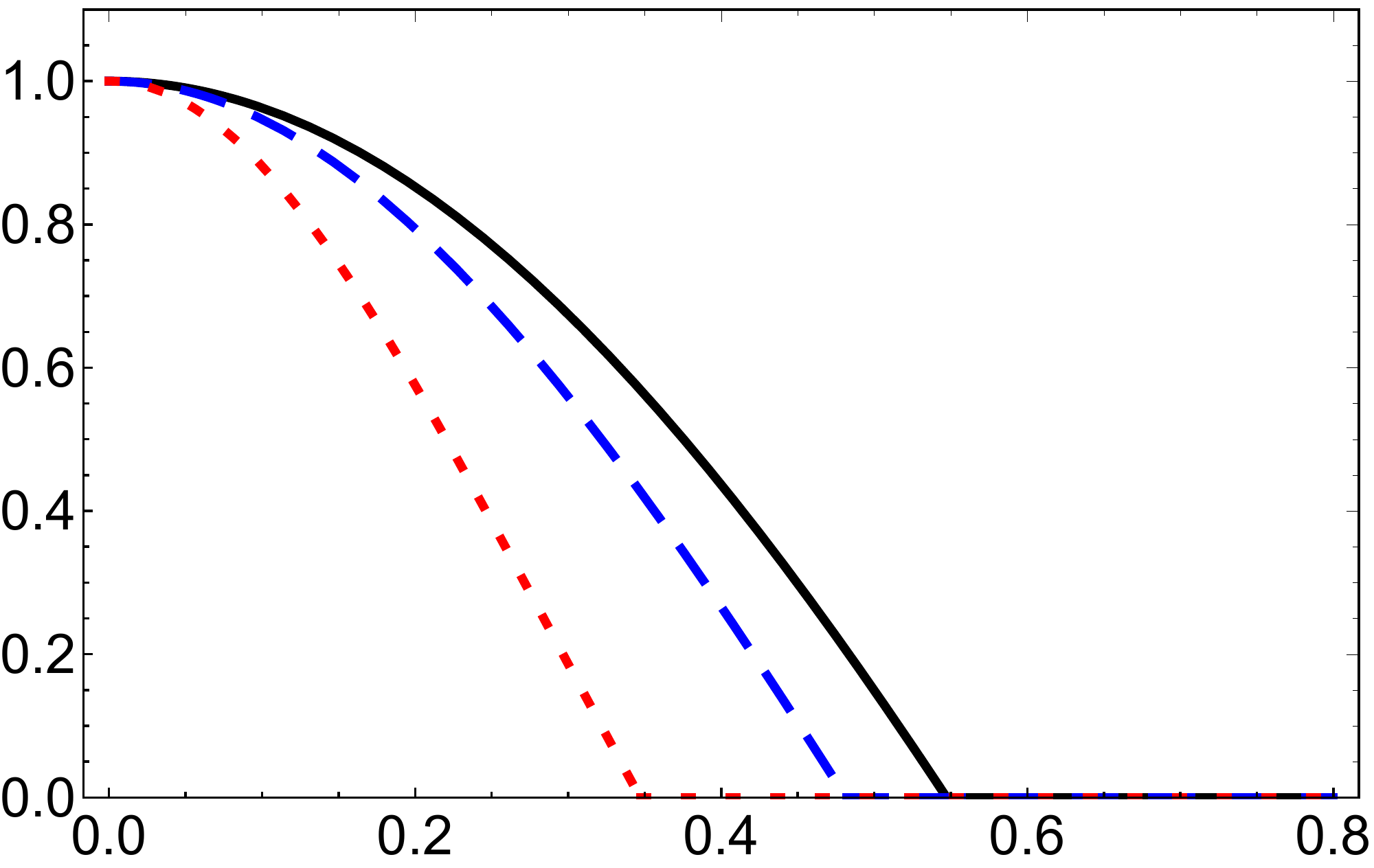}
\put(-160,100){($ c $)}\put(-160,60){$\mathcal{N}_{La}$}
	\put(-70,-15){$r$}~~~\quad\quad
	\end{center}
	\caption{\label{coo13}The same as Fig.(\ref{coo8}) but the symmetric operator is applied on both qubits.}
\end{figure}
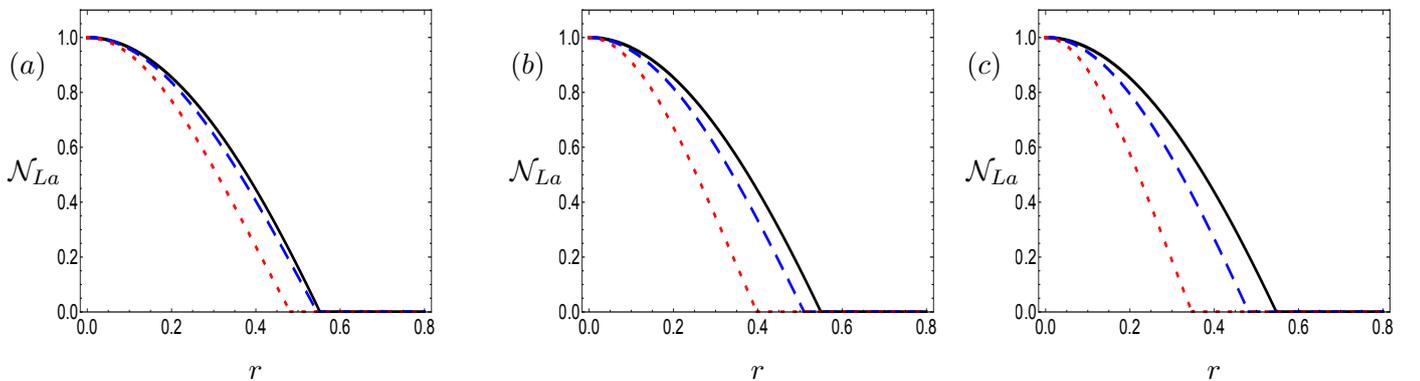
 Fig.(\ref{coo13}), we  shows the possibility of recovering the losses of the non-local coherent advantage, when the  $\mathcal{PT}$-symmetric operator is applied on both qubits, where different  values  of the initial strength are considered.  The results that displayed in this figure show that, $\mathcal{N}_{La}$ is much better than that depicted in Fig.(\ref{coo11}), where the maximum values of $\mathcal{N}_{La}$  are much better. However, as one increases the $\mathcal{PT}$-symmetric strength, the decay rate of $\mathcal{N}_{La}$  is larger than that  displayed at small values of $\alpha$. Moreover, the vanishing phenomena of $\mathcal{N}_{La}$  is depicted at small values of initial accelerations.

Figs.(\ref{coo11}) and (\ref{coo13}),  show that the possibility of imroving/recovering the non-local coherent advantage increases as the $\mathcal{PT}$-symmetric operator is applied on both qubits, where the maximum values are much larger. The only disadvantage that predicted  when applying the $\mathcal{PT}$ operator on both qubits, is that the non-local coherent quantum advantage  vanishes at small acceleration.

\begin{figure}[!h]
	\begin{center}
		\includegraphics[width=0.3\textwidth, height=125px]{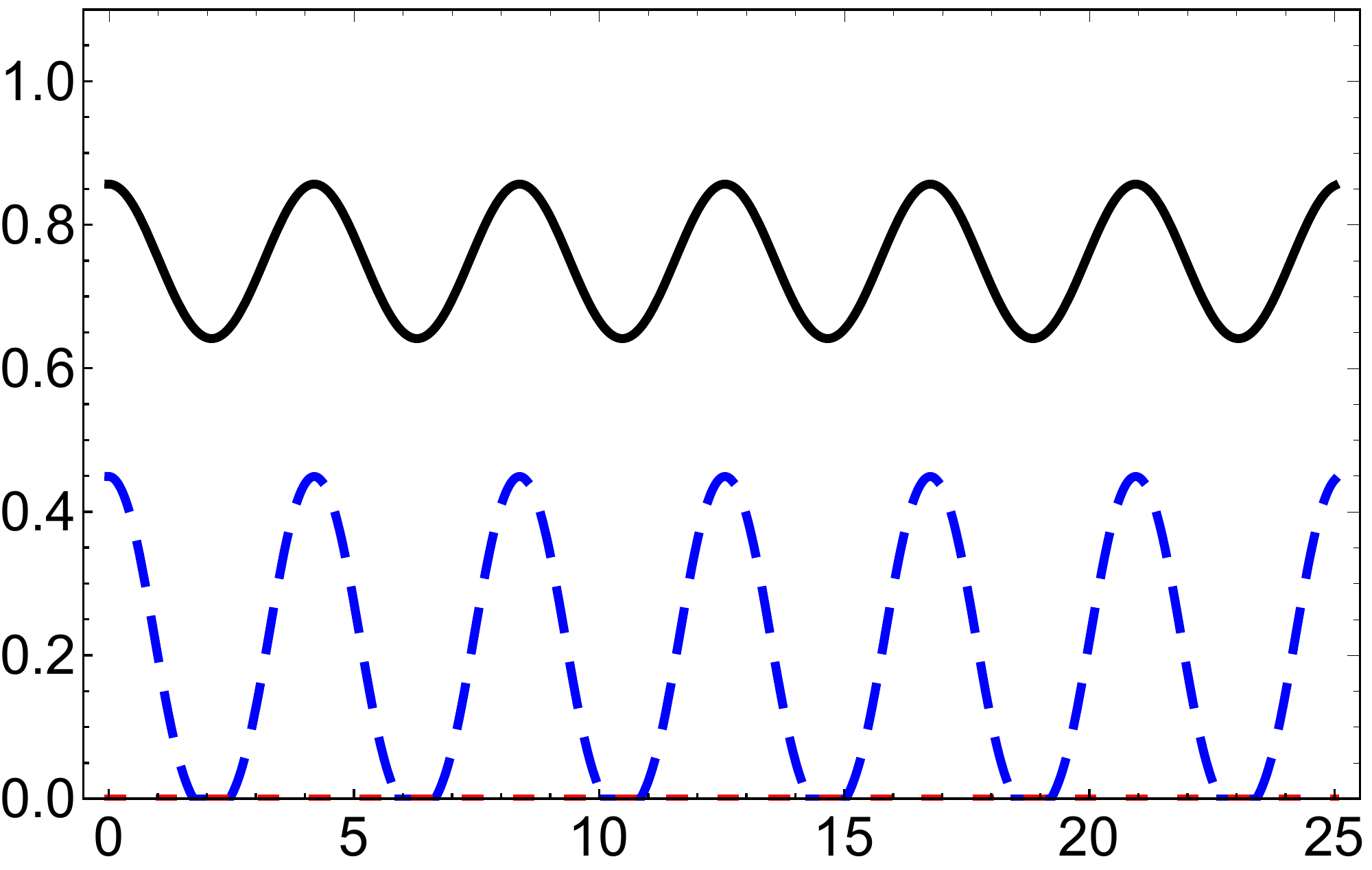}
\put(-160,100){($ a $)}\put(-160,60){$\mathcal{N}_{La}$}
	\put(-70,-15){$t$}~~~\quad\quad
	\includegraphics[width=0.3\textwidth, height=125px]{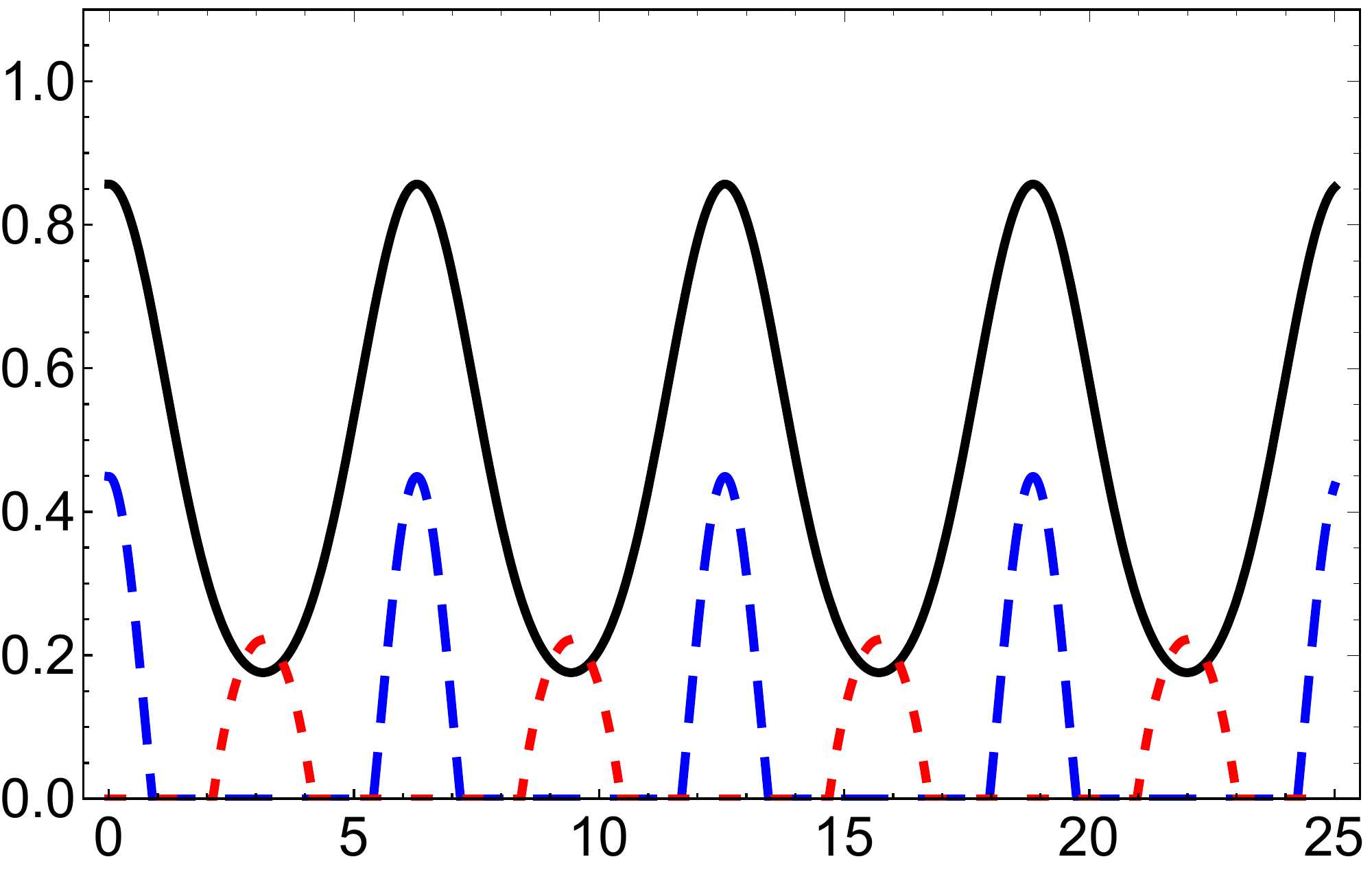}
\put(-160,100){($ b $)}\put(-160,60){$\mathcal{N}_{La}$}
	\put(-70,-15){$t$}~~~\quad\quad
			\includegraphics[width=0.3\textwidth, height=125px]{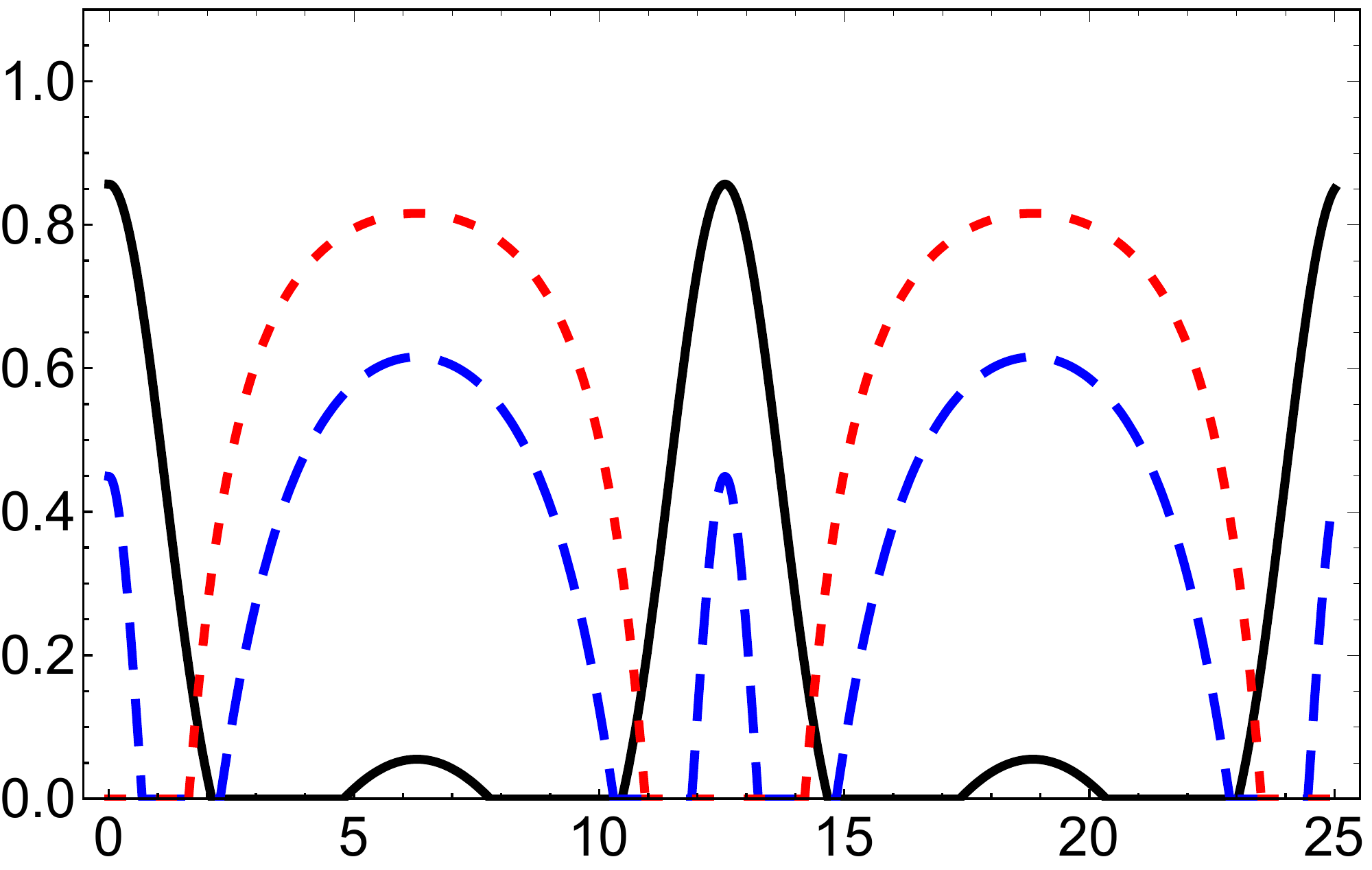}
\put(-160,100){($ c $)}\put(-160,60){$\mathcal{N}_{La}$}
	\put(-70,-15){$t$}~~~\quad\quad
	\end{center}
	\caption{\label{coo14}The same as Fig.(\ref{coo9}) but the symmetric operator is applied on both qubits.}
\end{figure}
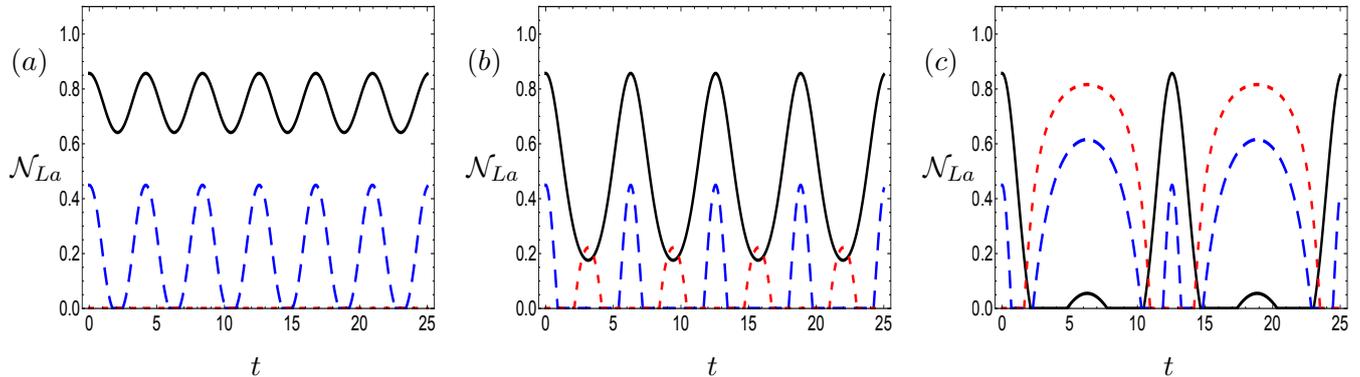
Fig.(\ref{coo14}) shows the behavior of  $\mathcal{N}_{La}(t)$, when the symmetric operator is applied on both qubits  at different accelerations. In general, the behavior is similar to that predicted in Fig.(\ref{coo12}), namely  $\mathcal{N}_{La}(t)$ oscillates between its maximum and minimum values. However,  as it is displayed in Fig.(\ref{coo14}), the amplitude of oscillations are smaller than those shown in Fig.(\ref{coo12}), which means that the non-local coherent advantage is improved.  The most important remark is that at large accelerations, $\mathcal{N}_{La}(t)$,  vanishes completely at small values of the strength $\alpha$. However, as the strength increases, the non-local coherent advantage re-births again and its maximum values increase as the strength of the symmetric operator $\alpha$  increases.

\section{Conclusion}\label{cohere6}

In this contribution, we investigate the possibility of recovering the loses of the entanglement and the non-local advantage  for  accelerated system initially prepared in the Bell state. As it is well known,  due to the acceleration, the entangled properties of the accelerated system loses its coherence and consequently, their efficiency to perform some quantum information tasks decrease.  One of the most important phenomena of the entanglement system, is the  non-local  coherent advantage of the accelerated system. We have used the local symmetric operator to improve/ recover the coherence of the non-local coherent advantage. In this context, two different cases are considered;  one (both) qubits are accelerated and the symmetric operator is applied on  one or both subsystems.

Due to the acceleration process, the entanglement, as quantified by means of  its  negativity, decreases. Possibility of improving the entanglement is  suggested   by using the symmetric operator. It is shown that, it can be increase instantaneously at small values of the symmetric-operator strength, where the  amplitudes of the negativity oscillations decrease, and consequently its minimum values increase.
The effect of the symmetric-operator on the accelerated systems is much better when it is applied on both qubits during the interaction.
 Moreover, this effect increases, when the symmetric operator is applied on both qubits. On the other hand, the possibility of improving and recovering  the losses of entanglement increase, if only one qubit  is  accelerated and the symmetric-operator is  applied on both
qubits.

The behavior of the non-local coherent advantage decays as the acceleration increases and the decay rate is large when both qubit are accelerated. We examined  the behavior of this  physical phenomena  at different interaction time, where different initial strength of the symmetric operator are considered.  It is shown that, as the symmetric operator is applied on only one qubit, the maximum values of  $\mathcal{N}_{La}(r)$  increase and  vanish   at large acceleration. The periodic behavior of the non-local coherent advantage is predicated, where the  periodic time increases at small values of the operator strength. In addition, also $\mathcal{N}_{La}(t)$, vanishes periodically, where the disappearing interval time  increases as the initial acceleration increases.  However, the small  values of the strength of the  symmetric operator recover the losses of the $\mathcal{N}_{La}(t)$ and  prevent its  disappearance. Although the non-local coherent advantage fluctuates fast, the amplitudes of the oscillations are small, and in turn its minimum values are improved. Moreover, the improvement of the non-local coherent advantage is clearly displayed  when only one qubit is accelerated and the local symmetric operator is applied on both qubits, where at small values of the symmetric strength  $\mathcal{N}_{La}(r)$ increase  and never vanishes.

The effect of the symmetric operator  on the non-local coherent advantage when both qubits are accelerated is discussed. Due to the accelerating process the decay rate of   $\mathcal{N}_{La}(r)$ is large.  However, if the symmetric operator is applied  on only one qubit, the $\mathcal{N}_{La}(r)$ increases  as  one decreases the  strength of the symmetric operator. Similarly, the periodic behavior of  $\mathcal{N}_{La}(t)$  is predicated with  small amplitudes of fluctuations. These means that, the minimum values are improved. The time periodicity  time increases  as the strength of $\mathcal{PT}$ increases, which means  that the non-local coherent advantage survives for a longer time.

Moreover, the effect of the symmetric operator on both qubits improve and recover the losses of   $\mathcal{N}_{La}(r)$, where, it reaches its maximum value even at large acceleration. Therefore, if the initial acceleration is large, the decay of
 $\mathcal{N}_{La}(r)$ can be less  by decreasing the symmetric operator strength. The amplitudes of oscillations reduce and consequently  the minimum values of $\mathcal{N}_{La}(t)$ are improved. One of the important  results,  we obtained is:  by applying the symmetric operator on both qubits, the phenomena of re-birthing the non-local coherent advantage  appears. Further, the  maximum values of the re-birthed non-local coherent advantage is much better than that obtained for small acceleration.

In conclusion, we  examined the possibility of improving and recovering the losses of entanglement and the non-local coherent advantage   by using the local symmetric-operator. The improvement efficiency may be increased by applying the symmetric operator on both qubits. The recovering process  of both phenomenon   is   exhibited  clearly when only one qubit is accelerated and the symmetric operator is applied on both qubits. It is shown that for  large acceleration,  the non-local coherent advantage may be re-birthed by using this symmetric operator

\section*{Acknowledgments}
We would like to thank the referees for their important remarks which helped us to
improve our manuscript.

‏

\begin{thebibliography}{99}
	\bibitem{cohe1}
Engel, G. S., Calhoun, T. R., et al. (2007). Evidence for wavelike energy transfer through quantum coherence in photosynthetic systems. Nature, 446(7137), 782.
\bibitem{cohe2}
Romero, E., Augulis, R., et al. (2014). Quantum coherence in photosynthesis for efficient solar-energy conversion. Nature physics, 10(9), 676.‏
\bibitem{cohe3}
Lostaglio, M., Jennings, D. and Rudolph, T. (2015). Description of quantum coherence in thermodynamic processes requires constraints beyond free energy. Nature communications, 6, 6383.
\bibitem{cohe4}
Lostaglio, M., Korzekwa, K., Jennings, D. and Rudolph, T. (2015). Quantum coherence, time-translation symmetry, and thermodynamics. Physical review X, 5(2), 021001.
\bibitem{cohe5}
Baumgratz, T., Cramer, M. and Plenio, M. B. (2014). Quantifying coherence. Physical review letters, 113(14), 140401.
\bibitem{cohe6}
Pires, D. P., Céleri, L. C. and Soares-Pinto, D. O. (2015). Geometric minimum bound for a quantum coherence measure. Physical Review A, 91(4), 042330.‏
\bibitem{cohe7}
Streltsov, A., Singh, U., Dhar, H. S., Bera, M. N. and Adesso, G. (2015). Measuring quantum coherence with entanglement. Physical review letters, 115(2), 020403.‏
\bibitem{cohe8}
Rana, S., Parashar, P. and Lewenstein, M. (2016). Trace-distance measure of coherence. Physical Review A, 93(1), 012110.‏
\bibitem{cohe9}
Hu, M. L. and Fan, H. (2018). Nonlocal advantage of quantum coherence in high-dimensional states. Physical Review A, 98(2), 022312.
\bibitem{cohe10}
Birrell, N. D., Birrell, N. D., Davies, P. C. W. and Davies, P. (1984). Quantum fields in curved space (No. 7). Cambridge university press.
\bibitem{cohe11}
Metwally, N. (2013). Usefulness classes of traveling entangled channels in noninertial frames. International Journal of Modern Physics B, 27(28), 1350155.
\bibitem{cohe12}
Metwally, N. and Sagheer, A. (2014). Quantum coding in non-inertial frames. Quantum information processing, 13(3), 771-780.
\bibitem{cohe13}
Bender, C. M. and Boettcher, S. (1998). Real spectra in non-Hermitian Hamiltonians having P T symmetry. Physical Review Letters, 80(24), 5243.
\bibitem{cohe14}‏‏‏‏
Chen, S. L., Chen, G. Y. and Chen, Y. N. (2014). Increase of entanglement by local $\mathcal{PT}$-symmetric operations. Physical Review A, 90(5), 054301.
\bibitem{cohe15}‏‏‏‏
Guo, Y. N., Fang, M. F., Wang, G. Y., Hang, J. and Zeng, K. (2017). Enhancing parameter estimation precision by non-Hermitian operator process. Quantum Information Processing, 16(12), 301.
\bibitem{cohe17}
 Martín-Martínez, E. and Leon, J. (2010). Quantum correlations through event horizons: Fermionic versus bosonic entanglement. Physical Review A, 81(3), 032320.
 \bibitem{cohe18}
Alsing, P. M., Fuentes-Schuller, I., Mann, R. B. and Tessier, T. E. (2006). Entanglement of Dirac fields in noninertial frames. Physical Review A, 74(3), 032326.

\bibitem{PhysRevLett.91.180404}
Alsing, Paul M. and Milburn, G. J., Teleportation with a Uniformly Accelerated Partner, Phys. Rev. A.{\bf91} 18, 180404 (2003).
\bibitem{cohe23}
Vidal, G. and Werner, R. F. (2002). Computable measure of entanglement. Physical Review A, 65(3), 032314.‏
 \bibitem{cohe21}
R. Horodecki, M. Horodecki and P. Horodecki, Phys. Lett. A 222 (1996) 1;
 K. Zyczkowski, P. Horodecki, A. Sanpera and M. Lewenstein, Phys. Rev. A 58 (1998) 883.
\bibitem{cohe22}
Mohammed A.R. and El-Shahat T.M. (2017) . Study the Entanglement Dynamics of an
Anisotropic Two-Qubit Heisenberg XYZ System in a Magnetic Field. Journal of Quantum”
Information Science, 7, 160.
 \bibitem{cohe19}
Mondal, D., Pramanik, T. and Pati, A. K. (2017). Nonlocal advantage of quantum coherence. Physical Review A, 95(1), 010301.
 \bibitem{cohe20}
 Hu, M. L., Wang, X. M. and Fan, H. (2018). Hierarchy of the nonlocal advantage of quantum coherence and Bell nonlocality. Physical Review A, 98(3), 032317.



 ‏‏‏
‏‏
\end{thebibliography}
\end{document}